\documentclass[12pt]{iopart}

\usepackage{graphicx}

\begin{document}


\thispagestyle{empty}

\begin{flushright}
CERN-PH-TH/2005-263\\
hep-ph/0512253
\end{flushright}

\vspace{1.6truecm}
\begin{center}
\boldmath
\large\bf Highlights of the $B$-Physics Landscape
\unboldmath
\end{center}

\vspace{0.9truecm}
\begin{center}
Robert Fleischer\\[0.1cm]
{\sl CERN, Department of Physics, Theory Division\\
CH-1211 Geneva 23, Switzerland}
\end{center}

\vspace{1.3truecm}

\begin{center}
{\bf Abstract}
\end{center}

{\small
\vspace{0.2cm}\noindent
The exploration of the quark-flavour sector of the Standard Model is one of the hot 
topics in particle physics of this decade. In these studies, which show a fruitful interplay
between theory and experiment, the $B$-meson system offers a particularly interesting
laboratory. After giving an introduction to quark-flavour mixing and CP violation as well 
as to the theoretical tools to deal with non-leptonic $B$ decays, we discuss popular 
avenues for new physics to enter the roadmap of quark-flavour physics. This allows us 
to have a detailed look at the $B$-factory benchmark modes $B^0_d\to J/\psi K_{\rm S}$,
$B^0_d\to \phi K_{\rm S}$ and $B^0_d\to\pi^+\pi^-$, with a particular emphasis of the 
impact of new physics. We then perform an analysis of the $B\to\pi K$ puzzle, which 
may indicate new sources of CP violation in the electroweak penguin sector, and 
discuss its implications for rare $B$ and $K$ decays. The next topic is given by
$b\to d$ penguin processes, which are now starting to become accessible
at the $B$ factories, thereby representing a new territory of the $B$-physics 
landscape. Finally, we discuss the prospects for $B$-decay studies at the 
Large Hadron Collider, where the $B^0_s$-meson system plays an outstanding 
r\^ole.
}

\vspace{1.5truecm}

\begin{center}
{\sl Invited topical review for Journal of Physics G: Nuclear and Particle Physics}
\end{center}

\vfill
\noindent
CERN-PH-TH/2005-263\\
December 2005

\newpage
\thispagestyle{empty}
\vbox{}
\newpage
 
\setcounter{page}{1}


\title[Highlights of the B-Physics Landscape]{Highlights of
the B-Physics Landscape}

\author{R Fleischer}

\address{CERN, Department of Physics, Theory Division, CH-1211 Geneva 23,
Switzerland}
\ead{robert.fleischer@cern.ch}
\begin{abstract}
The exploration of the quark-flavour sector of the Standard Model is one of the hot 
topics in particle physics of this decade. In these studies, which show a fruitful interplay
between theory and experiment, the $B$-meson system offers a particularly interesting
laboratory. After giving an introduction to quark-flavour mixing and CP violation as well 
as to the theoretical tools to deal with non-leptonic $B$ decays, we discuss popular 
avenues for new physics to enter the roadmap of quark-flavour physics. This allows us 
to have a detailed look at the $B$-factory benchmark modes $B^0_d\to J/\psi K_{\rm S}$,
$B^0_d\to \phi K_{\rm S}$ and $B^0_d\to\pi^+\pi^-$, with a particular emphasis of the 
impact of new physics. We then perform an analysis of the $B\to\pi K$ puzzle, which 
may indicate new sources of CP violation in the electroweak penguin sector, and 
discuss its implications for rare $B$ and $K$ decays. The next topic is given by
$b\to d$ penguin processes, which are now starting to become accessible
at the $B$ factories, thereby representing a new territory of the $B$-physics 
landscape. Finally, we discuss the prospects for $B$-decay studies at the 
Large Hadron Collider, where the $B^0_s$-meson system plays an outstanding 
r\^ole.
\end{abstract}

\maketitle

\section{Introduction}\label{sec:intro}
The history of CP violation, i.e.\ the non-invariance of the weak interactions
with respect to a combined charge-conjugation (C) and parity (P) 
transformation, goes back to the year 1964, where this phenomenon was
discovered through the observation of $K_{\rm L}\to\pi^+\pi^-$ decays
\cite{CP-obs}, which exhibit a branching ratio at the $10^{-3}$ level. This 
surprising effect is a manifestation of {\it indirect} CP violation, which arises 
from the fact that the mass eigenstates $K_{\rm L,S}$ of the neutral kaon 
system, which shows $K^0$--$\bar K^0$ mixing, are not eigenstates of the 
CP operator. In particular, the $K_{\rm L}$ state is governed by the CP-odd 
eigenstate, but has also a tiny admixture of the CP-even eigenstate, which 
may decay through CP-conserving interactions into the $\pi^+\pi^-$ final state. 
These CP-violating effects are described by the following observable:
\begin{equation}\label{epsK}
\varepsilon_K=(2.280\pm0.013)\times10^{-3}\times e^{i\pi/4}.
\end{equation}
On the other hand, CP-violating effects may also arise directly at the decay-amplitude
level, thereby yielding {\it direct} CP violation. This phenomenon, which leads to a
non-vanishing value of a quantity Re$(\varepsilon_K'/\varepsilon_K)$, could 
eventually be established in 1999 through the NA48 (CERN) and KTeV 
(FNAL) collaborations \cite{eps-prime}; the final results of the corresponding 
measurements are given by
\begin{equation}\label{epsp-eps-final}
\mbox{Re}(\varepsilon_K'/\varepsilon_K)=\left\{\begin{array}{ll}
(14.7\pm2.2)\times10^{-4}&\mbox{(NA48 \cite{NA48-final})}\\
(20.7\pm2.8)\times10^{-4}&\mbox{(KTeV \cite{KTeV-final}).}
\end{array}
\right.
\end{equation}

In this decade, there are huge experimental efforts to further 
explore CP violation and the quark-flavour sector of the Standard Model 
(SM). In these studies, the main actor is
the $B$-meson system, where we distinguish between charged and neutral 
$B$ mesons, which are characterized by the following valence-quark contents:
\begin{equation}
B^+\sim u \bar b, \quad B^+_c\sim c \bar b, \quad B^0_d\sim d \bar b, \quad
B^0_s\sim s \bar b.
\end{equation}
The asymmetric $e^+e^-$ $B$ factories at SLAC and KEK with their detectors 
BaBar and Belle, respectively, can only produce $B^+$ and $B^0_d$ mesons 
(and their anti-particles) since they operate at the $\Upsilon(4S)$ resonance, 
and have already collected ${\cal O}(10^8)$ $B\bar B$ pairs of this kind.
Moreover, first $B$-physics results from run II of the Tevatron were 
reported from the CDF and D0 collaborations, including also $B^+_c$
and $B^0_s$ studies, and second-generation $B$-decay
studies will become possible at the Large Hadron Collider (LHC)
at CERN, in particular thanks to the LHCb experiment, starting in the 
autumn of 2007. For the more distant future,
an $e^+$--$e^-$ ``super-$B$ factory'' is under consideration, with an
increase of luminosity by up to two orders of magnitude with respect to the 
currently operating machines. Moreover, there are plans to measure the 
very ``rare" kaon decays $K^+\to\pi^+\nu\bar\nu$ and $K_{\rm L}\to\pi^0\nu\bar\nu$, which are absent at the tree level within the SM, at CERN and KEK/J-PARC. 

In 2001, CP-violating effects were discovered in the $B$-meson system
with the help of $B_d\to J/\psi K_{\rm S}$ decays by the BaBar and Belle 
collaborations \cite{CP-B-obs}, representing the first observation of CP violation 
outside the kaon system. This particular kind of CP violation originates
from the interference between $B^0_d$--$\bar B^0_d$ mixing and
$B^0_d\to J/\psi K_{\rm S}$, $\bar B^0_d\to J/\psi K_{\rm S}$ decay processes,
and is referred to as ``mixing-induced" CP violation. In the summer of 2004,
also direct CP violation could be detected in $B_d\to\pi^\mp K^\pm$ decays
\cite{CP-B-dir}, thereby complementing the measurement of a non-zero
value of $\mbox{Re}(\varepsilon_K'/\varepsilon_K)$.

Studies of CP violation and flavour physics are particularly interesting since
``new physics" (NP), i.e.\ physics lying beyond the SM, typically leads to
new sources of flavour and CP violation. Furthermore, the origin of the
fermion masses, flavour mixing, CP violation etc.\ lies completely in the
dark and is expected to involve NP, too. Interestingly, CP violation
offers also a link to cosmology. One of the key features of our Universe is the
cosmological baryon asymmetry of ${\cal O}(10^{-10})$. As was pointed
out by Sakharov \cite{sach}, the necessary conditions for the generation of 
such an asymmetry include also the requirement that elementary interactions
violate CP (and C). Model calculations of the baryon asymmetry indicate, however,
that the CP violation present in the SM seems to be too small to generate
the observed asymmetry  \cite{shapos}. On the one hand, the required new sources 
of CP violation could be associated with very high energy scales, as in 
``leptogenesis", where new CP-violating effects appear in decays of heavy
Majorana neutrinos \cite{LG-rev}. On the other hand, new sources of
CP violation could also be accessible in the laboratory, as they arise 
naturally when going beyond the SM. 

Before searching for NP, it is essential to understand first the picture of
flavour physics and CP violation arising in the framework of the SM,
where the Cabibbo--Kobayashi--Maskawa (CKM) matrix -- the
quark-mixing matrix -- plays the key r\^ole \cite{cab,KM}. The 
corresponding phenomenology is extremely rich \cite{CKM-book}. In general,
the key problem for the theoretical interpretation is related to strong
interactions, i.e.\ to ``hadronic" uncertainties. A famous example is
$\mbox{Re}(\varepsilon_K'/\varepsilon_K)$, where we
have to deal with a subtle interplay between different contributions
which largely cancel \cite{epsp-rev}. Although the non-vanishing value of this
quantity has unambiguously ruled out ``superweak" models of
CP violation \cite{superweak}, it does currently not allow a stringent
test of the SM. 

In the $B$-meson system, there are various strategies to eliminate
the hadronic uncertainties in the exploration of CP violation (simply
speaking, there are many $B$ decays). Moreover, we may also search
for relations and/or correlations that hold in the SM but could well be
spoiled by NP. These topics will be the focus of this review. The 
outline is as follows: in Section~\ref{sec:CKM}, we discuss the quark 
mixing in the SM by having a closer look at the CKM matrix and the
associated unitarity triangles. The main actor of this review -- the
$B$-meson system -- will then be introduced in Section~\ref{sec:B}.
There we turn to the formalism of $B^0_q$--$\bar B^0_q$ mixing
($q\in\{d,s\}$), give an introduction to non-leptonic $B$ decays, which play 
the key r\^ole for CP violation, and discuss popular avenues for NP to enter 
the strategies to explore this phenomenon. In Section~\ref{sec:bench}, we 
then apply these considerations to the $B$-factory benchmark modes 
$B^0_d\to J/\psi K_{\rm S}$, $B^0_d\to \phi K_{\rm S}$ and $B^0_d\to\pi^+\pi^-$,
and address the possible impact of NP. Since the data for certain $B\to\pi K$
decays show a puzzling pattern for several years, we have devoted 
Section~\ref{sec:BpiK-puzzle} to a detailed discussion of this ``$B\to\pi K$
puzzle" and its interplay with rare $K$ and $B$ decays. In Section~\ref{sec:bd-pengs}, 
we focus on $b\to d$ penguin processes, which are now coming within experimental 
reach at the $B$ factories, thereby offering an exciting new playground. Finally, 
in Section~\ref{sec:LHC}, we discuss $B$-decay studies at the  LHC, where the 
physics potential of the $B^0_s$-meson system can be fully exploited. 
The conclusions and a brief outlook are given in Section~\ref{sec:concl}.

For textbooks dealing with CP violation, the reader is referred to
Refs.~\cite{BLS-textbook}--\cite{KK-textbook}, while a selection of
alternative recent reviews can be found in Refs.~\cite{nir-rev}--\cite{ali-ichep}.

\section{Quark Mixing in the Standard Model}\label{sec:CKM}
\subsection{The CKM Matrix}
In the SM, CP-violating phenomena may originate from the 
charged-current interaction processes of the quarks, $D\to U W^-$, 
where $D\in\{d,s,b\}$ and $U\in\{u,c,t\}$ denote the down- and up-type quark 
flavours, respectively, and the $W^-$ is the usual $SU(2)_{\rm L}$ 
gauge boson. The generic ``coupling strengths'' $V_{UD}$ of these
processes are the elements of a $3\times 3$ matrix, the CKM matrix 
\cite{cab,KM}. It connects the electroweak states $(d',s',b')$ of the down, 
strange and bottom quarks with their mass eigenstates $(d,s,b)$ through 
the following unitary transformation:
\begin{equation}\label{ckm}
\left(\begin{array}{c}
d'\\
s'\\
b'
\end{array}\right)=\left(\begin{array}{ccc}
V_{ud}&V_{us}&V_{ub}\\
V_{cd}&V_{cs}&V_{cb}\\
V_{td}&V_{ts}&V_{tb}
\end{array}\right)\cdot
\left(\begin{array}{c}
d\\
s\\
b
\end{array}\right)
\equiv \hat V_{\rm CKM} \cdot
\left(\begin{array}{c}
d\\
s\\
b
\end{array}\right),
\end{equation}
and is, therefore, a unitary matrix. Since this feature ensures the absence 
of flavour-changing neutral-current (FCNC) processes at the tree level in the SM, 
it is at the basis of the Glashow--Iliopoulos--Maiani (GIM) mechanism \cite{GIM}. 
Expressing the non-leptonic charged-current interaction Lagrangian 
in terms of the mass eigenstates (\ref{ckm}), we obtain
\begin{equation}\label{cc-lag2}
{\cal L}_{\mbox{{\scriptsize int}}}^{\mbox{{\scriptsize CC}}}=
-\frac{g_2}{\sqrt{2}}\left(\begin{array}{ccc}
\bar u_{\mbox{{\scriptsize L}}},& \bar c_{\mbox{{\scriptsize L}}},
&\bar t_{\mbox{{\scriptsize L}}}\end{array}\right)\gamma^\mu\,\hat
V_{\mbox{{\scriptsize CKM}}}
\left(\begin{array}{c}
d_{\mbox{{\scriptsize L}}}\\
s_{\mbox{{\scriptsize L}}}\\
b_{\mbox{{\scriptsize L}}}
\end{array}\right)W_\mu^\dagger\,\,+\,\,\mbox{h.c.,}
\end{equation}
where $g_2$ is the $SU(2)_{\mbox{{\scriptsize L}}}$ gauge coupling, 
and $W_\mu^{(\dagger)}$ the field of the charged $W$ bosons. 

Since the CKM matrix elements governing a $D\to U W^-$ transition and its 
CP conjugate $\bar D\to \bar U W^+$ are related to each other through
\begin{equation}\label{CKM-CP}
V_{UD}\stackrel{{ CP}}{\longrightarrow}V_{UD}^\ast,
\end{equation}
we observe that CP violation is associated with complex phases
of the CKM matrix.

\subsection{The Phase Structure of the CKM Matrix}
We have the freedom of redefining the up- and down-type quark fields as follows:
\begin{equation}
U\to \exp(i\xi_U)U,\quad D\to \exp(i\xi_D)D. 
\end{equation}
Performing such transformations in (\ref{cc-lag2}), the invariance 
of the charged-current interaction Lagrangian implies the following transformations
of the CKM matrix elements:
\begin{equation}\label{CKM-trafo}
V_{UD}\to\exp(i\xi_U)V_{UD}\exp(-i\xi_D).
\end{equation}
If we consider a general $N\times N$ quark-mixing matrix, where $N$ denotes the 
number of fermion generations, and eliminate unphysical phases through these transformations, we are left with the following quantities to parametrize the
quark-mixing matrix:
\begin{equation}
\underbrace{\frac{1}{2}N(N-1)}_{\mbox{Euler angles}} \, + \,
\underbrace{\frac{1}{2}(N-1)(N-2)}_{\mbox{complex phases}}=
(N-1)^2.
\end{equation}

Applying this expression to $N=2$ generations, we observe
that only one rotation angle -- the Cabibbo angle
$\theta_{\rm C}$ \cite{cab} -- is required for the parametrization of the $2\times2$
quark-mixing matrix, which can be written as
\begin{equation}\label{Cmatrix}
\hat V_{\rm C}=\left(\begin{array}{cc}
\cos\theta_{\rm C}&\sin\theta_{\rm C}\\
-\sin\theta_{\rm C}&\cos\theta_{\rm C}
\end{array}\right),
\end{equation}
where the value of $\sin\theta_{\rm C}=0.22$ follows from the experimental 
data for $K\to\pi\ell\bar\nu_\ell$ decays. On the other hand, in the case of $N=3$ generations, the 
parametrization of the corresponding $3\times3$ quark-mixing matrix involves 
three Euler-type angles and a single {\it complex} phase. This complex phase 
allows us to accommodate CP violation in the SM, as was pointed out by 
Kobayashi and Maskawa in 1973 \cite{KM}. The corresponding picture
is referred to as the Kobayashi--Maskawa (KM) mechanism of CP violation.

The Particle Data Group advocates the following ``standard parametrization'' 
\cite{PDG}:
\begin{equation}\label{standard}
\hspace*{-1.5truecm}\hat V_{\rm CKM}=
\left(\begin{array}{ccc}
c_{12}c_{13}&s_{12}c_{13}&s_{13}e^{-i\delta_{13}}\\ -s_{12}c_{23}
-c_{12}s_{23}s_{13}e^{i\delta_{13}}&c_{12}c_{23}-
s_{12}s_{23}s_{13}e^{i\delta_{13}}&
s_{23}c_{13}\\ s_{12}s_{23}-c_{12}c_{23}s_{13}e^{i\delta_{13}}&-c_{12}s_{23}
-s_{12}c_{23}s_{13}e^{i\delta_{13}}&c_{23}c_{13}
\end{array}\right),
\end{equation}
with $c_{ij}\equiv\cos\theta_{ij}$ and $s_{ij}\equiv\sin\theta_{ij}$. 
If we redefine the quark-field phases appropriately, $\theta_{12}$, $\theta_{23}$ 
and $\theta_{13}$ can all be made to lie in the first quadrant. The advantage of 
this parametrization is that the mixing between two generations $i$ and $j$ vanishes 
if $\theta_{ij}$ is set to zero. In particular, for 
$\theta_{23}=\theta_{13}=0$, the third generation decouples, and the
submatrix describing the mixing between the first and 
second generations takes the same form as (\ref{Cmatrix}).

\subsection{The Wolfenstein Parametrization}\label{ssec:wolf}
The experimental data for the charged-current interactions of the quarks exhibit an interesting hierarchy \cite{PDG}: transitions within the same 
generation involve CKM matrix elements of ${\cal O}(1)$, those between the first and the second generation are associated with CKM elements of ${\cal O}(10^{-1})$, 
those between the second and the third generation are related to CKM elements of 
${\cal O}(10^{-2})$, and those between the first and third generation are described by 
CKM matrix elements of ${\cal O}(10^{-3})$. It would be useful for phenomenological applications to have a parametrization of the CKM matrix available that makes this 
pattern explicit \cite{wolf}. To this end, we introduce a set of new parameters, 
$\lambda$, $A$, $\rho$ and $\eta$, by imposing the following 
relations \cite{blo}:
\begin{equation}\label{set-rel}
s_{12}\equiv\lambda=0.22,\quad s_{23}\equiv A\lambda^2,\quad 
s_{13}e^{-i\delta_{13}}\equiv A\lambda^3(\rho-i\eta).
\end{equation}
If we go back to the standard parametrization (\ref{standard}), we 
obtain an exact parametrization of the CKM matrix in terms of
$\lambda$ (and $A$, $\rho$, $\eta$), which allows us to expand each CKM 
element in powers of the small parameter $\lambda$. Neglecting terms of 
${\cal O}(\lambda^4)$ yields the famous ``Wolfenstein 
parametrization'' \cite{wolf}:
\begin{equation}\label{W-par}
\hat V_{\mbox{{\scriptsize CKM}}} =\left(\begin{array}{ccc}
1-\frac{1}{2}\lambda^2 & \lambda & A\lambda^3(\rho-i\eta) \\
-\lambda & 1-\frac{1}{2}\lambda^2 & A\lambda^2\\
A\lambda^3(1-\rho-i\eta) & -A\lambda^2 & 1
\end{array}\right)+{\cal O}(\lambda^4).
\end{equation}
On the other hand, also higher-order terms of the expansion in $\lambda$
can straightforwardly be included by following the recipe described above.

\begin{figure}[t]
\centerline{
\begin{tabular}{ll}
   {\small(a)} & {\small(b)} \\
   \qquad \includegraphics[width=6.3truecm]{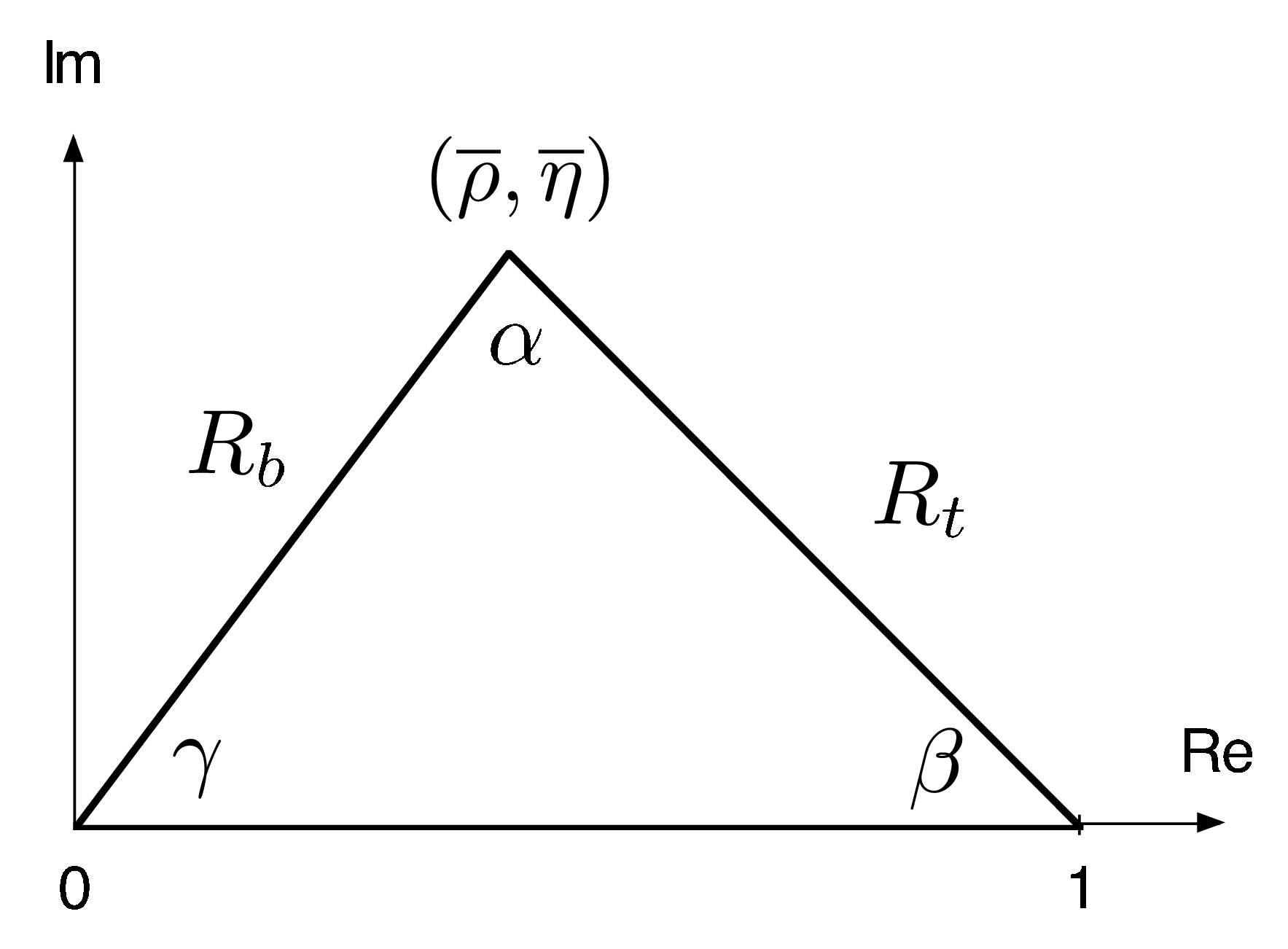}
&
\qquad \includegraphics[width=6.3truecm]{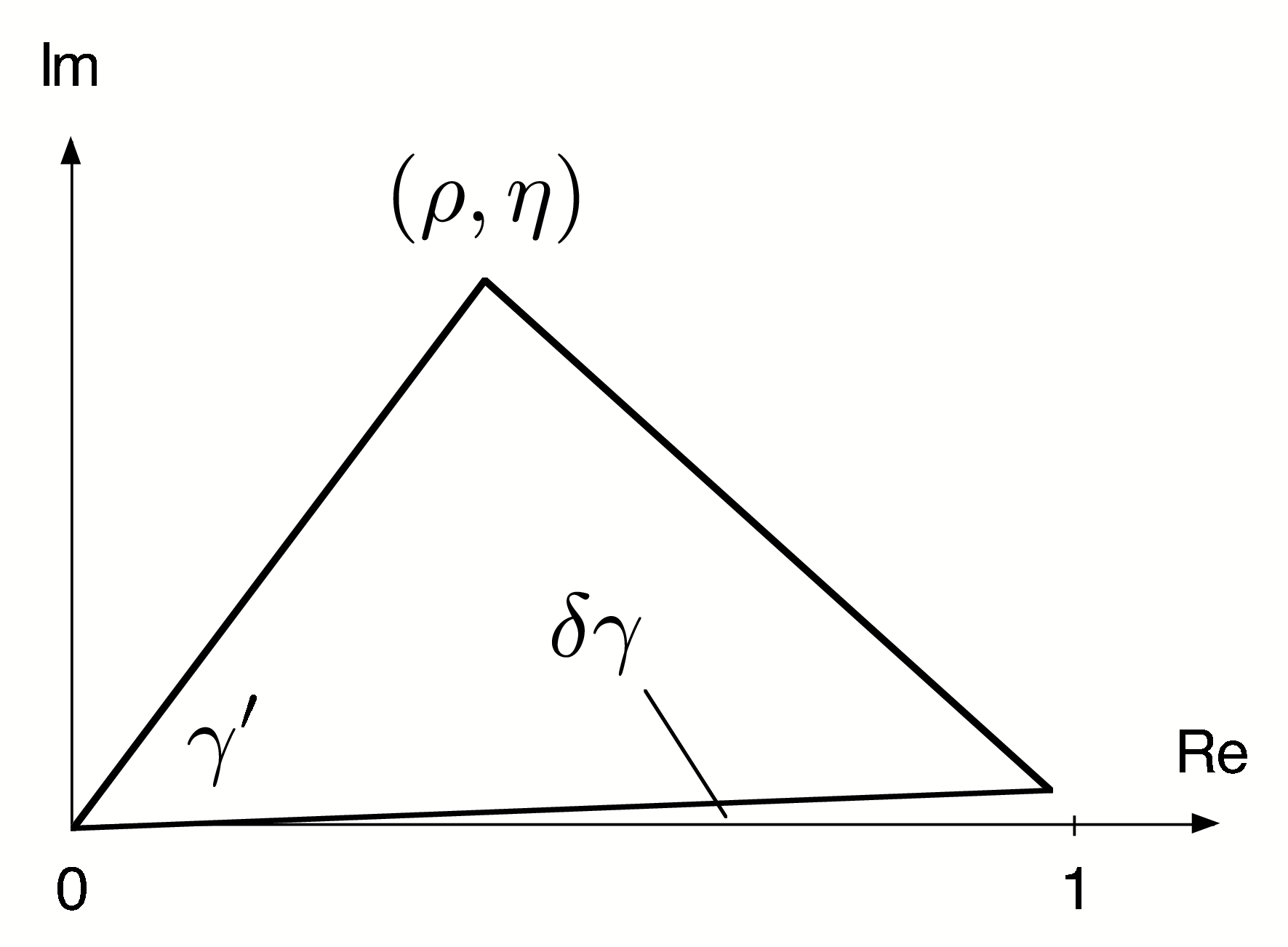}
 \end{tabular}}
 \vspace*{-0.2truecm}
\caption{The two non-squashed unitarity triangles of the CKM matrix: 
(a) and (b) correspond to the orthogonality relations (\ref{UT1}) and (\ref{UT2}), respectively. In Asia, the notation $\phi_1\equiv\beta$,
$\phi_2\equiv\alpha$ and $\phi_3\equiv\gamma$ is used for the angles of the
triangle shown in (a).}
\label{fig:UT}
\end{figure}

\subsection{The Unitarity Triangles of the CKM Matrix}
Since the CKM matrix is a unitary matrix, it satisfies
\begin{equation}
\hat V_{\mbox{{\scriptsize CKM}}}^{\,\,\dagger}\cdot\hat 
V_{\mbox{{\scriptsize CKM}}}=
\hat 1=\hat V_{\mbox{{\scriptsize CKM}}}\cdot\hat V_{\mbox{{\scriptsize 
CKM}}}^{\,\,\dagger},
\end{equation}
leading to a set of 12 equations, which consist of 6 normalization 
and 6 orthogonality relations. The latter can be represented as 6 
triangles in the complex plane, which have all the same area. The 
Wolfenstein parametrization of the CKM matrix allows us straightforwardly
to explore the generic shape of these triangles: we find two triangles, where  
one side is suppressed with respect to the others by a factor of ${\cal O}(\lambda^2)$,
and another set of two triangles, where one side is even suppressed with respect 
to the others by a factor of ${\cal O}(\lambda^4)$; however, there are also two
triangles, where all three sides are of the same order of magnitude. They
are described by the following orthogonality relations:
\begin{eqnarray}
V_{ud}V_{ub}^\ast+V_{cd}V_{cb}^\ast+V_{td}V_{tb}^\ast & = &
0\label{UT1}\\
V_{ud}^\ast V_{td}+V_{us}^\ast V_{ts}+V_{ub}^\ast V_{tb}
& = & 0.\label{UT2}
\end{eqnarray}
If we keep just the leading, non-vanishing terms of the expansion in $\lambda$, 
these relations give actually the same result, which is given by
\begin{equation}
\left[(\rho+i\eta)+(1-\rho-i\eta)+(-1)\right]A\lambda^3=0,
\end{equation}
and describes {\it the} unitarity triangle of the CKM matrix. 

Following the procedure described in Subsection~\ref{ssec:wolf}, we may also 
include the next-to-leading order corrections in the $\lambda$ expansion \cite{blo}.
The degeneracy between the leading-order triangles corresponding to 
(\ref{UT1}) and (\ref{UT2}) is then lifted, and we arrive at the situation
illustrated in Fig.~\ref{fig:UT}. The triangle sketched in Fig.~\ref{fig:UT} (a)
is a straightforward generalization of the leading-order case. Its apex takes
the following coordinates \cite{blo}:
\begin{equation}\label{rho-eta-bar}
\bar\rho\equiv\rho\left(1-\frac{1}{2}\lambda^2\right),\quad
\bar\eta\equiv\eta\left(1-\frac{1}{2}\lambda^2\right),
\end{equation}
which correspond to the triangle sides
\begin{equation}\label{Rb-Rt-def}
R_b
=\left(1-\frac{\lambda^2}{2}\right)\frac{1}{\lambda}\left|\frac{V_{ub}}{V_{cb}}\right|,
\quad
R_t
=\frac{1}{\lambda}\left|\frac{V_{td}}{V_{cb}}\right|.
\end{equation}
This triangle is usually the one considered in the literature, and whenever referring
to {\it a} unitarity triangle (UT) in the following discussion, also we shall always 
mean this triangle. The characteristic feature of the second triangle shown in 
Fig.~\ref{fig:UT} (b) is the small angle between the basis of the triangle and the 
real axis, satisfying 
\begin{equation}
\delta\gamma\equiv \gamma-\gamma'=\lambda^2\eta={\cal O}(1^\circ).
\end{equation}
As we will see below, this triangle is of particular interest for the LHCb 
experiment.

\begin{figure}[t]
\centerline{
\begin{tabular}{ll}
  \includegraphics[width=7.0truecm]{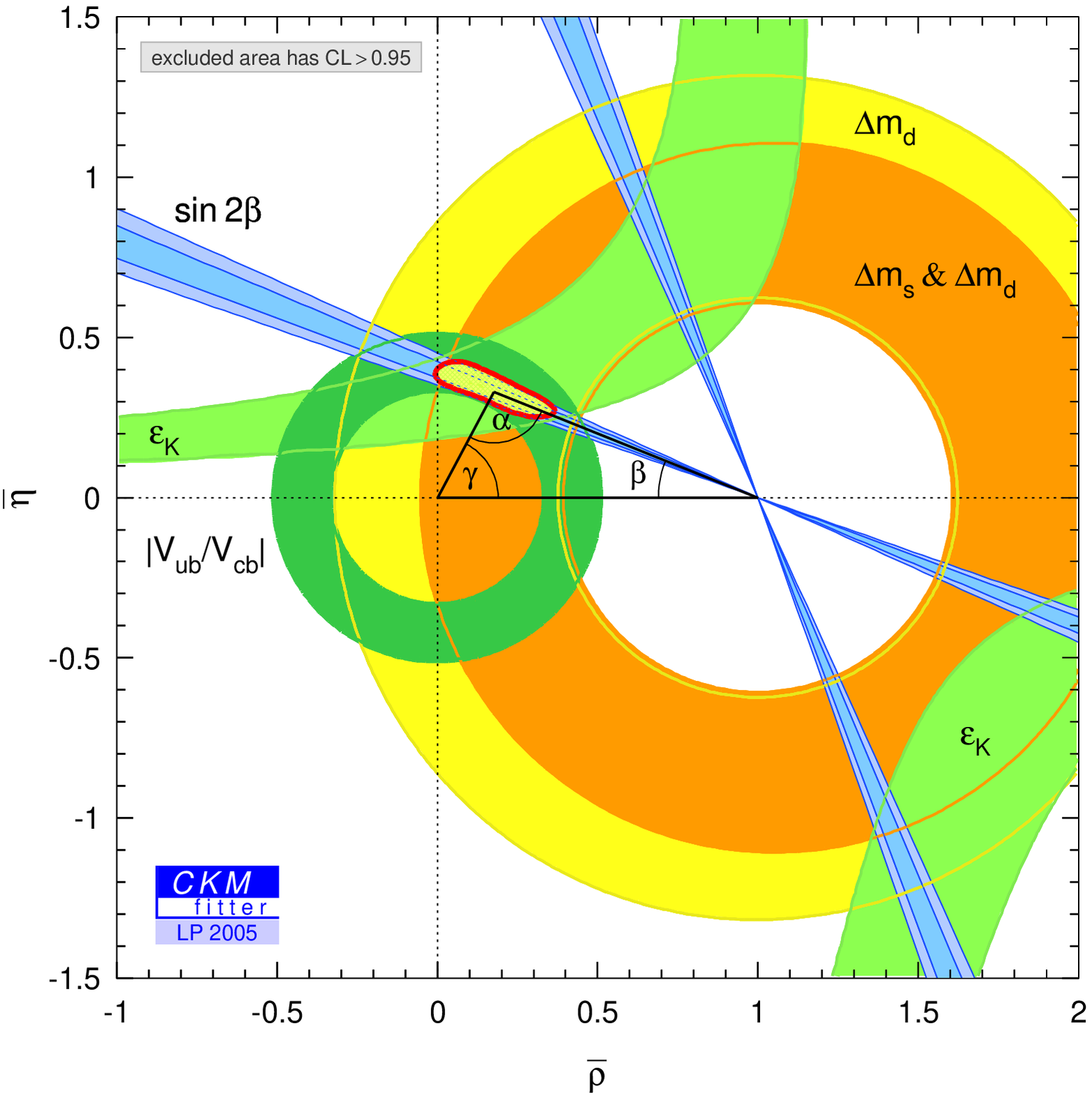} &
\includegraphics[width=9.0truecm]{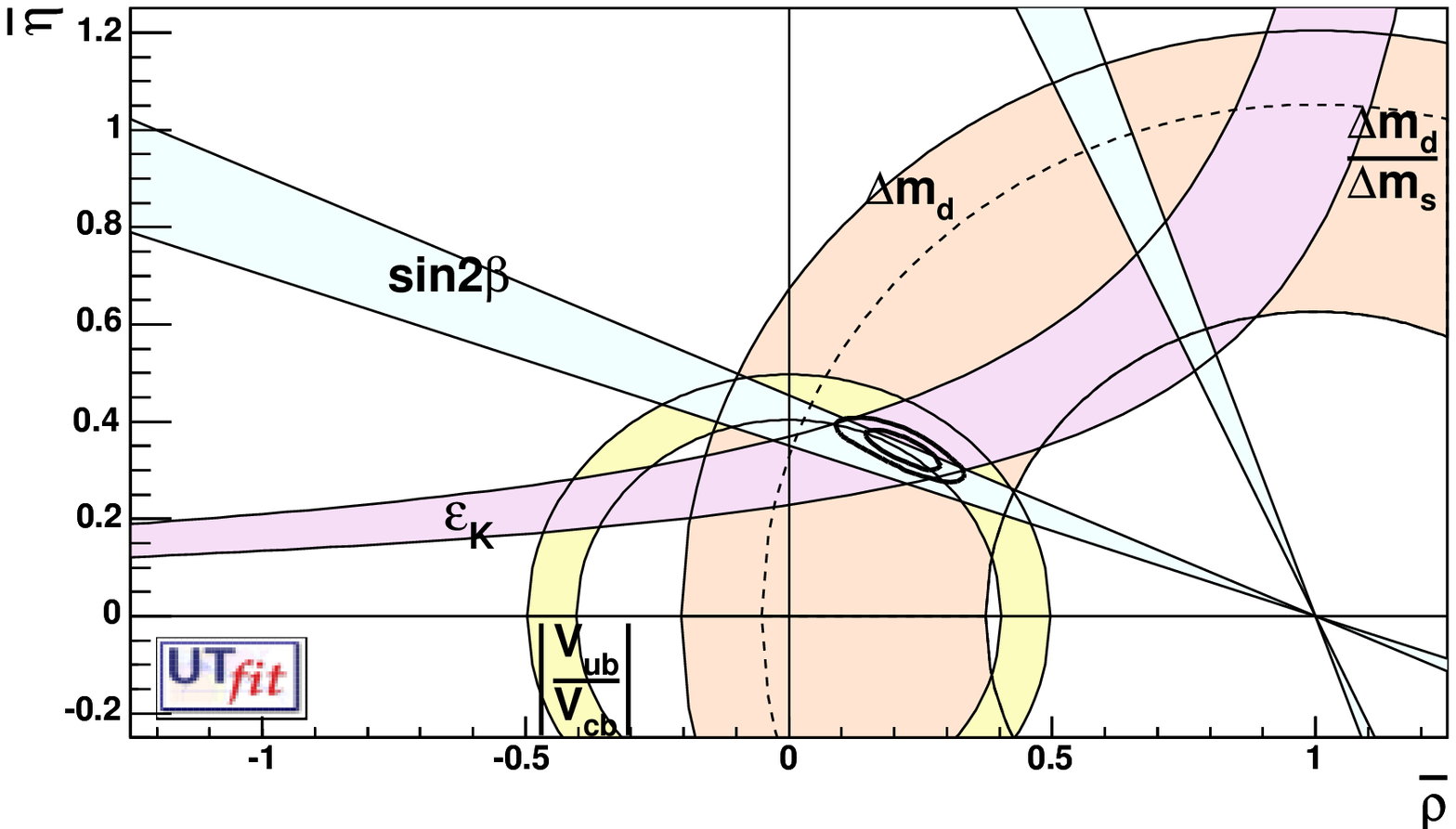}
 \end{tabular}}
\caption{The most recent analyses of the CKMfitter and 
UTfit  collaborations \cite{CKMfitter,UTfit}.}\label{fig:UTfits}
\end{figure}

\subsection{The Determination of the Unitarity Triangle}\label{ssec:UT}
The next obvious question is how to determine the UT. There are two 
conceptually different avenues that we may follow to this end:
\begin{itemize}
\item[(i)] In the ``CKM fits'', theory is used to convert 
experimental data into contours in the $\bar\rho$--$\bar\eta$ plane. In particular, 
semi-leptonic $b\to u \ell \bar\nu_\ell$, $c \ell \bar\nu_\ell$ decays and 
$B^0_q$--$\bar B^0_q$ mixing ($q\in\{d,s\}$) allow us to determine the UT sides 
$R_b$ and $R_t$, respectively, i.e.\ to fix two circles in the $\bar\rho$--$\bar\eta$ 
plane. Furthermore, the indirect CP violation in the neutral kaon system
described by $\varepsilon_K$ can be transformed into a hyperbola. 
\item[(ii)] Theoretical considerations allow us to convert measurements of 
CP-violating effects in $B$-meson decays into direct information on the UT angles. 
The most prominent example is the determination of $\sin2\beta$ through 
CP violation in $B^0_d\to J/\psi K_{\rm S}$ decays, but several other strategies 
were proposed.
\end{itemize}
The goal is to ``overconstrain'' the UT as much as possible. In the future, 
additional contours can be fixed in the $\bar\rho$--$\bar\eta$ plane through 
the measurement of rare decays. 

In Fig.~\ref{fig:UTfits}, we show the most recent results of the comprehensive
analyses of the UT that were performed by the ``CKM Fitter Group'' \cite{CKMfitter}
and the ``UTfit collaboration''~\cite{UTfit}. In these figures, we can nicely see the
circles that are determined through the semi-leptonic $B$ decays and the 
$\varepsilon_K$ hyperbolas. Moreover, also the straight lines following from the 
direct measurement of $\sin 2\beta$ with the help of $B^0_d\to J/\psi K_{\rm S}$ 
modes are shown. We observe that the global consistency is very good. However,
looking closer, we also see that the most recent average for 
$(\sin 2\beta)_{\psi K_{\rm S}}$ is now on the lower side, so that the situation in 
the $\bar\rho$--$\bar\eta$ plane is no longer ``perfect". Moreover, as we
shall discuss in detail in the course of this review, there are certain puzzles in the
$B$-factory data, and several important aspects could not yet be addressed
experimentally and are hence still essentially unexplored. Consequently, we may hope
that flavour studies will eventually establish deviations from the SM description 
of CP violation. Since $B$ mesons play a key r\^ole in these explorations, let us 
next have a closer look at them.

\boldmath
\section{The Main Actor: The $B$-Meson System}\label{sec:B}
\unboldmath
\subsection{A Closer Look at $B_q^0$--$\bar B_q^0$ Mixing}\label{ssec:Bmix}
\begin{figure}
\centerline{
 \includegraphics[width=5.5truecm]{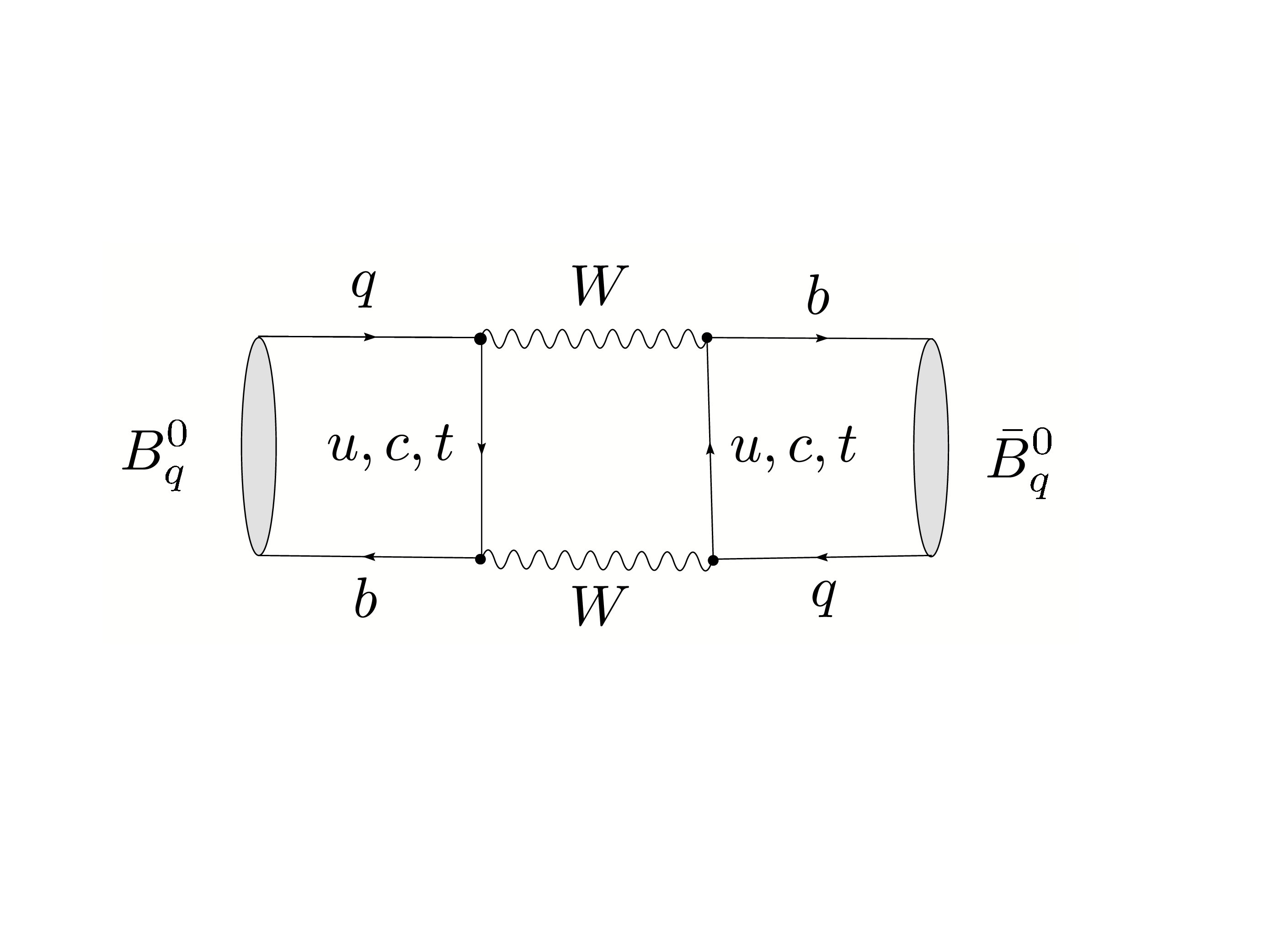}
 \hspace*{0.5truecm}
 \includegraphics[width=5.5truecm]{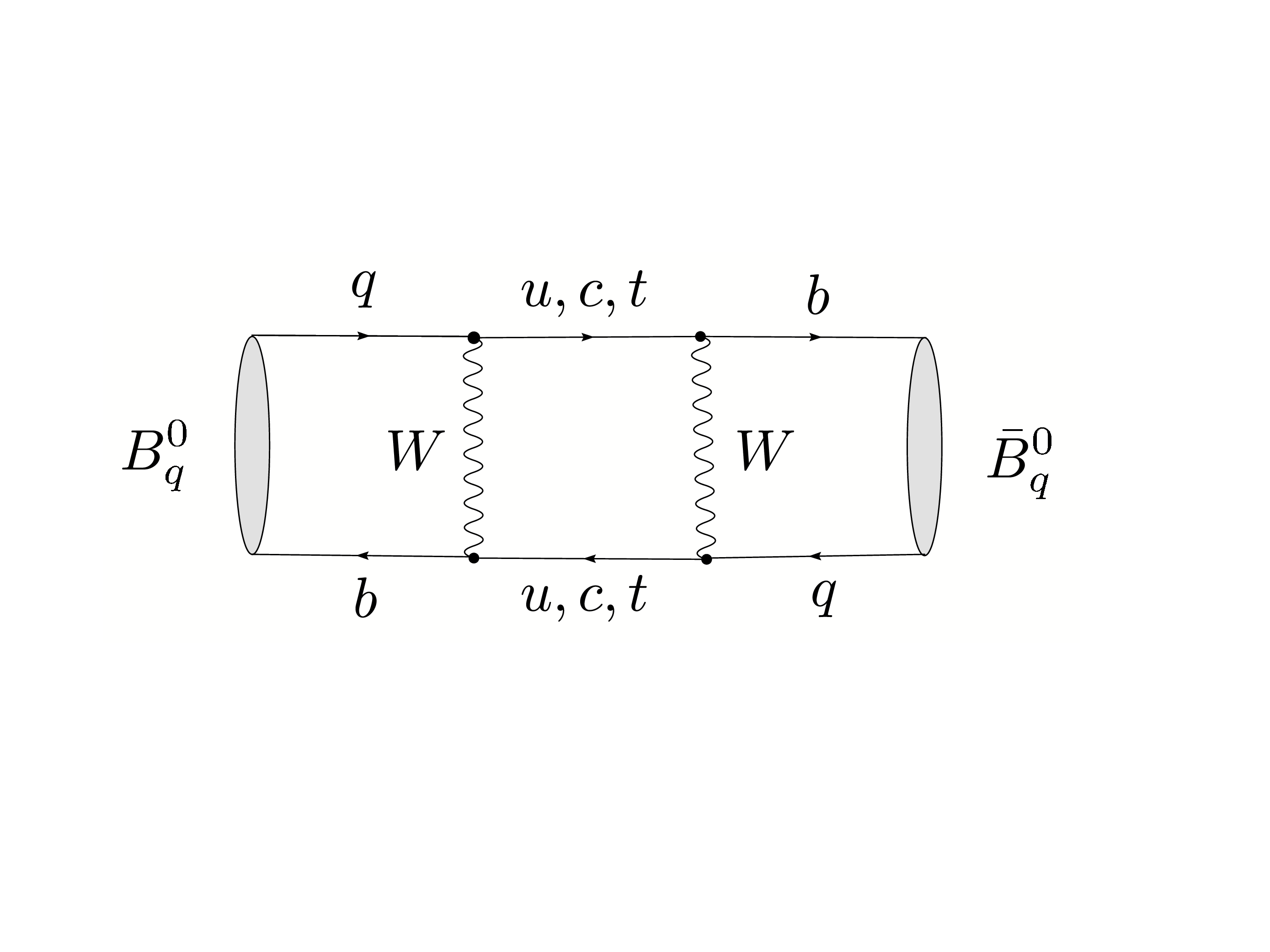}  
 }
 \caption{Box diagrams contributing to $B^0_q$--$\bar B^0_q$ mixing in the
 SM ($q\in\{d,s\}$).}
   \label{fig:boxes}
\end{figure}

In contrast to their charged counterparts, the neutral $B_q$ ($q\in \{d,s\}$) 
mesons show $B_q^0$--$\bar B_q^0$ mixing, which we encountered already
in the determination of the UT discussed in Subsection~\ref{ssec:UT}. This 
phenomenon is the counterpart of $K^0$--$\bar K^0$ mixing, and originates,
in the SM, from box diagrams, as illustrated in Fig.~\ref{fig:boxes}. Thanks to
$B_q^0$--$\bar B_q^0$ mixing, an initially, i.e.\ at time $t=0$, present 
$B^0_q$-meson state evolves into a time-dependent linear combination 
of $B^0_q$ and $\bar B^0_q$ states:
\begin{equation}
|B_q(t)\rangle=a(t)|B^0_q\rangle + b(t)|\bar B^0_q\rangle,
\end{equation}
where $a(t)$ and $b(t)$ are governed by a Schr\"odinger equation of 
the following form:
\begin{equation}\label{SG-OSZ}
i\,\frac{{\rm d}}{{\rm d} t}\left(\begin{array}{c} a(t)\\ b(t)
\end{array}\right)=
\Biggl[\underbrace{\left(\begin{array}{cc}
M_{0}^{(q)} & M_{12}^{(q)}\\ M_{12}^{(q)\ast} & M_{0}^{(q)}
\end{array}\right)}_{\mbox{mass matrix}}-
\frac{i}{2}\underbrace{\left(\begin{array}{cc}
\Gamma_{0}^{(q)} & \Gamma_{12}^{(q)}\\
\Gamma_{12}^{(q)\ast} & \Gamma_{0}^{(q)}
\end{array}\right)}_{\mbox{decay matrix}}\Biggr]
\cdot\left(\begin{array}{c}
a(t)\\ b(t)\nonumber
\end{array}
\right).
\end{equation}
The special form $H_{11}=H_{22}$ of the Hamiltonian $H$ is an implication 
of the CPT theorem, i.e.\ of the invariance under combined CP and 
time-reversal (T) transformations.

In the SM, the mass and decay matrices can be calculated through the
dispersive and absorptive parts of the box diagrams in Fig.~\ref{fig:boxes},
respectively, where the former is dominated by top-quark exchanges. Following 
these lines, we arrive at
\begin{equation}\label{Gam12M12-rat}
\frac{\Gamma_{12}^{(q)}}{M_{12}^{(q)}}\approx
-\frac{3\pi}{2S_0(x_{t})}\left(\frac{m_b^2}{M_W^2}\right)
={\cal O}(m_b^2/m_t^2)\ll 1,
\end{equation}
where $S_0(x_t\equiv m_t^2/M_W^2)$ is one of the Inami--Lim functions 
\cite{IL}, describing the dependence on the top-quark mass $m_t$. The ratio
in (\ref{Gam12M12-rat}) can be probed experimentally through the following 
``wrong-charge'' lepton asymmetries:
\begin{equation}\label{ASL}
{\cal A}^{(q)}_{\mbox{{\scriptsize SL}}}\equiv
\frac{\Gamma(B^0_q(t)\to \ell^-\bar\nu X)-\Gamma(\bar B^0_q(t)\to
\ell^+\nu X)}{\Gamma(B^0_q(t)\to \ell^-\bar \nu X)+
\Gamma(\bar B^0_q(t)\to \ell^+\nu X)}\approx\left|
\frac{\Gamma_{12}^{(q)}}{M_{12}^{(q)}}\right|
\sin\delta\Theta^{(q)}_{M/\Gamma},
\end{equation}
which are a measure of CP violation in $B^0_q$--$\bar B^0_q$ oscillations. In
this expression, we have neglected second-order terms in 
$\Gamma_{12}^{(q)}/M_{12}^{(q)}$, and have introduced
\begin{equation}
\delta\Theta_{M/\Gamma}^{(q)}\equiv
\Theta_{M_{12}}^{(q)}-\Theta_{\Gamma_{12}}^{(q)},
\end{equation}
with $M_{12}^{(q)}\equiv e^{i\Theta_{M_{12}}^{(q)}}\vert
M_{12}^{(q)}\vert$ and $\Gamma_{12}^{(q)}\equiv
e^{i\Theta_{\Gamma_{12}}^{(q)}}\vert\Gamma_{12}^{(q)}\vert$.
Because of the strong suppression of (\ref{Gam12M12-rat}) and
$\sin\delta\Theta^{(q)}_{M/\Gamma}\propto m_c^2/m_b^2$,
the asymmetry ${\cal A}^{(q)}_{\mbox{{\scriptsize SL}}}$ is suppressed by 
a factor of $m_c^2/m_t^2={\cal O}(10^{-4})$ and is hence tiny in the SM.
However, this observable may be enhanced through NP effects, thereby
representing an interesting probe for physics beyond the SM \cite{LLNP,BBLN-CFLMT}.
The current experimental average for the $B_d$-meson system compiled by 
the ``Heavy Flavour Averaging Group" \cite{HFAG} is given by
\begin{equation}
{\cal A}^{(d)}_{\mbox{{\scriptsize SL}}}=0.0030\pm0.0078, 
\end{equation}
and does not
indicate any non-vanishing effect.

In the following discussion, we neglect the tiny CP-violating effects
in the $B^0_q$--$\bar B^0_q$ oscillations that are descirbed by 
(\ref{ASL}). The solution of (\ref{SG-OSZ}) yields then the following
time-dependent rates for decays of initially, i.e.\ at time $t=0$, 
present $B^0_q$ or $\bar B^0_q$ mesons:
\begin{equation}\label{rates}
\hspace*{-1.2truecm}
\Gamma(\stackrel{{\mbox{\tiny (--)}}}{B^0_q}(t)\to f)
=\tilde\Gamma_f
\left[|g_\mp^{(q)}(t)|^2+|\xi_f^{(q)}|^2|g_\pm^{(q)}(t)|^2-
2\mbox{\,Re}\left\{\xi_f^{(q)}
g_\pm^{(q)}(t)g_\mp^{(q)}(t)^\ast\right\}\right].
\end{equation}
Here the time-independent rate $\tilde\Gamma_f$ corresponds to the 
``unevolved'' decay amplitude $A(B^0_q\to f)$, and can be calculated by 
performing the usual phase-space integrations. The time dependence enters
through the functions
\begin{eqnarray}
|g^{(q)}_{\pm}(t)|^2&=&\frac{1}{4}\left[e^{-\Gamma_{\rm L}^{(q)}t}+
e^{-\Gamma_{\rm H}^{(q)}t}\pm2\,e^{-\Gamma_q t}\cos(\Delta M_qt)
\right]\label{g-funct-1}\\
g_-^{(q)}(t)\,g_+^{(q)}(t)^\ast&=&\frac{1}{4}\left[e^{-\Gamma_{\rm L}^{(q)}t}-
e^{-\Gamma_{\rm H}^{(q)}t}+2\,i\,e^{-\Gamma_q t}\sin(\Delta M_qt)
\right],\label{g-funct-2}
\end{eqnarray}
where the $\Gamma_{\rm H}^{(q)}$ and $\Gamma_{\rm L}^{(q)}$
 are the decay widths of the ``heavy'' and ``light'' mass eigenstates of the 
 $B_q$-meson system, respectively, and 
\begin{equation}\label{DeltaMq-def}
\Delta M_q\equiv M_{\rm H}^{(q)}-M_{\rm L}^{(q)}=2|M_{12}^{(q)}|>0
\end{equation}
denotes the corresponding mass difference. The rates into the CP-conjugate 
final state $\bar f$ can straightforwardly be obtained from those in
(\ref{rates}) by making the substitutions
\begin{equation}
\tilde\Gamma_f  \,\,\,\to\,\,\, 
\tilde\Gamma_{\bar f},
\quad\,\,\xi_f^{(q)} \,\,\,\to\,\,\, 
\xi_{\bar f}^{(q)},
\end{equation}
where
\begin{equation}\label{xi-def}
\xi_f^{(q)}\equiv e^{-i\Theta_{M_{12}}^{(q)}}
\frac{A(\bar B_q^0\to f)}{A(B_q^0\to f)},\quad
\xi_{\bar f}^{(q)}\equiv e^{-i\Theta_{M_{12}}^{(q)}}
\frac{A(\bar B_q^0\to \bar f)}{A(B_q^0\to \bar f)}
\end{equation}
describe the interference effects between $B_q^0$--$\bar B_q^0$ mixing
and decay processes. Finally, 
\begin{equation}\label{theta-def}
\Theta_{M_{12}}^{(q)}=\pi+2\mbox{arg}(V_{tq}^\ast V_{tb})-
\phi_{\mbox{{\scriptsize CP}}}(B_q),
\end{equation}
where the CKM factor can be read off from the box diagrams in Fig.~\ref{fig:boxes}
with top-quark exchanges, and $\phi_{\mbox{{\scriptsize CP}}}(B_q)$ is a 
convention-dependent phase, which is introduced through 
\begin{equation}\label{CP-def}
({\cal CP})\vert B^{0}_q\rangle=
e^{i\phi_{\mbox{{\scriptsize CP}}}(B_q)}
\vert\bar B^{0}_q\rangle.
\end{equation}
This quantity is cancelled in (\ref{xi-def}) through the amplitude ratios, so that
$\xi_f^{(q)}$ and $\xi_{\bar f}^{(q)}$ are actually physical observables, as we
will see explicitly in Subsection~\ref{ssec:CP-strat}.

In the literature, the ``mixing parameter"
\begin{equation}\label{mix-par}
x_q\equiv\frac{\Delta M_q}{\Gamma_q}=\left\{\begin{array}{ll}
\quad\, 0.774\pm0.008&(q=d)\\
> 19.9 \mbox{ @ 95\% C.L.} & (q=s)
\end{array}\right.
\end{equation}
is frequently considered (for the numerical values, see \cite{HFAG}), where
\begin{equation}
\Gamma_q\equiv\frac{\Gamma^{(q)}_{\rm H}+\Gamma^{(q)}_{\rm L}}{2}=
\Gamma^{(q)}_0. 
\end{equation}
It is complemented by the width difference
\begin{equation}
\Delta\Gamma_q\equiv\Gamma_{\rm H}^{(q)}-\Gamma_{\rm L}^{(q)}=
\frac{4\mbox{\,Re}\left[M_{12}^{(q)}\Gamma_{12}^{(q)\ast}\right]}{\Delta M_q},
\end{equation}
which satisfies
\begin{equation}\label{DGoG}
\frac{\Delta\Gamma_q}{\Gamma_q}\approx-\frac{3\pi}{2S_0(x_t)}
\left(\frac{m_b^2}{M_W^2}\right)x_q=-{\cal O}(10^{-2})\times x_q.
\end{equation}
Consequently, $\Delta\Gamma_d/\Gamma_d\sim 10^{-2}$ is 
negligibly small, while $\Delta\Gamma_s/\Gamma_s\sim 10^{-1}$ is expected to
be sizeable. Although $B^0_d$--$\bar B^0_d$ mixing is now an experimentally 
well-established phenomenon, its counterpart in the $B_s$-meson system has not 
yet been observed, and is one of the key targets of the $B$-physics studies at hadron colliders,  as we will see in Section~\ref{sec:LHC}, where we shall also have a 
closer look at the width difference $\Delta\Gamma_s$.

\begin{figure}
   \centerline{
   \begin{tabular}{lc}
     {\small(a)} & \\
    &     \includegraphics[width=3.3truecm]{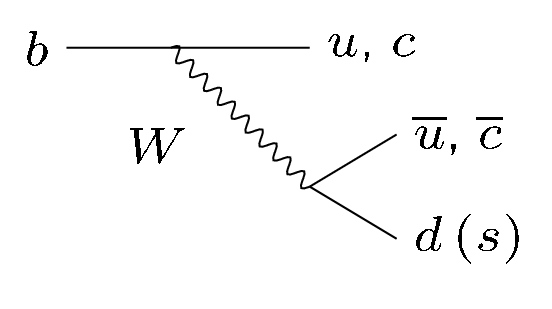}\\
     {\small(b)} & \\
    &  \includegraphics[width=5.0truecm]{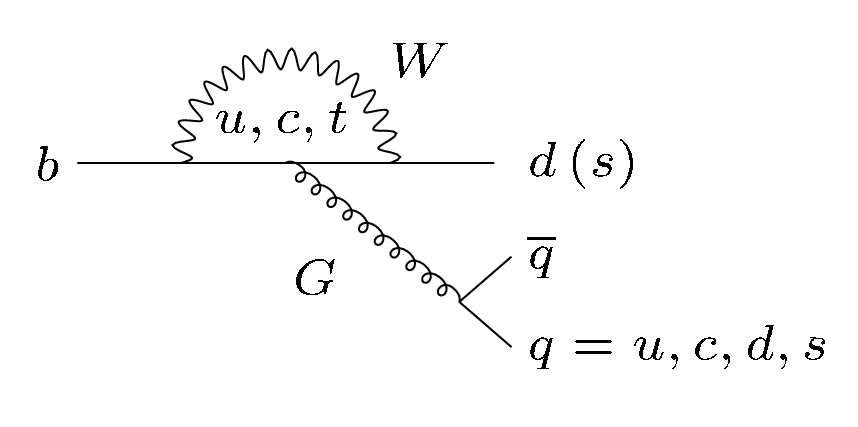}\\
     {\small(c)} & \\
    &     \includegraphics[width=8.3truecm]{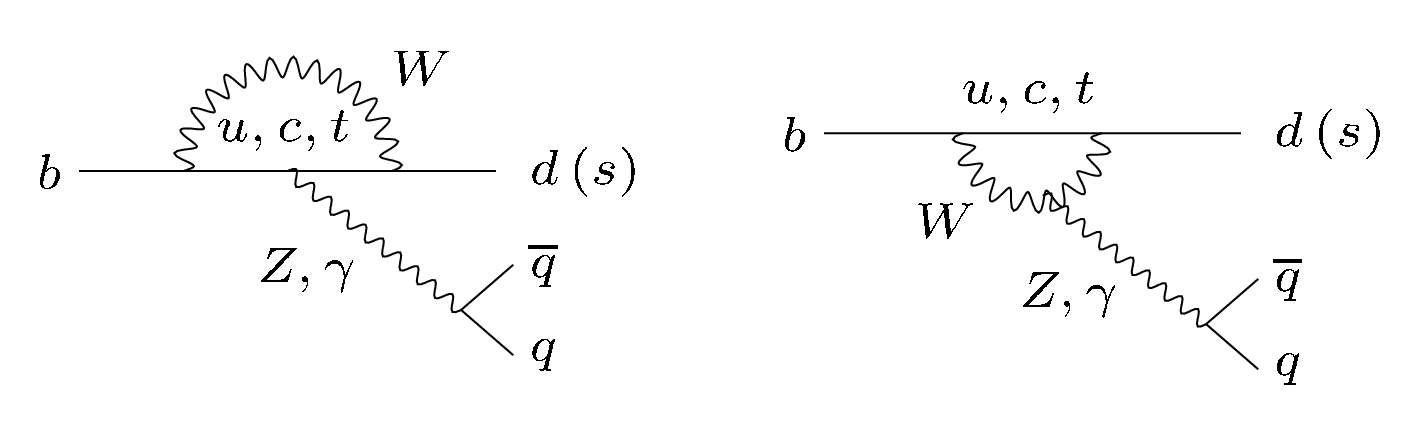} 
     \end{tabular}}
     \caption{Feynman diagrams of the topologies characterizing non-leptonic 
     $B$ decays: trees (a), QCD penguins (b), and electroweak penguins 
     (c).}\label{fig:topol}
\end{figure}

\subsection{Non-Leptonic $B$ Decays}\label{ssec:non-lept}
As far as the exploration of CP violation is concerned, non-leptonic $B$ decays 
play the key r\^ole. In such processes, CP-violating asymmetries can be generated
through certain interference effects, as we will see below. The final states of 
non-leptonic transitions consist only of quarks, and they originate from 
$b\to q_1 \bar q_2 d (s)$ quark-level processes, with $q_1,q_2\in\{u,d,c,s\}$. 
There are two kinds of topologies contributing to such decays: ``tree'' and 
``penguin'' topologies. The latter consist of gluonic (QCD) and electroweak (EW) 
penguins. In Fig.~\ref{fig:topol}, we show the corresponding 
leading-order Feynman diagrams. Depending on the flavour content of their final 
states, non-leptonic $b\to q_1 \bar q_2 d (s)$ decays can be classified as follows:
\begin{itemize}
\item $q_1\not=q_2\in\{u,c\}$: {\it only} tree diagrams contribute.
\item $q_1=q_2\in\{u,c\}$: tree {\it and} penguin diagrams contribute.
\item $q_1=q_2\in\{d,s\}$: {\it only} penguin diagrams contribute.
\end{itemize}

For the analysis of non-leptonic $B$ decays, low-energy effective Hamiltonians 
offer the appropriate tool, yielding transition amplitudes of the following structure:
\begin{equation}\label{LE-Ham}
\langle f|{\cal H}_{\rm eff}|i\rangle=\frac{G_{\rm F}}{\sqrt{2}}
\lambda_{\rm CKM}\sum_k C_k(\mu)\langle f|Q_k(\mu)|i\rangle.
\end{equation}
As usual, $G_{\rm F}$ denotes Fermi's constant, $\lambda_{\rm CKM}$ is an
appropriate CKM factor, and $\mu$ a renormalization scale. The technique of 
the operator product expansion allows us to separate the short-distance 
contributions to this transition amplitude from the long-distance ones, 
which are described by perturbative quantities $C_k(\mu)$ (``Wilson 
coefficient functions'') and non-perturbative quantities 
$\langle f|Q_k(\mu)|i\rangle$ (``hadronic matrix elements''), respectively. 
The $Q_k$ are local operators, which are generated through the
electroweak interactions and the interplay with QCD, and govern 
``effectively'' the considered decay. The Wilson coefficients are -- simply 
speaking -- the scale-dependent couplings of the vertices described by the $Q_k$,
and contain in particular the information about the heavy degrees of freedom,
which are ``integrated out" from appearing explicitly in (\ref{LE-Ham}).
The $C_k(\mu)$ are calculated with the help of renormalization-group improved perturbation theory, which allows us to systematically sum up terms of the 
following structure:
\begin{equation}
\alpha_s^n\left[\log\left(\frac{\mu}{M_W}\right)\right]^n 
\,\,\mbox{(LO)},\quad\,\,\alpha_s^n\left[\log\left(\frac{\mu}{M_W}\right)
\right]^{n-1}\,\,\mbox{(NLO)},\quad ...\quad;
\end{equation}
detailed discussions of these rather technical aspects can be found in
\cite{B-LH98}.

For the phenomenology of CP violation, non-leptonic $B$ decays with 
$\Delta C=\Delta U=0$ play the key r\^ole. As can be seen in
Fig.~\ref{fig:topol}, transitions of this kind receive contributions both from
tree and from penguin topologies. Consequently, these decays involve, in
the SM, two heavy degrees of freedom, the $W$ boson and the
top quark. Once the corresponding fields are integrated out, their presence
is only felt through the initial conditions of the renormalization group evolution
from $\mu={\cal O}(M_W,m_t)$ down to $\mu={\cal O}(m_b)$. The corresponding
initial Wilson coefficients depend on certain Inami--Lim functions \cite{IL}, 
in analogy to the case of $B^0_q$--$\bar B^0_q$ mixing, where $S_0(x_t)$ enters.
Because of the unitarity of the CKM matrix, the following relation is implied:
\begin{equation}\label{UT-rel}
V_{ur}^\ast V_{ub}+V_{cr}^\ast V_{cb}+
V_{tr}^\ast V_{tb}=0,
\end{equation}
where the label $r=d,s$ distinguishes between $b\to d,s$ transitions. 
Consequently, only {\it two} independent weak amplitudes contribute to 
any given decay of this category. Using (\ref{UT-rel}) to eliminate 
$V_{tr}^\ast V_{tb}$, we obtain an effective Hamiltonian of the following form:
\begin{equation}\label{e4}
{\cal H}_{\mbox{{\scriptsize eff}}}=\frac{G_{\mbox{{\scriptsize 
F}}}}{\sqrt{2}}\Biggl[\sum\limits_{j=u,c}V_{jr}^\ast V_{jb}\biggl\{
\sum\limits_{k=1}^2C_k(\mu)\,Q_k^{jr}+\sum\limits_{k=3}^{10}
C_k(\mu)\,Q_k^{r}\biggr\}\Biggr].
\end{equation}
Here we have introduced another quark-flavour label $j\in\{u,c\}$,
and the four-quark operators $Q_k^{jr}$ can be divided as follows:
\begin{itemize}
\item Current--current operators:
\begin{equation}
\begin{array}{rcl}
Q_{1}^{jr}&=&(\bar r_{\alpha}j_{\beta})_{\mbox{{\scriptsize V--A}}}
(\bar j_{\beta}b_{\alpha})_{\mbox{{\scriptsize V--A}}}\\
Q_{2}^{jr}&=&(\bar r_\alpha j_\alpha)_{\mbox{{\scriptsize 
V--A}}}(\bar j_\beta b_\beta)_{\mbox{{\scriptsize V--A}}}.
\end{array}
\end{equation}
\item QCD penguin operators:
\begin{equation}\label{qcd-penguins}
\begin{array}{rcl}
Q_{3}^r&=&(\bar r_\alpha b_\alpha)_{\mbox{{\scriptsize V--A}}}\sum_{q'}
(\bar q'_\beta q'_\beta)_{\mbox{{\scriptsize V--A}}}\\
Q_{4}^r&=&(\bar r_{\alpha}b_{\beta})_{\mbox{{\scriptsize V--A}}}
\sum_{q'}(\bar q'_{\beta}q'_{\alpha})_{\mbox{{\scriptsize V--A}}}\\
Q_{5}^r&=&(\bar r_\alpha b_\alpha)_{\mbox{{\scriptsize V--A}}}\sum_{q'}
(\bar q'_\beta q'_\beta)_{\mbox{{\scriptsize V+A}}}\\
Q_{6}^r&=&(\bar r_{\alpha}b_{\beta})_{\mbox{{\scriptsize V--A}}}
\sum_{q'}(\bar q'_{\beta}q'_{\alpha})_{\mbox{{\scriptsize V+A}}}.
\end{array}
\end{equation}
\item EW penguin operators, where the $e_{q'}$ denote the
electrical quark charges:
\begin{equation}
\begin{array}{rcl}
Q_{7}^r&=&\frac{3}{2}(\bar r_\alpha b_\alpha)_{\mbox{{\scriptsize V--A}}}
\sum_{q'}e_{q'}(\bar q'_\beta q'_\beta)_{\mbox{{\scriptsize V+A}}}\\
Q_{8}^r&=&
\frac{3}{2}(\bar r_{\alpha}b_{\beta})_{\mbox{{\scriptsize V--A}}}
\sum_{q'}e_{q'}(\bar q_{\beta}'q'_{\alpha})_{\mbox{{\scriptsize V+A}}}\\
Q_{9}^r&=&\frac{3}{2}(\bar r_\alpha b_\alpha)_{\mbox{{\scriptsize V--A}}}
\sum_{q'}e_{q'}(\bar q'_\beta q'_\beta)_{\mbox{{\scriptsize V--A}}}\\
Q_{10}^r&=&
\frac{3}{2}(\bar r_{\alpha}b_{\beta})_{\mbox{{\scriptsize V--A}}}
\sum_{q'}e_{q'}(\bar q'_{\beta}q'_{\alpha})_{\mbox{{\scriptsize V--A}}}.
\end{array}
\end{equation}
\end{itemize}
Here $\alpha$, $\beta$ are $SU(3)_{\rm C}$ indices, ${\rm V \pm A}$ refers
to $\gamma_\mu(1\pm\gamma_5)$, and $q'\in\{u,d,c,s,b\}$ runs over the
active quark flavours at $\mu={\cal O}(m_b)$. For such a renormalization
scale, the Wilson coefficients of the 
current--current operators are $C_1(\mu)={\cal O}(10^{-1})$ and 
$C_2(\mu)={\cal O}(1)$, whereas those of the penguin operators are found to
be at most of ${\cal O}(10^{-2})$ \cite{B-LH98}. 

The short-distance part of (\ref{e4}) is nowadays under full control. On the other 
hand, the long-distance piece suffers still from large theoretical uncertainties.
For a given non-leptonic decay $\bar B \to \bar f$, it is described by the hadronic 
matrix elements $\langle \bar f|Q_k(\mu)|\bar B\rangle$ of the
four-quark operators. A popular way of dealing with these quantities is to 
assume that they ``factorize'' into the product of the matrix elements of
two quark currents at some ``factorization scale'' $\mu=\mu_{\rm F}$. 
This procedure can be justified in the large-$N_{\rm C}$ approximation 
\cite{largeN}, where $N_{\rm C}$ is the number of $SU(N_{\rm C})$ quark colours, 
and there are decays, where this concept is suggested by
``colour transparency" arguments \cite{QCDF-old}. However, it is in general 
not on solid ground. Interesting theoretical progress could be made 
through the development of the ``QCD factorization" (QCDF) \cite{BBNS} 
and ``perturbative QCD" (PQCD) \cite{PQCD} approaches, and most recently
through the ``soft collinear effective theory'' (SCET) \cite{SCET}. Moreover,
also QCD light-cone sum-rule techniques were applied to non-leptonic
$B$ decays \cite{LCSR}. An important target of these analyses is given by 
$B\to\pi\pi$ and $B\to\pi K$ decays. Thanks to the $B$ factories, the 
corresponding theoretical results can now be confronted with experiment. 
Since the data indicate large non-factorizable corrections 
\cite{BFRS}--\cite{CGRS}, the long-distance contributions to these decays 
remain a theoretical challenge.

\subsection{Strategies for the Exploration of CP Violation}\label{ssec:CP-strat}
Let us consider a non-leptonic decay $\bar B\to\bar f$ that is described by
the low-energy effective Hamiltonian  in (\ref{e4}). The corresponding
decay amplitude is then given as follows:
\begin{eqnarray}
\lefteqn{\hspace*{-1.3truecm}A(\bar B\to \bar f)=\langle \bar f\vert
{\cal H}_{\mbox{{\scriptsize eff}}}\vert\bar B\rangle}\nonumber\\
&&\hspace*{-1.3truecm}=\frac{G_{\mbox{{\scriptsize F}}}}{\sqrt{2}}\left[
\sum\limits_{j=u,c}V_{jr}^\ast V_{jb}\left\{\sum\limits_{k=1}^2
C_{k}(\mu)\langle \bar f\vert Q_{k}^{jr}(\mu)\vert\bar B\rangle
+\sum\limits_{k=3}^{10}C_{k}(\mu)\langle \bar f\vert Q_{k}^r(\mu)
\vert\bar B\rangle\right\}\right].~~~\mbox{}\label{Bbarfbar-ampl}
\end{eqnarray}
Concerning the CP-conjugate process $B\to\ f$, we have
\begin{eqnarray}
\lefteqn{\hspace*{-1.3truecm}A(B \to f)=\langle f|
{\cal H}_{\mbox{{\scriptsize 
eff}}}^\dagger|B\rangle}\nonumber\\
&&\hspace*{-1.3truecm}=\frac{G_{\mbox{{\scriptsize F}}}}{\sqrt{2}}
\left[\sum\limits_{j=u,c}V_{jr}V_{jb}^\ast \left\{\sum\limits_{k=1}^2
C_{k}(\mu)\langle f\vert Q_{k}^{jr\dagger}(\mu)\vert B\rangle
+\sum\limits_{k=3}^{10}C_{k}(\mu)\langle f\vert Q_k^{r\dagger}(\mu)
\vert B\rangle\right\}\right].~~~\mbox{}\label{Bf-ampl}
\end{eqnarray}
If we use now that strong interactions are invariant under CP transformations 
(omitting the ``strong CP problem" \cite{strong-CP}, which leads to negligible 
effects in the processes considered here), insert $({\cal CP})^\dagger({\cal CP})=\hat 1$ 
both after the $\langle f|$ and in front of the $|B\rangle$, and take the 
relation $({\cal CP})Q_k^{jr\dagger}({\cal CP})^\dagger=Q_k^{jr}$
into account, we arrive at
\begin{eqnarray}
\lefteqn{\hspace*{-1.3truecm}A(B \to f)=
e^{i[\phi_{\mbox{{\scriptsize CP}}}(B)-\phi_{\mbox{{\scriptsize CP}}}(f)]}}\nonumber\\
&&\hspace*{-1.3truecm}\times\frac{G_{\mbox{{\scriptsize F}}}}{\sqrt{2}}
\left[\sum\limits_{j=u,c}V_{jr}V_{jb}^\ast\left\{\sum\limits_{k=1}^2
C_{k}(\mu)\langle \bar f\vert Q_{k}^{jr}(\mu)\vert\bar B\rangle
+\sum\limits_{k=3}^{10}C_{k}(\mu)
\langle \bar f\vert Q_{k}^r(\mu)\vert\bar B\rangle\right\}\right],
\end{eqnarray}
where the convention-dependent phases $\phi_{\mbox{{\scriptsize CP}}}(B)$ 
and $\phi_{\mbox{{\scriptsize CP}}}(f)$ are defined in analogy to (\ref{CP-def}).
Consequently, we may write
\begin{eqnarray}
A(\bar B\to\bar f)&=&e^{+i\varphi_1}
|A_1|e^{i\delta_1}+e^{+i\varphi_2}|A_2|e^{i\delta_2}\label{par-ampl}\\
A(B\to f)&=&e^{i[\phi_{\mbox{{\scriptsize CP}}}(B)-\phi_{\mbox{{\scriptsize CP}}}(f)]}
\left[e^{-i\varphi_1}|A_1|e^{i\delta_1}+e^{-i\varphi_2}|A_2|e^{i\delta_2}
\right].\label{par-ampl-CP}
\end{eqnarray}
Here the CP-violating phases $\varphi_{1,2}$ originate from the CKM factors 
$V_{jr}^\ast V_{jb}$, and the CP-conserving ``strong'' amplitudes
$|A_{1,2}|e^{i\delta_{1,2}}$ involve the hadronic matrix elements of the 
four-quark operators. In fact, these expressions are the most general forms
of any non-leptonic $B$-decay amplitude in the SM, i.e.\ they do not only
refer to the $\Delta C=\Delta U=0$ case described by (\ref{e4}). 
Using (\ref{par-ampl}) and (\ref{par-ampl-CP}), we obtain
the following CP asymmetry:
\begin{eqnarray}
{\cal A}_{\rm CP}&\equiv&\frac{\Gamma(B\to f)-
\Gamma(\bar B\to\bar f)}{\Gamma(B\to f)+\Gamma(\bar B
\to \bar f)}=\frac{|A(B\to f)|^2-|A(\bar B\to \bar f)|^2}{|A(B\to f)|^2+
|A(\bar B\to \bar f)|^2}\nonumber\\
&=&\frac{2|A_1||A_2|\sin(\delta_1-\delta_2)
\sin(\varphi_1-\varphi_2)}{|A_1|^2+2|A_1||A_2|\cos(\delta_1-\delta_2)
\cos(\varphi_1-\varphi_2)+|A_2|^2}.\label{direct-CPV}
\end{eqnarray}
We observe that a non-vanishing value can be generated through the 
interference between the two weak amplitudes, provided both a non-trivial 
weak phase difference $\varphi_1-\varphi_2$ and a non-trivial strong phase 
difference $\delta_1-\delta_2$ are present. This kind of
CP violation is referred to as ``direct'' CP violation, as it originates 
directly at the amplitude level of the considered decay. It is the 
$B$-meson counterpart of the effect that is probed through 
$\mbox{Re}(\varepsilon_K'/\varepsilon_K)$ in the neutral kaon 
system, and could recently be established with the help of 
$B_d\to\pi^\mp K^\pm$ decays \cite{CP-B-dir}, as we will see in
Subsection~\ref{ssec:Bpi+pi-}.

Since $\varphi_1-\varphi_2$ is in general given by one of the UT angles  -- 
usually $\gamma$ -- the goal is to extract this quantity from the measured 
value of ${\cal A}_{\rm CP}$. Unfortunately, hadronic uncertainties affect this
determination through the poorly known hadronic matrix elements in 
(\ref{Bbarfbar-ampl}). In order to deal with this problem, we may proceed along 
one of the following two avenues:
\begin{itemize}
\item[(i)] Amplitude relations can be used to eliminate the 
hadronic matrix elements. We distinguish between exact relations, 
using pure ``tree'' decays  of the kind $B\to KD$ \cite{gw,ADS} or 
$B_c\to D_sD$ \cite{fw}, and relations, which follow from the flavour symmetries 
of strong interactions, i.e.\ isospin or $SU(3)_{\rm F}$, and involve 
$B_{(s)}\to\pi\pi,\pi K,KK$ modes~\cite{GHLR}. 
\item[(ii)] In decays of neutral $B_q$ mesons ($q\in\{d,s\}$), interference effects 
between $B^0_q$--$\bar B^0_q$ mixing and decay processes may induce
 ``mixing-induced CP violation''. If a single CKM amplitude governs the decay, 
 the hadronic matrix elements cancel in the corresponding
CP asymmeties; otherwise we have to use again amplitude relations.
The most important example is the decay $B^0_d\to J/\psi K_{\rm S}$ \cite{bisa}.
\end{itemize}

As neutral $B_q$  mesons play an outstanding r\^ole for the exploration
of CP violation, let us have a closer look at their CP asymmetries.  
A particularly simple -- but also very interesting  -- situation arises 
if we restrict ourselves to decays into final states $f$ that are eigenstates of 
the CP operator, i.e.\ satisfy the relation 
\begin{equation}\label{CP-eigen}
({\cal CP})|f\rangle=\pm |f\rangle. 
\end{equation}
Looking at (\ref{xi-def}), we see that $\xi_f^{(q)}=\xi_{\bar f}^{(q)}$ in this case.
If we use the decay rates in (\ref{rates}), we arrive at a time-dependent CP 
asymmetry of the following structure:
\begin{eqnarray}
{\cal A}_{\rm CP}(t)&\equiv&\frac{\Gamma(B^0_q(t)\to f)-
\Gamma(\bar B^0_q(t)\to f)}{\Gamma(B^0_q(t)\to f)+
\Gamma(\bar B^0_q(t)\to f)}\nonumber\\
&=&\left[\frac{{\cal A}_{\rm CP}^{\rm dir}(B_q\to f)\,\cos(\Delta M_q t)+
{\cal A}_{\rm CP}^{\rm mix}(B_q\to f)\,\sin(\Delta 
M_q t)}{\cosh(\Delta\Gamma_qt/2)-{\cal A}_{\rm 
\Delta\Gamma}(B_q\to f)\,\sinh(\Delta\Gamma_qt/2)}\right],\label{time-dep-CP}
\end{eqnarray}
where
\begin{equation}\label{CPV-OBS}
{\cal A}^{\mbox{{\scriptsize dir}}}_{\mbox{{\scriptsize CP}}}(B_q\to f)\equiv
\frac{1-\bigl|\xi_f^{(q)}\bigr|^2}{1+\bigl|\xi_f^{(q)}\bigr|^2},\qquad
{\cal A}^{\mbox{{\scriptsize mix}}}_{\mbox{{\scriptsize
CP}}}(B_q\to f)\equiv\frac{2\,\mbox{Im}\,\xi^{(q)}_f}{1+
\bigl|\xi^{(q)}_f\bigr|^2}.
\end{equation}
Since we may write
\begin{equation}
{\cal A}^{\mbox{{\scriptsize dir}}}_{\mbox{{\scriptsize CP}}}(B_q\to f)=
\frac{|A(B^0_q\to f)|^2-|A(\bar B^0_q\to \bar f)|^2}{|A(B^0_q\to f)|^2+
|A(\bar B^0_q\to \bar f)|^2},
\end{equation}
we see that this quantity measures the direct CP violation in the decay
$B_q\to f$, which originates from the interference between different
weak amplitudes (see (\ref{direct-CPV})). On the other hand, the interesting 
new aspect of (\ref{time-dep-CP}) is given by
${\cal A}^{\mbox{{\scriptsize mix}}}_{\mbox{{\scriptsize
CP}}}(B_q\to f)$, which is generated through the interference between 
$B_q^0$--$\bar B_q^0$ mixing and decay processes, thereby describing
``mixing-induced'' CP violation. Finally, the width difference $\Delta\Gamma_q$,
which is expected to be sizeable in the $B_s$-meson system, provides another
observable:
\begin{equation}\label{ADGam}
{\cal A}_{\rm \Delta\Gamma}(B_q\to f)\equiv
\frac{2\,\mbox{Re}\,\xi^{(q)}_f}{1+\bigl|\xi^{(q)}_f\bigr|^2}.
\end{equation}
Because of the relation
\begin{equation}\label{Obs-rel}
\Bigl[{\cal A}_{\rm CP}^{\rm dir}(B_q\to f)\Bigr]^2+
\Bigl[{\cal A}_{\rm CP}^{\rm mix}(B_q\to f)\Bigr]^2+
\Bigl[{\cal A}_{\Delta\Gamma}(B_q\to f)\Bigr]^2=1,
\end{equation}
it is, however, not independent  from ${\cal A}^{\mbox{{\scriptsize 
dir}}}_{\mbox{{\scriptsize CP}}}(B_q\to f)$ and 
${\cal A}^{\mbox{{\scriptsize mix}}}_{\mbox{{\scriptsize CP}}}(B_q\to f)$.

In order to calculate $\xi_f^{(q)}$, we use the general expressions in 
(\ref{par-ampl}) and (\ref{par-ampl-CP}), where 
$e^{-i\phi_{\mbox{{\scriptsize CP}}}(f)}=\pm1$ because of (\ref{CP-eigen}), and 
$\phi_{\mbox{{\scriptsize CP}}}(B)=\phi_{\mbox{{\scriptsize CP}}}(B_q)$. If we insert
these amplitude parametrizations into (\ref{xi-def}) and take (\ref{theta-def}) into 
account, we observe that the phase-convention-dependent 
quantity $\phi_{\mbox{{\scriptsize CP}}}(B_q)$ cancels, and finally 
arrive at
\begin{equation}\label{xi-re}
\xi_f^{(q)}=\mp\, e^{-i\phi_q}\left[
\frac{e^{+i\varphi_1}|A_1|e^{i\delta_1}+
e^{+i\varphi_2}|A_2|e^{i\delta_2}}{
e^{-i\varphi_1}|A_1|e^{i\delta_1}+
e^{-i\varphi_2}|A_2|e^{i\delta_2}}\right],
\end{equation}
where
\begin{equation}\label{phiq-def}
\phi_q\equiv 2\,\mbox{arg} (V_{tq}^\ast V_{tb})=\left\{\begin{array}{cl}
+2\beta&\mbox{($q=d$)}\\
-2\delta\gamma&\mbox{($q=s$)}\end{array}\right.
\end{equation}
is associated with the CP-violating weak $B_q^0$--$\bar B_q^0$ mixing
phase arising in the SM; $\beta$ and $\delta\gamma$ refer to the corresponding
angles in the unitarity triangles shown in Fig.\ \ref{fig:UT}.

In analogy to (\ref{direct-CPV}), the caclulation
of $\xi_f^{(q)}$ is -- in general -- also affected by large hadronic uncertainties. 
However, if one CKM amplitude plays the dominant r\^ole in the $B_q\to f$
transition, we obtain
\begin{equation}\label{xi-si}
\xi_f^{(q)}=\mp\, e^{-i\phi_q}\left[
\frac{e^{+i\phi_f/2}|M_f|e^{i\delta_f}}{e^{-i\phi_f/2}|M_f|e^{i\delta_f}}
\right]=\mp\, e^{-i(\phi_q-\phi_f)},
\end{equation}
and observe that the hadronic matrix element $|M_f|e^{i\delta_f}$ 
cancels in this expression. Since the requirements for 
direct CP violation discussed above are no longer satisfied, direct CP violation 
vanishes in this important special case, i.e.\ 
${\cal A}^{\mbox{{\scriptsize dir}}}_{\mbox{{\scriptsize CP}}}
(B_q\to f)=0$. On the other hand, this is {\it not} the case for the mixing-induced 
CP asymmetry. In particular, 
\begin{equation}\label{Amix-simple}
{\cal A}^{\rm mix}_{\rm CP}(B_q\to f)=\pm\sin\phi
\end{equation}
is now governed by the CP-violating weak phase difference 
$\phi\equiv\phi_q-\phi_f$ and is not affected by hadronic 
uncertainties. The corresponding time-dependent CP asymmetry
takes then the simple form
\begin{equation}\label{Amix-t-simple}
\left.\frac{\Gamma(B^0_q(t)\to f)-
\Gamma(\bar B^0_q(t)\to \bar f)}{\Gamma(B^0_q(t)\to f)+
\Gamma(\bar B^0_q(t)\to \bar f)}\right|_{\Delta\Gamma_q=0}
=\pm\sin\phi\,\sin(\Delta M_q t),
\end{equation}
and allows an elegant determination of $\sin\phi$.

\begin{figure}
   \centering
   \includegraphics[width=10.0truecm]{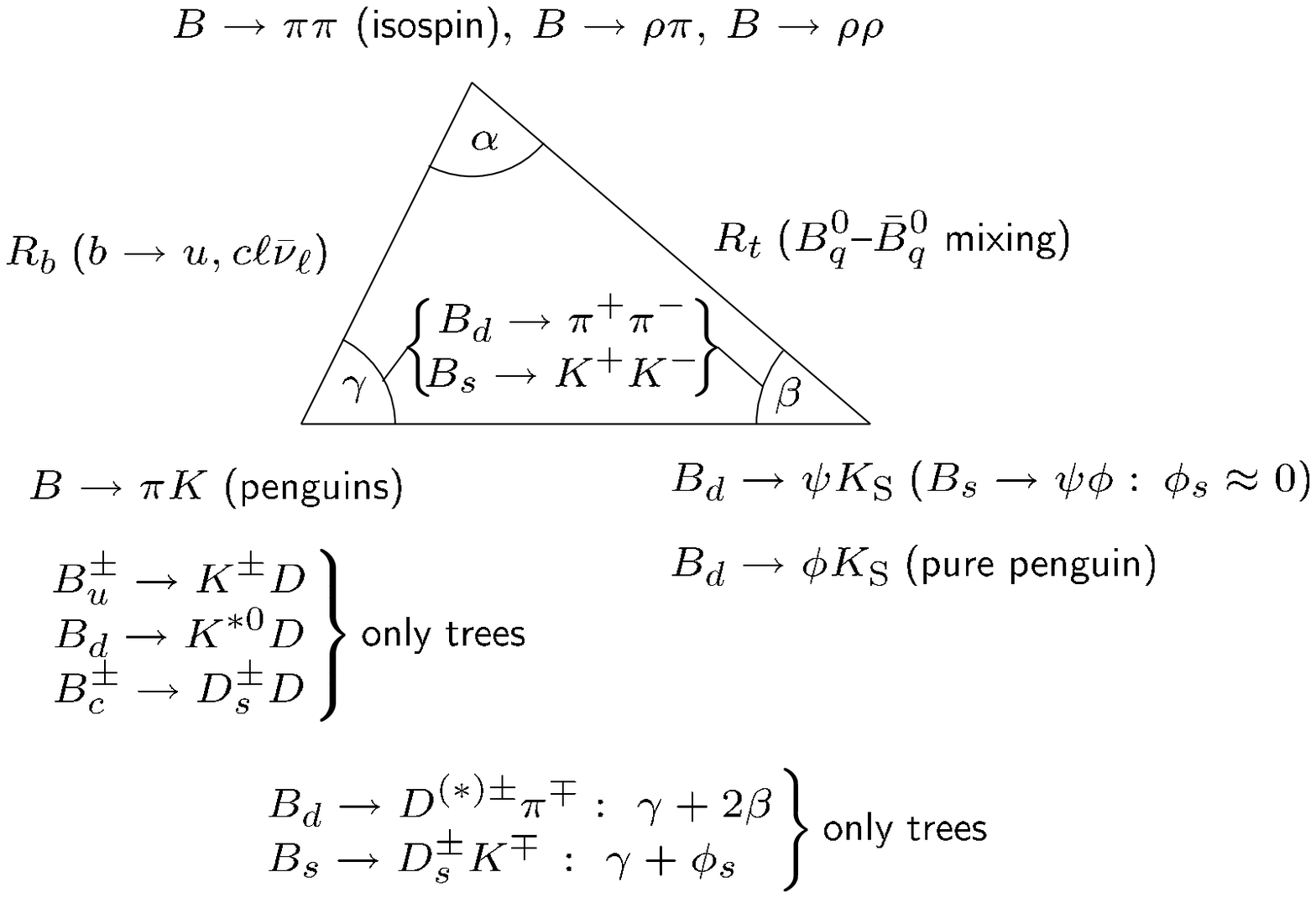} 
   \caption{A brief roadmap of $B$-decay strategies for the exploration of
   CP violation.}
   \label{fig:flavour-map}
\end{figure}

\subsection{How Could New Physics Enter?}\label{sec:NP}
Using the concept of the low-energy effective Hamiltonians introduced
in Subsection~\ref{ssec:non-lept}, we may address this important question
in a systematic manner \cite{buras-NP}:
\begin{itemize}
\item[(i)] NP may modify the ``strength" of the SM operators through new
short-distance functions which depend on the NP parameters, such as the masses 
of charginos, squarks, charged Higgs particles and $\tan\bar\beta\equiv v_2/v_1$ 
in the ``minimal supersymmetric SM'' (MSSM). The NP particles may enter in 
box and penguin topologies, and are ``integrated out'' as the $W$ boson and 
top quark in the SM. Consequently, the initial conditions for the
renormalization-group evolution take the following form:
\begin{equation}\label{WC-NP}
C_k \to C_k^{\rm SM} + C_k^{\rm NP}.
\end{equation}
It should be emphasized that the NP pieces $C_k^{\rm NP}$ may also involve 
new CP-violating phases which are {\it not} related to the CKM matrix.
\item[(ii)] NP may enhance the operator basis:
\begin{equation}
\{Q_k\} \to \{Q_k^{\rm SM}, Q_l^{\rm NP}\},
\end{equation}
so that operators which are not present (or strongly suppressed) in the 
SM may actually play an important r\^ole. In this case, we encounter, 
in general, also new sources for flavour and CP violation.
\end{itemize}
The $B$-meson system offers a variety of processes and strategies for the
exploration of CP violation \cite{CKM-book,RF-Phys-Rep}, as we have illustrated in 
Fig.~\ref{fig:flavour-map} through a collection of prominent examples. 
We see that there are processes with a very {\it different} dynamics that 
are -- in the SM -- sensitive to the {\it same} angles of the UT. 
Moreover, rare $B$- and $K$-meson decays \cite{rare}, 
which originate from loop effects in the SM, provide complementary insights 
into flavour physics and interesting correlations with the CP-B sector; key 
examples are $B\to X_s\gamma$ and the exclusive modes
$B\to K^\ast \gamma$, $B\to\rho\gamma$, as well as $B_{s,d}\to \mu^+\mu^-$ 
and $K^+\to\pi^+\nu\bar\nu$, $K_{\rm L}\to\pi^0\nu\bar\nu$. 

In the presence of NP contributions, the subtle interplay between the different 
processes could well be disturbed. There are two popular avenues for NP to 
enter the roadmap of quark-flavour physics:
\begin{itemize}
\item[(i)]{\it $B^0_q$--$\bar B^0_q$ mixing:} NP could enter through the exchange
of new particles in the box diagrams, or through new contributions at the
tree level, thereby leading to 
\begin{equation}\label{Dm-Phi-NP}
\Delta M_q=\Delta M_q^{\rm SM}+\Delta M_q^{\rm NP}, \quad
\phi_q=\phi_q^{\rm SM}+\phi_q^{\rm NP}.
\end{equation}
Whereas $\Delta M_q^{\rm NP}$ would affect the determination of the UT 
side $R_t$, $\phi_q^{\rm NP}$ would manifest itself through mixing-induced
CP asymmetries. Using dimensional arguments borrowed from effective field 
theory \cite{FM-BpsiK,FIM}, it can be shown that 
$\Delta M_q^{\rm NP}/\Delta M_q^{\rm SM}\sim1$ and
$\phi_q^{\rm NP}/\phi_q^{\rm SM}\sim1$ could -- in principle -- be possible
for a NP scale $\Lambda_{\rm NP}$ in the TeV regime; such a pattern may 
also arise in specific NP scenarios. Thanks to the $B$-factory data, dramatic
NP effects of this kind are already ruled out in the $B_d$-meson system,
although the new world average for $(\sin 2\beta)_{\psi K_{\rm S}}$ 
could be interpreted in terms of $\phi_q^{\rm NP}\sim-8^\circ$.
On the other hand, the $B_s$ sector is still essentially unexplored, thereby 
leaving a lot of hope for the LHC.
\item[(ii)]{\it Decay amplitudes:} NP has typically a small effect if SM tree processes
play the dominant r\^ole. However, NP could well have a significant impact on 
the FCNC sector: new particles may enter in penguin or box diagrams, or new 
FCNC contributions may even be generated at the tree level. In fact, sizeable 
contributions arise generically in field-theoretical estimates with 
$\Lambda_{\rm NP}\sim\mbox{TeV}$ \cite{FM-BphiK}, as well as in specific 
NP models. Interestingly, there are hints in the $B$-factory data that this may 
actually be the case. 
\end{itemize}
Concerning model-dependent NP analyses, in particular SUSY
scenarios have received a lot of attention; for a selection of recent studies, see
Refs.~\cite{GOSST}--\cite{GHK}. Examples of other fashionable NP scenarios 
are left--right-symmetric models \cite{LR-sym}, scenarios with extra dimensions
\cite{extra-dim}, models with an extra $Z'$ \cite{Z-prime}, ``little Higgs'' 
scenarios \cite{little-higgs}, and models with a fourth generation \cite{hou-4}.

The simplest extension of the SM is given by models with ``minimal flavour violation'' (MFV). Following the characterization given in Ref.~\cite{MFV-1}, 
the flavour-changing processes are here still governed by the CKM matrix -- in 
particular there are no new sources for CP violation --  and the only relevant 
operators are those present in the SM (for an alternative definition, see 
Ref.~\cite{MFV-2}). Specific examples are the Two-Higgs Doublet Model II,
the MSSM without new sources of flavour violation and $\tan\bar\beta$ not
too large, models with one extra universal dimension and the simplest
little Higgs models. Due to their simplicity, the extensions of the SM with
MFV show several correlations between various observables, 
thereby allowing for powerful tests of this scenario \cite{buras-MFV}. A 
systematic discussion of  models with ``next-to-minimal flavour violation" was 
recently given in Ref.~\cite{NMFV}.

There are other fascinating probes for the search of NP. Important examples are 
the $D$-meson system \cite{petrov}, electric dipole moments \cite{PR}, or 
flavour-violating charged lepton decays \cite{CEPRT}. Since a discussion of
these topics is beyond the scope of this review, the interested reader should consult 
the corresponding references. Let us next have a closer look at prominent $B$ 
decays, with a particular emphasis of the impact of NP.

\begin{figure}[t]
\centerline{
 \includegraphics[width=5.7truecm]{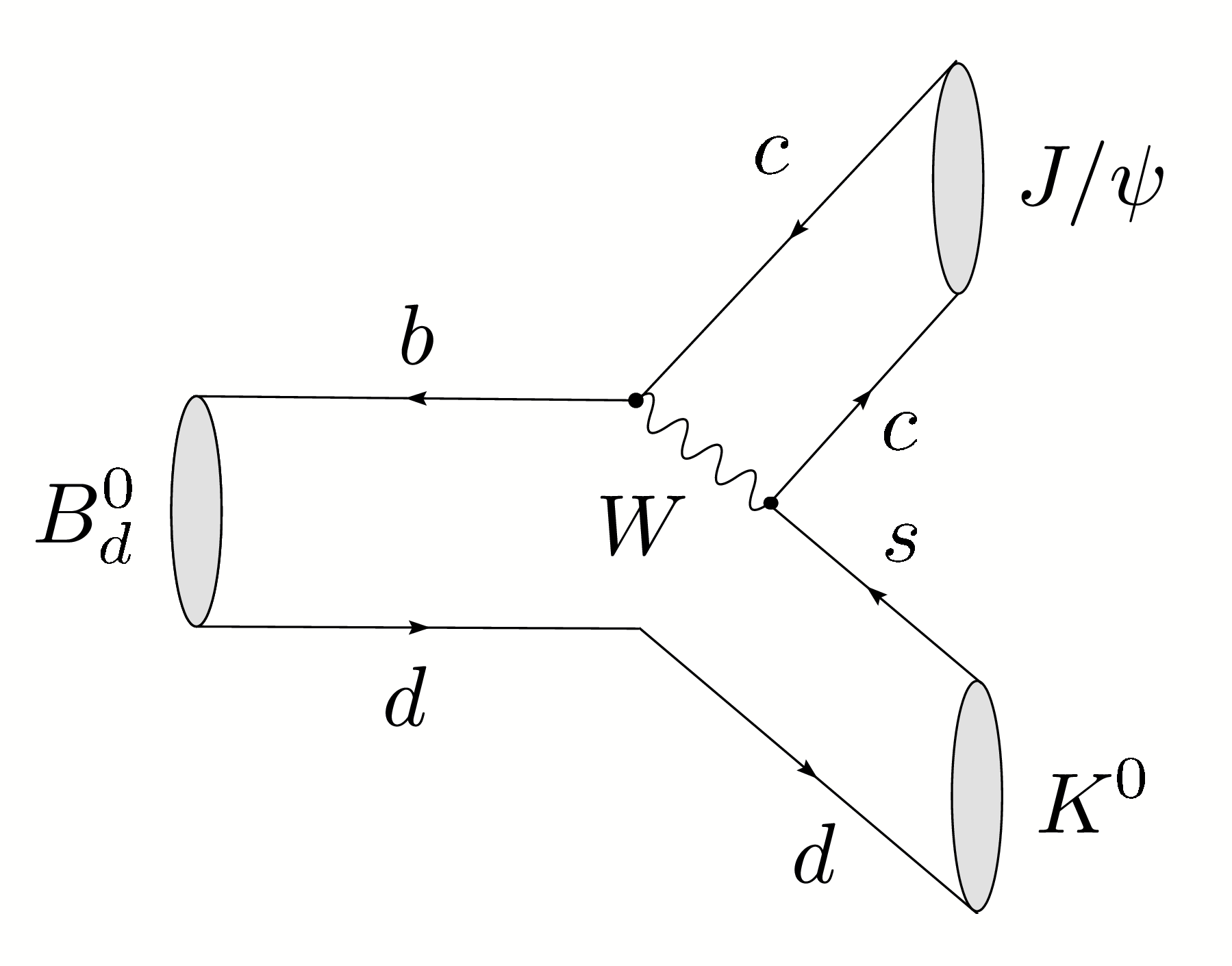}
 \hspace*{0.5truecm}
 \includegraphics[width=5.7truecm]{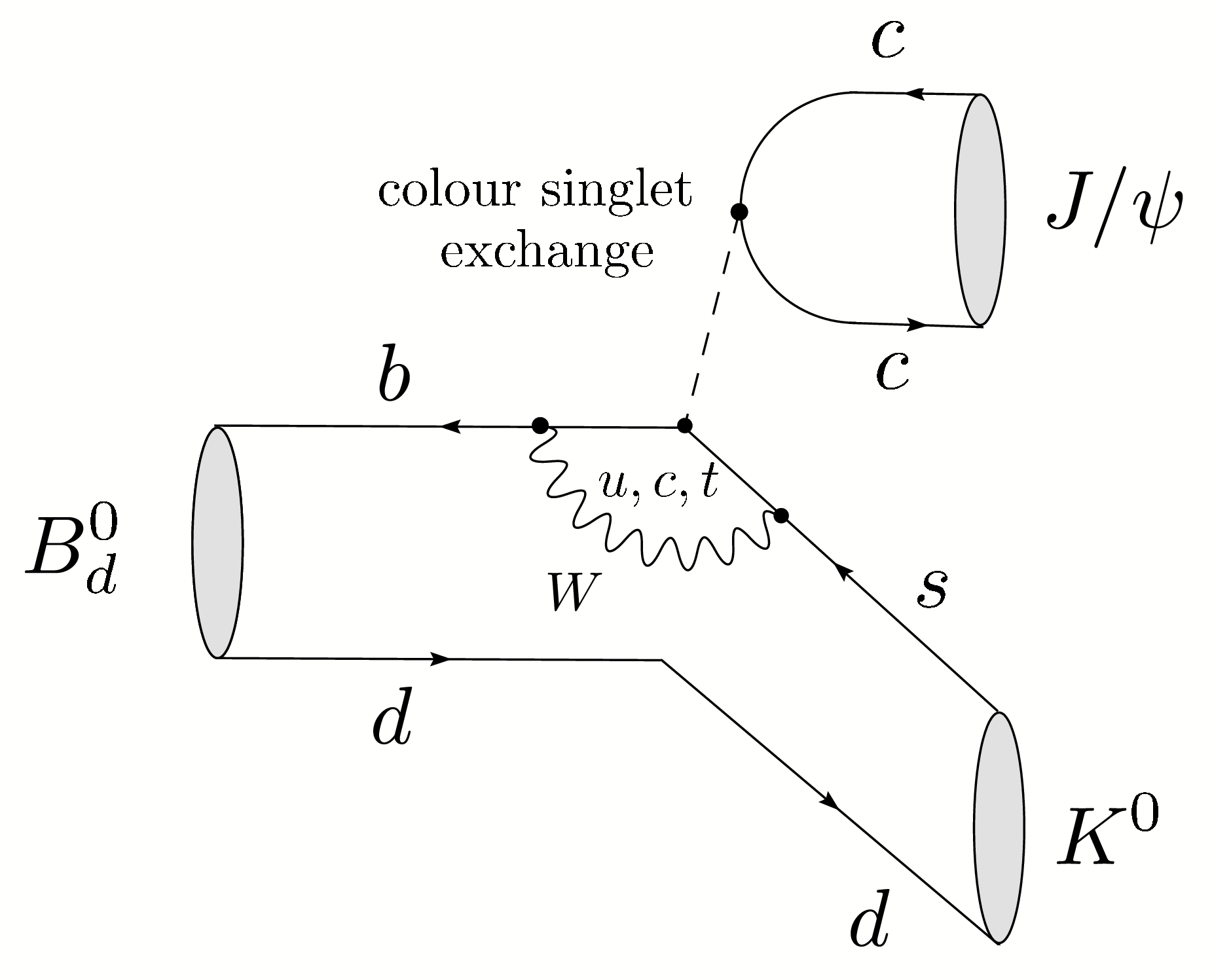}  
 }
 \vspace*{-0.3truecm}
\caption{Feynman diagrams contributing to $B^0_d\to J/\psi K^0$ 
decays.}\label{fig:BpsiK-diag}
\end{figure}

\boldmath
\section{Status of Important $B$-Factory Benchmark Modes}\label{sec:bench}
\unboldmath
\subsection{$B^0_d\to J/\psi K_{\rm S}$}\label{ssec:BpsiK}
This decay has a CP-odd final state, and originates from 
$\bar b\to\bar c c \bar s$ quark-level transitions. Consequently, as we
discussed in the context of the classification in Subsection~\ref{ssec:non-lept},
it receives contributions both from tree and from penguin topologies, 
as can be seen in Fig.~\ref{fig:BpsiK-diag}. In the SM, the decay 
amplitude can hence be written as follows \cite{RF-BdsPsiK}:
\begin{equation}\label{Bd-ampl1}
A(B_d^0\to J/\psi K_{\rm S})=\lambda_c^{(s)}\left(A_{\rm T}^{c'}+
A_{\rm P}^{c'}\right)+\lambda_u^{(s)}A_{\rm P}^{u'}
+\lambda_t^{(s)}A_{\rm P}^{t'}.
\end{equation}
Here the
\begin{equation}\label{lamqs-def}
\lambda_q^{(s)}\equiv V_{qs}V_{qb}^\ast
\end{equation}
are CKM factors, $A_{\rm T}^{c'}$ is the CP-conserving strong tree amplitude, 
while the $A_{\rm P}^{q'}$ describe the penguin topologies with internal 
$q$ quarks ($q\in\{u,c,t\})$, including QCD and EW penguins; 
the primes remind us that we are dealing with a $\bar b\to\bar s$ 
transition. If we eliminate now $\lambda_t^{(s)}$ through (\ref{UT-rel}) 
and apply the Wolfenstein parametrization, we obtain
\begin{equation}\label{BdpsiK-ampl2}
A(B_d^0\to J/\psi K_{\rm S})\propto\left[1+\lambda^2 a e^{i\theta}
e^{i\gamma}\right],
\end{equation}
where
\begin{equation}
a e^{i\vartheta}\equiv\left(\frac{R_b}{1-\lambda^2}\right)
\left[\frac{A_{\rm P}^{u'}-A_{\rm P}^{t'}}{A_{\rm T}^{c'}+
A_{\rm P}^{c'}-A_{\rm P}^{t'}}\right]
\end{equation}
is a hadronic parameter. Using now the formalism of 
Subsection~\ref{ssec:CP-strat} yields
\begin{equation}\label{xi-BdpsiKS}
\xi_{\psi K_{\rm S}}^{(d)}=+e^{-i\phi_d}\left[\frac{1+
\lambda^2a e^{i\vartheta}e^{-i\gamma}}{1+\lambda^2a e^{i\vartheta}
e^{+i\gamma}}\right].
\end{equation}
Unfortunately, $a e^{i\vartheta}$, which is a measure for the ratio of the
$B_d^0\to J/\psi K_{\rm S}$ penguin to tree contributions,
can only be estimated with large hadronic uncertainties. However, since 
this parameter enters (\ref{xi-BdpsiKS}) in a doubly Cabibbo-suppressed way, its 
impact on the CP-violating observables is practically negligible. We can put 
this important statement on a more quantitative basis by making the plausible
assumption that $a={\cal O}(\bar\lambda)={\cal O}(0.2)={\cal O}(\lambda)$,
where $\bar\lambda$ is a ``generic'' expansion parameter:
\begin{eqnarray}
{\cal A}^{\mbox{{\scriptsize dir}}}_{\mbox{{\scriptsize
CP}}}(B_d\to J/\psi K_{\mbox{{\scriptsize S}}})&=&0+
{\cal O}(\overline{\lambda}^3)\label{Adir-BdpsiKS}\\
{\cal A}^{\mbox{{\scriptsize mix}}}_{\mbox{{\scriptsize
CP}}}(B_d\to J/\psi K_{\mbox{{\scriptsize S}}})&=&-\sin\phi_d +
{\cal O}(\overline{\lambda}^3) \, \stackrel{\rm SM}{=} \, -\sin2\beta+
{\cal O}(\overline{\lambda}^3).\label{Amix-BdpsiKS}
\end{eqnarray}
Consequently, (\ref{Amix-BdpsiKS}) allows an essentially {\it clean}
determination of $\sin2\beta$ \cite{bisa}.

Since the CKM fits performed within the SM pointed to a large value of 
$\sin2\beta$, $B^0_d\to J/\psi K_{\rm S}$ offered the exciting perspective 
of exhibiting {\it large} mixing-induced CP violation. In 2001, the measurement of  
${\cal A}^{\mbox{{\scriptsize mix}}}_{\mbox{{\scriptsize CP}}}
(B_d\to J/\psi K_{\mbox{{\scriptsize S}}})$
allowed indeed the first observation of CP violation {\it outside} the 
$K$-meson system \cite{CP-B-obs}.
The most recent data are still not showing any signal for {\it direct} CP violation
in $B^0_d\to J/\psi K_{\rm S}$ within the current uncertainties, as is expected from 
(\ref{Adir-BdpsiKS}). The current world average reads as follows \cite{HFAG}:
\begin{equation}
{\cal A}_{\rm CP}^{\rm dir}(B_d\to J/\psi K_{\rm S})=0.026\pm0.041.
\end{equation}
As far as (\ref{Amix-BdpsiKS}) is concerned, we have
\begin{equation}\label{s2b-psiK-exp}
\hspace*{-2.0truecm}(\sin2\beta)_{\psi K_{\rm S}}\equiv 
-{\cal A}^{\mbox{{\scriptsize mix}}}_{\mbox{{\scriptsize
CP}}}(B_d\to J/\psi K_{\mbox{{\scriptsize S}}})
=\left\{
\begin{array}{ll}
0.722\pm0.040\pm0.023 & \mbox{(BaBar \cite{s2b-babar})}\\
0.652\pm0.039\pm0.020 & \mbox{(Belle \cite{s2b-belle}),}
\end{array}
\right.
\end{equation}
which gives the following world average \cite{HFAG}:
\begin{equation}\label{s2b-average}
(\sin 2\beta)_{\psi K_{\rm S}}=0.687\pm0.032.
\end{equation}
Within the SM, the theoretical uncertainties are generically expected to be
below the 0.01 level; significantly smaller effects are found in \cite{BMR}, 
whereas a fit performed in \cite{CPS} yields a theoretical penguin uncertainty 
comparable to the present experimental systematic error. A possibility
to control these uncertainties is provided by the $B^0_s\to J/\psi K_{\rm S}$ 
channel \cite{RF-BdsPsiK}, which can be explored at the LHC \cite{LHC-Book}.

In \cite{FM-BpsiK}, a set of observables was introduced, which allows us to search
systematically for NP contributions to the $B\to J/\psi K $ decay amplitudes. It
uses also the charged $B^\pm\to J/\psi K^\pm$ decay, and is given as follows:
\begin{equation}\label{BpsiK}
{\cal B}_{\psi K}\equiv \frac{1-{\cal A}_{\psi K}}{1+{\cal A}_{\psi K}}, 
\end{equation}
with
\begin{equation}\label{ApsiK-def}
{\cal A}_{\psi K}\equiv\left[\frac{\mbox{BR}(B^+\to J/\psi K^+)+
\mbox{BR}(B^-\to J/\psi K^-)}{\mbox{BR}(B^0_d\to J/\psi K^0)+
\mbox{BR}(\bar B^0_d\to J/\psi\bar K^0)}
\right]\left[\frac{\tau_{B^0_d}}{\tau_{B^+}}\right],
\end{equation}
and
\begin{equation}\label{DpmPsiK}
{\cal D}^\pm_{\psi K}\equiv\frac{1}{2}\left[
{\cal A}_{\rm CP}^{\rm dir}(B_d\to J/\psi K_{\rm S})\pm
{\cal A}_{\rm CP}^{\rm dir}(B^\pm\to J/\psi K^\pm)\right].
\end{equation}
As is discussed in detail in \cite{RF-Phys-Rep,FM-BpsiK}, 
the observables ${\cal B}_{\psi K}$ 
and ${\cal D}^-_{\psi K}$ are sensitive to NP in the $I=1$ isospin sector, 
whereas a non-vanishing value of ${\cal D}^+_{\psi K}$ would signal NP in the
$I=0$ isospin sector. Moreover, the NP contributions with $I=1$ are expected
to be dynamically suppressed with respect to the $I=0$ case because of their 
flavour structure. Using the most recent $B$-factory results, we obtain 
\begin{equation}\label{B-Dpm-PsiK-res}
\hspace*{-0.7truecm} {\cal B}_{\psi K} =-0.035\pm0.037,\quad
{\cal D}^-_{\psi K}=0.010\pm0.023, \quad
{\cal D}^+_{\psi K}=0.017\pm0.023.
\end{equation}
Consequently, NP effects of ${\cal O}(10\%)$ in the $I=1$ sector of the 
$B\to J/\psi K$ decay amplitudes are already disfavoured by the
data for ${\cal B}_{\psi K}$ and ${\cal D}^-_{\psi K}$. However, since a 
non-vanishing value of ${\cal D}^+_{\psi K}$ requires also a large CP-conserving 
strong phase, this observable still leaves room for sizeable NP 
contributions to  the $I=0$ sector. 

\begin{figure}[t]
\centerline{
\begin{tabular}{ll}
   {\small(a)} & {\small(b)} \\
   \includegraphics[width=7.3truecm]{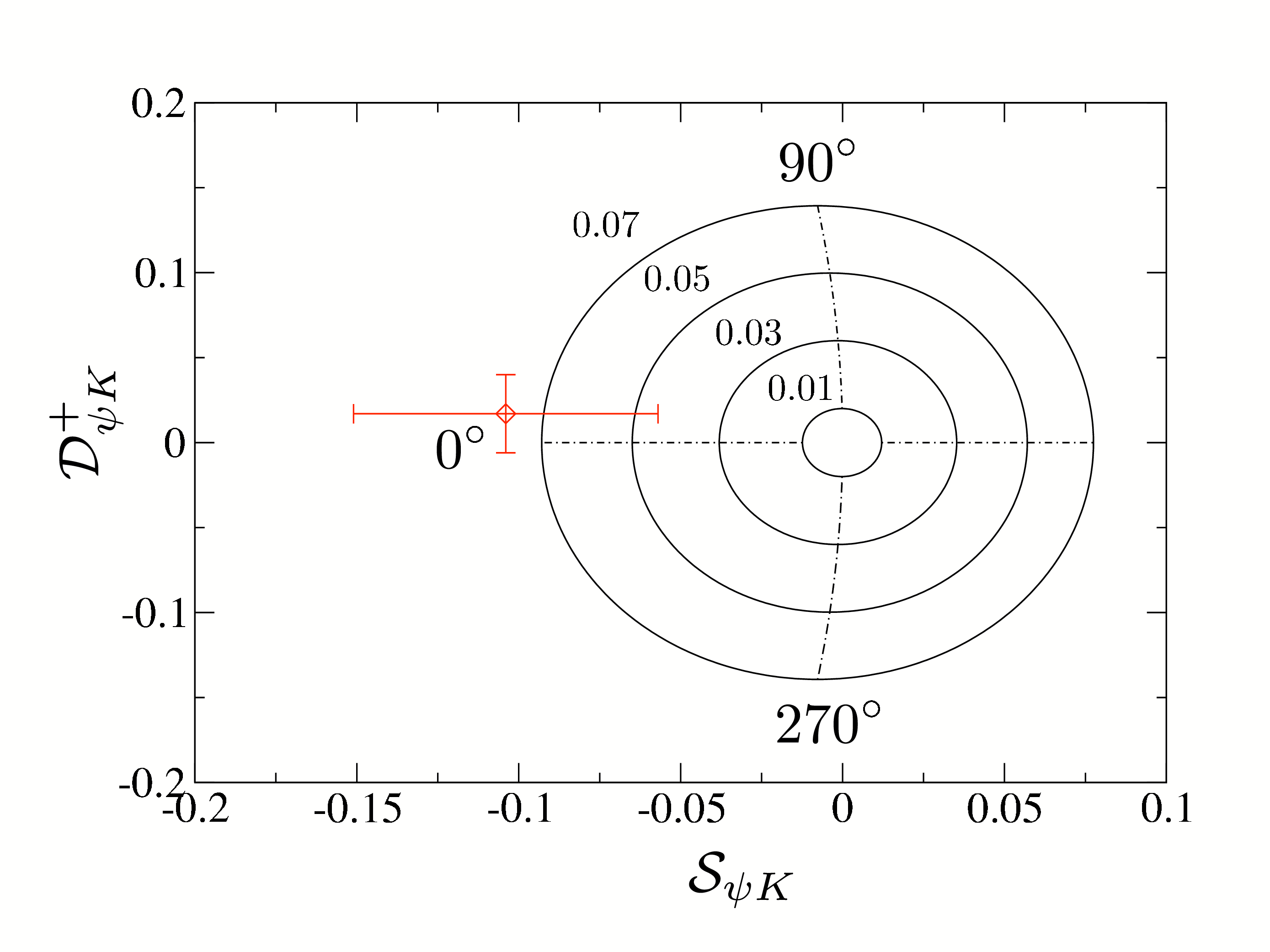}  &
   \includegraphics[width=7.3truecm]{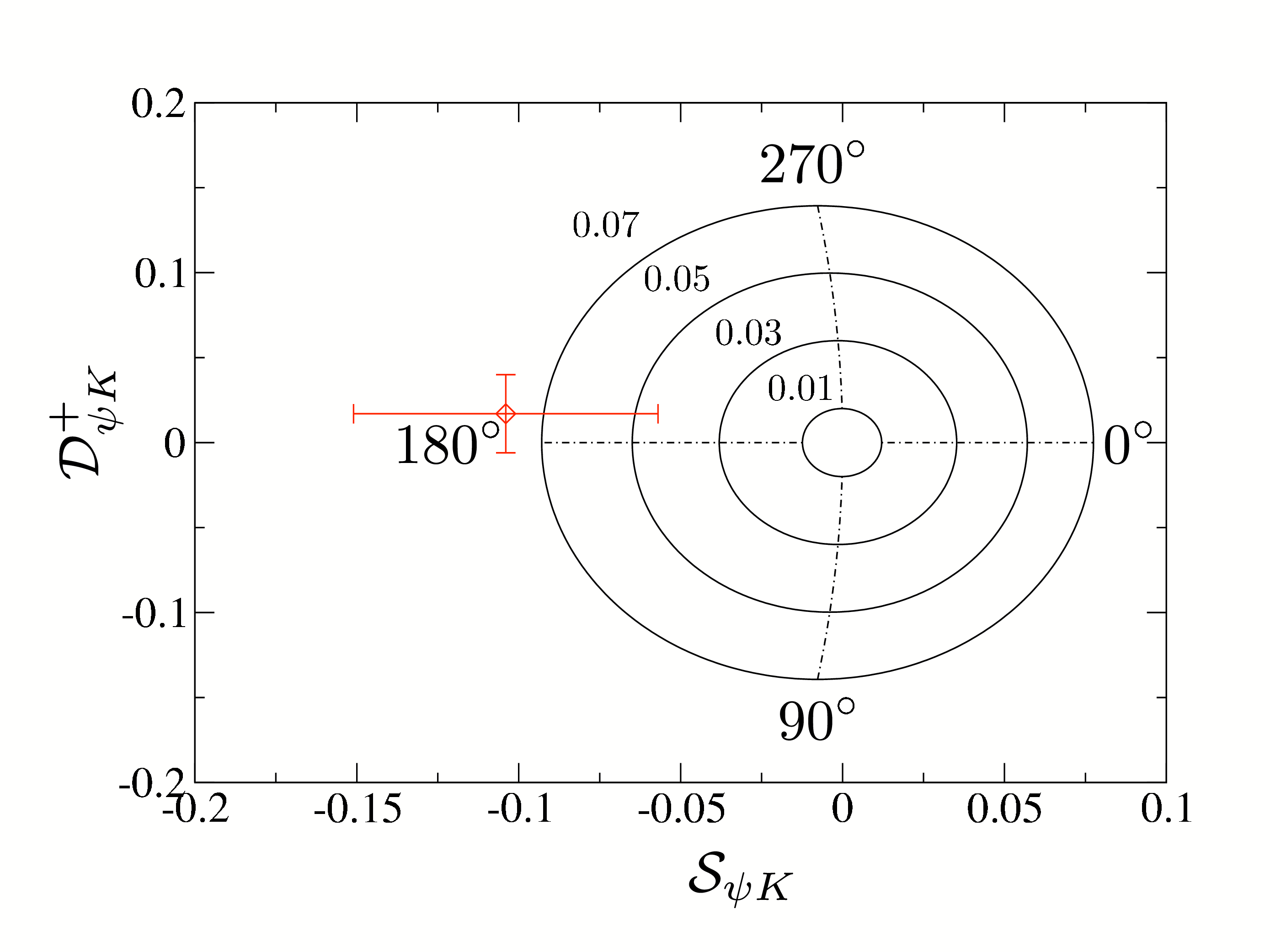}
 \end{tabular}}
 \vspace*{-0.4truecm}
\caption{The situation in the ${\cal S}_{\psi K}$--${\cal D}^+_{\psi K}$ plane for
NP contributions to the $B\to J/\psi K$ decay amplitudes in the $I=0$ isospin 
sector for NP phases $\phi_0=-90^\circ$ (a) and $\phi_0=+90^\circ$ (b). The 
diamonds with the error bars represent the averages of the current data, whereas 
the numbers correspond to the values of $\Delta_0$ and $v_0$.}\label{fig:Plot-BpsiK}
\end{figure}

Thanks to the new Belle result listed in (\ref{s2b-psiK-exp}), the average for
$(\sin2\beta)_{\psi K_{\rm S}}$ went down by about $1 \sigma$, which 
is a somewhat surprising development of this summer. Consequently, 
the comparison of (\ref{s2b-average}) with the CKM fits in the
$\bar\rho$--$\bar\eta$ plane does no longer look ``perfect", as we saw 
in Fig.~\ref{fig:UTfits}. In particular, if we use the value of the UT fits for 
$\sin2\beta$ that follow from the experimental information for the UT sides 
and $\varepsilon_K$, $(\sin 2\beta)_{\rm UT}=0.791\pm0.034$ \cite{UTfit}, 
we obtain
\begin{equation}\label{S-psi-K}
{\cal S}_{\psi K}\equiv (\sin 2\beta)_{\psi K_{\rm S}}-(\sin 2\beta)_{\rm UT}=
-0.104\pm0.047.
\end{equation}
The are two limiting cases of this possible discrepancy with the KM mechanism
of CP violation: NP contributions to the $B\to J/\psi K$ decay amplitudes, or 
NP effects entering through $B^0_d$--$\bar B^0_d$ mixing. Let us first illustrate 
the former case. Since the NP effects in the $I=1$ sector are expected to be 
dynamically suppressed, we consider only NP in the $I=0$ isospin sector,
which implies ${\cal B}_{\psi K} ={\cal D}^-_{\psi K}=0$, in accordance with
(\ref{B-Dpm-PsiK-res}). To simplify the discussion, we assume that there is 
effectively only a single NP contribution of this kind, so that we may write
\begin{equation}\label{ApsiK-NP}
A(B^0_d\to J/\psi K^0)=A_0\left[1+v_0e^{i(\Delta_0+\phi_0)}\right]=A(B^+\to J/\psi K^+).
\end{equation}
Here $v_0$ and the CP-conserving strong phase $\Delta_0$ are hadronic parameters,
whereas $\phi_0$ denotes a CP-violating phase originating beyond the SM. 
An interesting specific scenario falling into this category arises if the NP effects
enter through EW penguins. This kind of NP has recently received a lot of attention 
in the context of the $B\to\pi K$ puzzle, which we shall discuss in 
Section~\ref{sec:BpiK-puzzle}. Also within the SM, where $\phi_0$ vanishes, 
EW penguins have a sizeable impact on the $B\to J/\psi K$ system \cite{RF-EWP-rev}.
Using factorization, the following estimate can be obtained \cite{BFRS}:
\begin{equation}\label{v-SM}
\left. v_0e^{i\Delta_0}\right|_{\rm fact}^{\rm SM}\approx -0.03.
\end{equation}
In Figs.~\ref{fig:Plot-BpsiK} (a) and (b), we show the situation in the
${\cal S}_{\psi K}$--${\cal D}^+_{\psi K}$ plane for $\phi_0=-90^\circ$
and $\phi_0=+90^\circ$, respectively. The contours correspond to 
different values of $v_0$, and are obtained by varying $\Delta_0$ between 
$0^\circ$ and $360^\circ$; the experimental data are represented by the diamonds
with the error bars. Since factorization gives $\Delta_0=180^\circ$, as can be
seen in (\ref{v-SM}), the case of $\phi_0=-90^\circ$ is disfavoured. On the
other hand, in the case of $\phi_0=+90^\circ$, the experimental region can straightforwardly be reached for $\Delta_0$ not differing too much from the 
factorization result, although an enhancement of $v_0$ by a factor of 
${\cal O}(3)$ with respect to the SM estimate in (\ref{v-SM}), which suffers from 
large uncertainties, would simultaneously be required in order to reach the central 
experimental value. Consequently, NP contributions to the EW penguin sector 
could, in principle, be at the origin of the possible discrepancy indicated by 
(\ref{S-psi-K}). This scenario should be carefully monitored as the data improve.

Another explanation of  (\ref{S-psi-K}) is provided by CP-violating NP contributions to 
$B^0_d$--$\bar B^0_d$ mixing, which affect the corresponding mixing phase as follows:
\begin{equation}\label{ref-phid}
\phi_d=\phi_d^{\rm SM}+\phi_d^{\rm NP}=2\beta+\phi_d^{\rm NP}.
\end{equation}
If we assume that the NP contributions to the $B\to J/\psi K$ decay amplitudes
are negligible, the world average in (\ref{s2b-average}) implies
\begin{equation}\label{phid-exp}
\phi_d=(43.4\pm2.5)^\circ \quad\lor\quad (136.6\pm2.5)^\circ.
\end{equation}
Here the latter solution would be in dramatic conflict with the CKM fits, and
would require a large NP contribution to $B^0_d$--$\bar B^0_d$ 
mixing \cite{FIM,FlMa}. Both solutions can be distinguished through the 
measurement of the sign of $\cos\phi_d$, where a positive value would 
select the SM-like branch. Using an angular analysis of the decay products of
$B_d\to J/\psi[\to\ell^+\ell^-] K^\ast[\to\pi^0K_{\rm S}]$ processes,
the BaBar collaboration finds \cite{babar-c2b}
\begin{equation}
\cos\phi_d =2.72^{+0.50}_{-0.79} \pm 0.27,
\end{equation}
thereby favouring the solution around $\phi_d=43^\circ$. Interestingly, this 
picture emerges also from the first data for CP-violating effects in 
$B_d\to D^{(*)\pm}\pi^\mp$ modes \cite{RF-gam-ca}, and an analysis of 
the $B\to\pi\pi,\pi K$ system \cite{BFRS}, although in an indirect manner.
Recently, a new method has been proposed, which makes use of 
the interference pattern in $D\to K_{\rm S}\pi^+\pi^-$ decays emerging
from $B_d\to D\pi^0$ and similar decays \cite{bo-ge}. The results of
this method are also consistent with the SM, so that a negative value
of $\cos\phi_d$ is now ruled out with greater than 95\% confidence 
\cite{gershon}.
Since the value of $(\sin2\beta)_{\rm UT}$ given before (\ref{S-psi-K}) corresponds 
to $\beta=(26.1\pm1.6)^\circ$, (\ref{ref-phid}) yields 
$\phi_d^{\rm NP}=-(8.9\pm4.1)^\circ$. Consequently, the $B$-factory data do not
leave too much space for CP-violating NP contributions to $B^0_d$--$\bar B^0_d$ 
mixing. On the other hand, such effects are still unexplored in 
$B^0_s$--$\bar B^0_s$ mixing, where they can nicely be probed through 
$B^0_s\to J/\psi \phi$ decays, which are very accessible at the LHC. For
NP models that are interesting in this context, see 
Refs.~\cite{JN,BKK,Z-prime}.

The possibility of having a non-zero value of (\ref{S-psi-K}) could of course just 
be due to a statistical fluctuation. However, should it be confirmed, 
it could be due to CP-violating NP contributions to the $B^0_d\to J/\psi K_{\rm S}$ 
decay amplitude or to $B^0_d$--$\bar B^0_d$ mixing, as we just saw.
A tool to distinguish between these avenues is provided by decays 
of the kind $B_d\to D\pi^0, D\rho^0, ...$, which are pure ``tree" decays, i.e.\
they do {\it not} receive any penguin contributions. If the neutral $D$ mesons
are observed through their decays into CP eigenstates $D_\pm$, these decays
allow extremely clean determinations of the ``true" value of $\sin2\beta$ 
\cite{RF-BdDpi0}, as we shall discuss in more detail in Subsection~\ref{ssec:BsDsK}. 
In view of (\ref{S-psi-K}), this would be very interesting, so that detailed 
feasibility studies for the exploration of the $B_d\to D\pi^0, D\rho^0, ...$\ modes 
at a super-$B$ factory are strongly encouraged.

\begin{figure}[t]
\centerline{
 \includegraphics[width=5.5truecm]{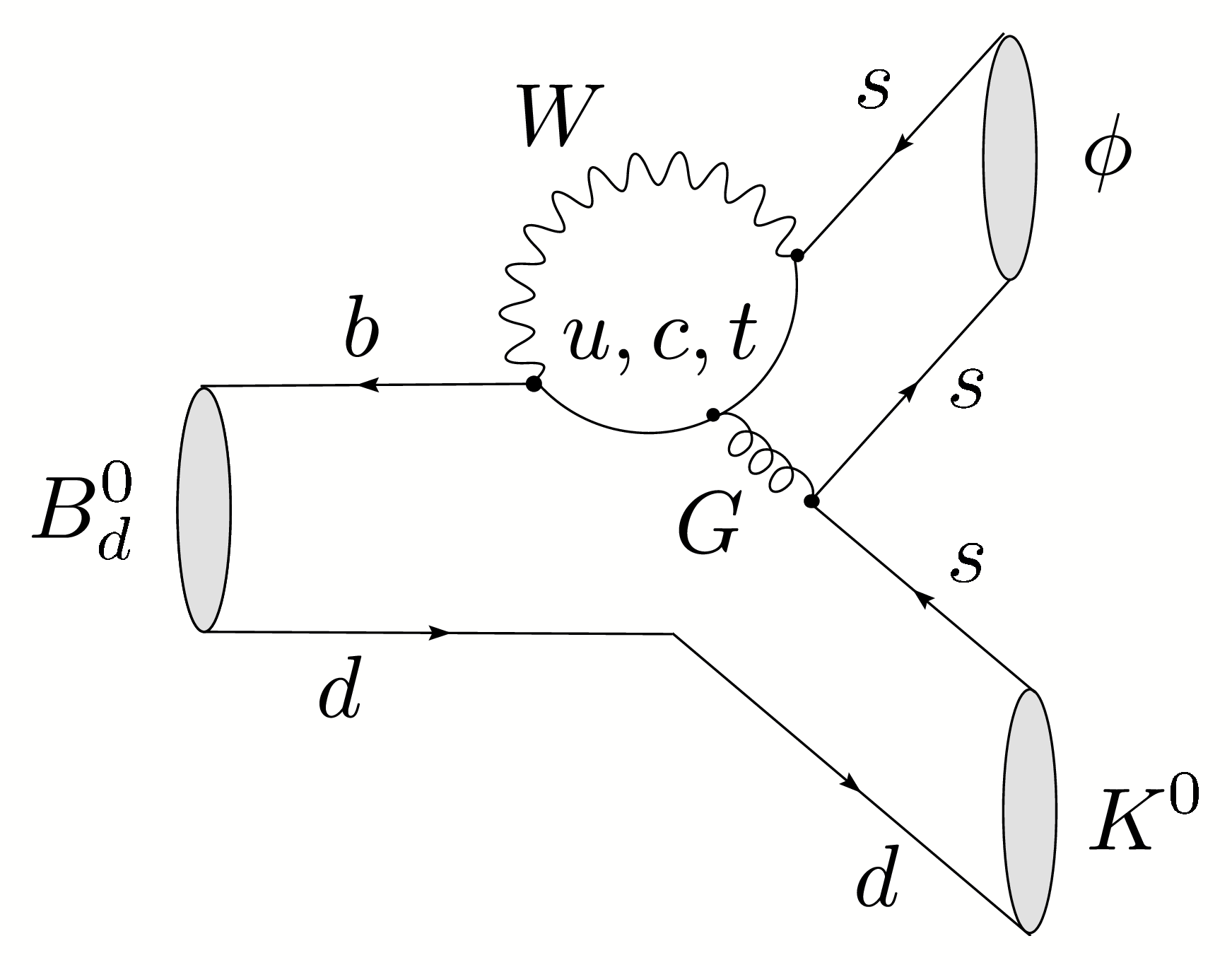} 
 }
 \vspace*{-0.3truecm}
\caption{Feynman diagrams contributing to $B^0_d\to \phi K^0$ 
decays.}\label{fig:BphiK-diag}
\end{figure}

\begin{figure}
\centerline{
\begin{tabular}{ll}
   {\small(a)} & {\small(b)} \\
   \includegraphics[width=7.3truecm]{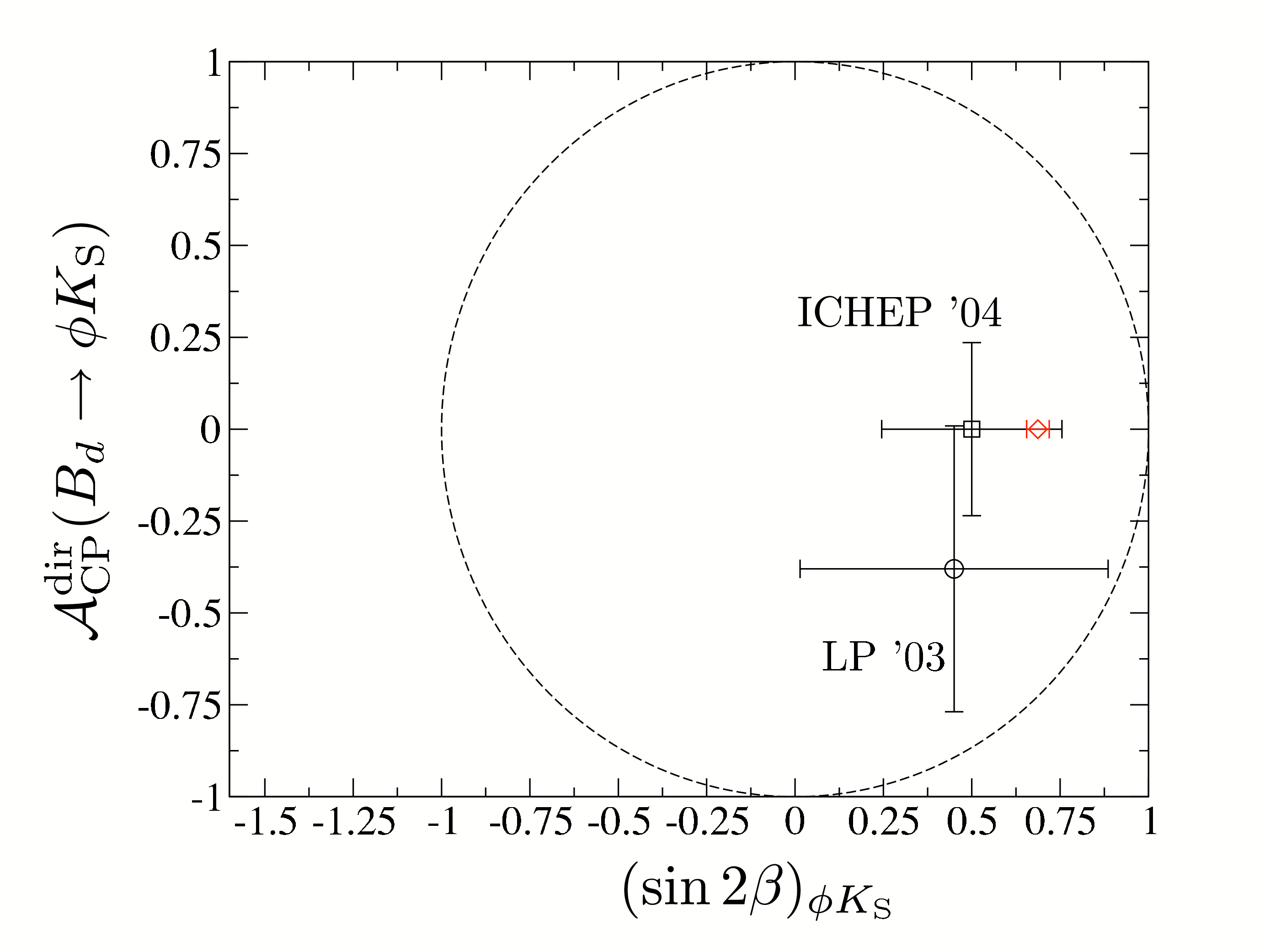}  &
   \includegraphics[width=7.3truecm]{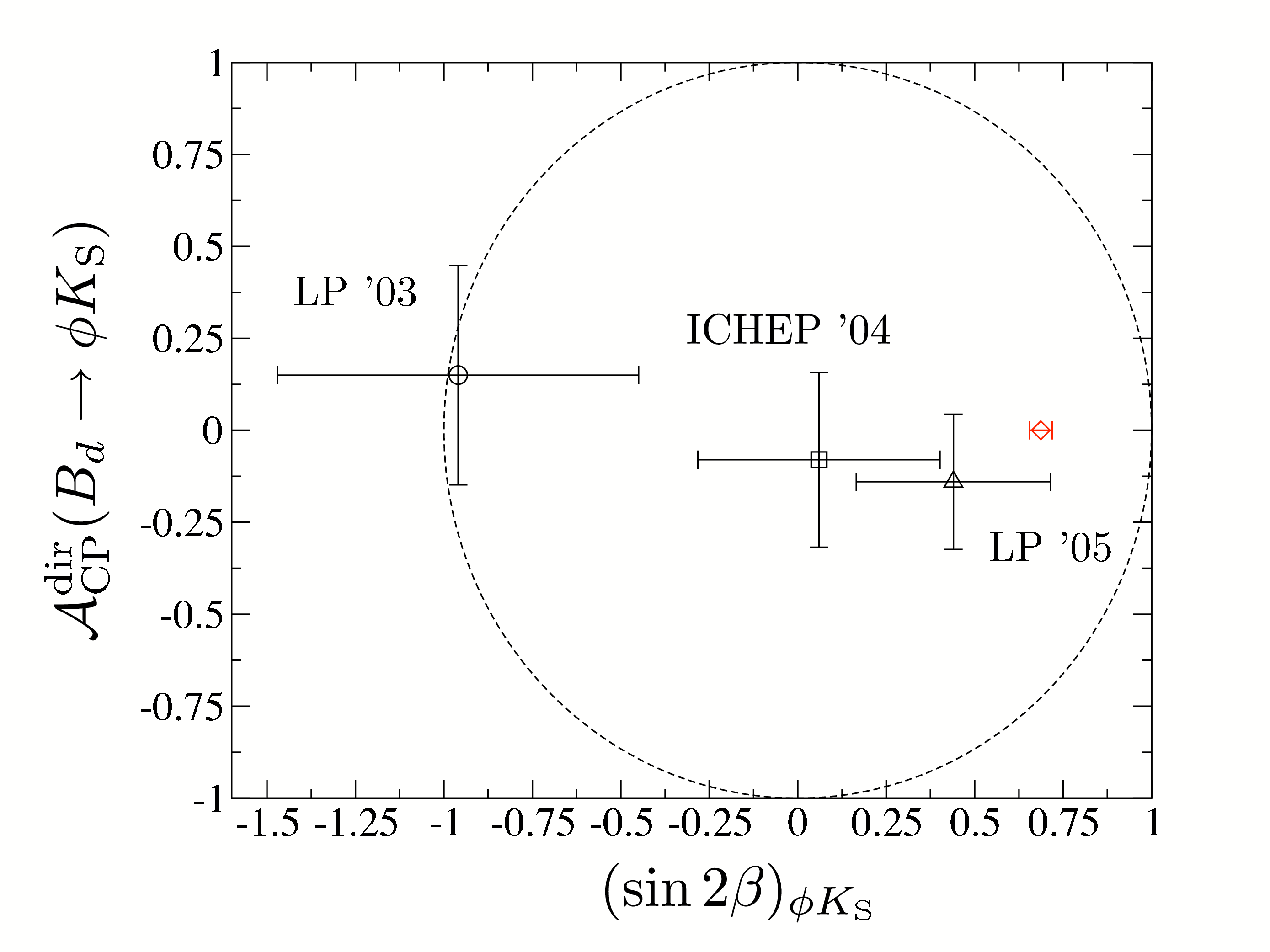}
 \end{tabular}
 }
 \vspace*{-0.4truecm}
   \caption{The time evolution of the BaBar (a)  and Belle (b) data for 
   the CP violation in $B_d\to \phi K_{\rm S}$. The diamonds represent the 
   SM relations (\ref{BphiK-rel1})--(\ref{Bd-phiKS-SM-rel}) with 
   (\ref{s2b-average}).}\label{fig:BphiK-data}
\end{figure}

\subsection{$B^0_d\to \phi K_{\rm S}$}\label{ssec:BphiK}
Another important probe for the testing of the KM mechanism is 
offered by $B_d^0\to \phi K_{\rm S}$, which is a 
decay into a CP-odd final state. As can be seen in Fig.~\ref{fig:BphiK-diag},
it originates from $\bar b\to \bar s s \bar s$ transitions and is, therefore, a 
pure penguin mode. This decay is described by the low-energy effective 
Hamiltonian in (\ref{e4}) with $r=s$, where the current--current operators 
may only contribute through penguin-like contractions, which describe the
penguin topologies with internal up- and charm-quark exchanges. The dominant
r\^ole is played by the QCD penguin operators \cite{BphiK-old}. However,
thanks to the large top-quark mass, EW penguins have a sizeable impact as 
well \cite{RF-EWP,DH-PhiK}. In the SM, we may write
\begin{equation}\label{B0phiK0-ampl}
A(B_d^0\to \phi K_{\rm S})=\lambda_u^{(s)}\tilde A_{\rm P}^{u'}
+\lambda_c^{(s)}\tilde A_{\rm P}^{c'}+\lambda_t^{(s)}\tilde A_{\rm P}^{t'},
\end{equation}
where we have applied the same notation as in Subsection~\ref{ssec:BpsiK}.
Eliminating the CKM factor $\lambda_t^{(s)}$ with the help of
(\ref{UT-rel}) yields
\begin{equation}
A(B_d^0\to \phi K_{\rm S})\propto
\left[1+\lambda^2 b e^{i\Theta}e^{i\gamma}\right],
\end{equation}
where 
\begin{equation}
b e^{i\Theta}\equiv\left(\frac{R_b}{1-\lambda^2}\right)\left[
\frac{\tilde A_{\rm P}^{u'}-\tilde A_{\rm P}^{t'}}{\tilde A_{\rm P}^{c'}-
\tilde A_{\rm P}^{t'}}\right].
\end{equation}
Consequently,  we obtain
\begin{equation}\label{xi-phiKS}
\xi_{\phi K_{\rm S}}^{(d)}=+e^{-i\phi_d}
\left[\frac{1+\lambda^2b e^{i\Theta}e^{-i\gamma}}{1+
\lambda^2b e^{i\Theta}e^{+i\gamma}}\right].
\end{equation}
The theoretical estimates of $b e^{i\Theta}$ 
suffer from large hadronic uncertainties. However, since this parameter enters 
(\ref{xi-phiKS}) in a doubly Cabibbo-suppressed way, we obtain the 
following expressions \cite{RF-EWP-rev}:
\begin{eqnarray}
{\cal A}_{\rm CP}^{\rm dir}(B_d\to \phi K_{\rm S})&=&0+
{\cal O}(\lambda^2)\label{BphiK-rel1}\\
{\cal A}_{\rm CP}^{\rm mix}(B_d\to \phi K_{\rm S})&=&-\sin\phi_d
+{\cal O}(\lambda^2),\label{BphiK-rel2}
\end{eqnarray}
where we made the plausible assumption that $b={\cal O}(1)$. On the other 
hand, the mixing-induced CP asymmetry of 
$B_d\to J/\psi K_{\rm S}$ measures also $-\sin\phi_d$, as we saw in
(\ref{Amix-BdpsiKS}). We arrive therefore at the following 
relation \cite{RF-EWP-rev,growo}:
\begin{equation}\label{Bd-phiKS-SM-rel}
-(\sin2\beta)_{\phi K_{\rm S}}\equiv
{\cal A}_{\rm CP}^{\rm mix}(B_d\to \phi K_{\rm S}) 
={\cal A}_{\rm CP}^{\rm mix}(B_d\to J/\psi K_{\rm S}) + 
{\cal O}(\lambda^2),
\end{equation}
which offers an interesting test of the SM. Since $B_d\to \phi K_{\rm S}$ is 
governed by penguin processes in the SM, this decay may well be affected by 
NP. In fact, if we assume that NP arises generically in the TeV regime, it can be 
shown through field-theoretical estimates that the NP contributions to  
$b\to s\bar s s$ transitions may well lead to sizeable violations of
(\ref{BphiK-rel1}) and (\ref{Bd-phiKS-SM-rel}) \cite{RF-Phys-Rep,FM-BphiK}. Moreover, 
this is also the case for several specific NP scenarios; for examples, see 
Refs.~\cite{CFMS,Ko,GHK,Z-prime-BpiK}.

\begin{figure}
\centerline{
\begin{tabular}{ll}
   {\small(a)} & {\small(b)} \\
   \includegraphics[width=7.3truecm]{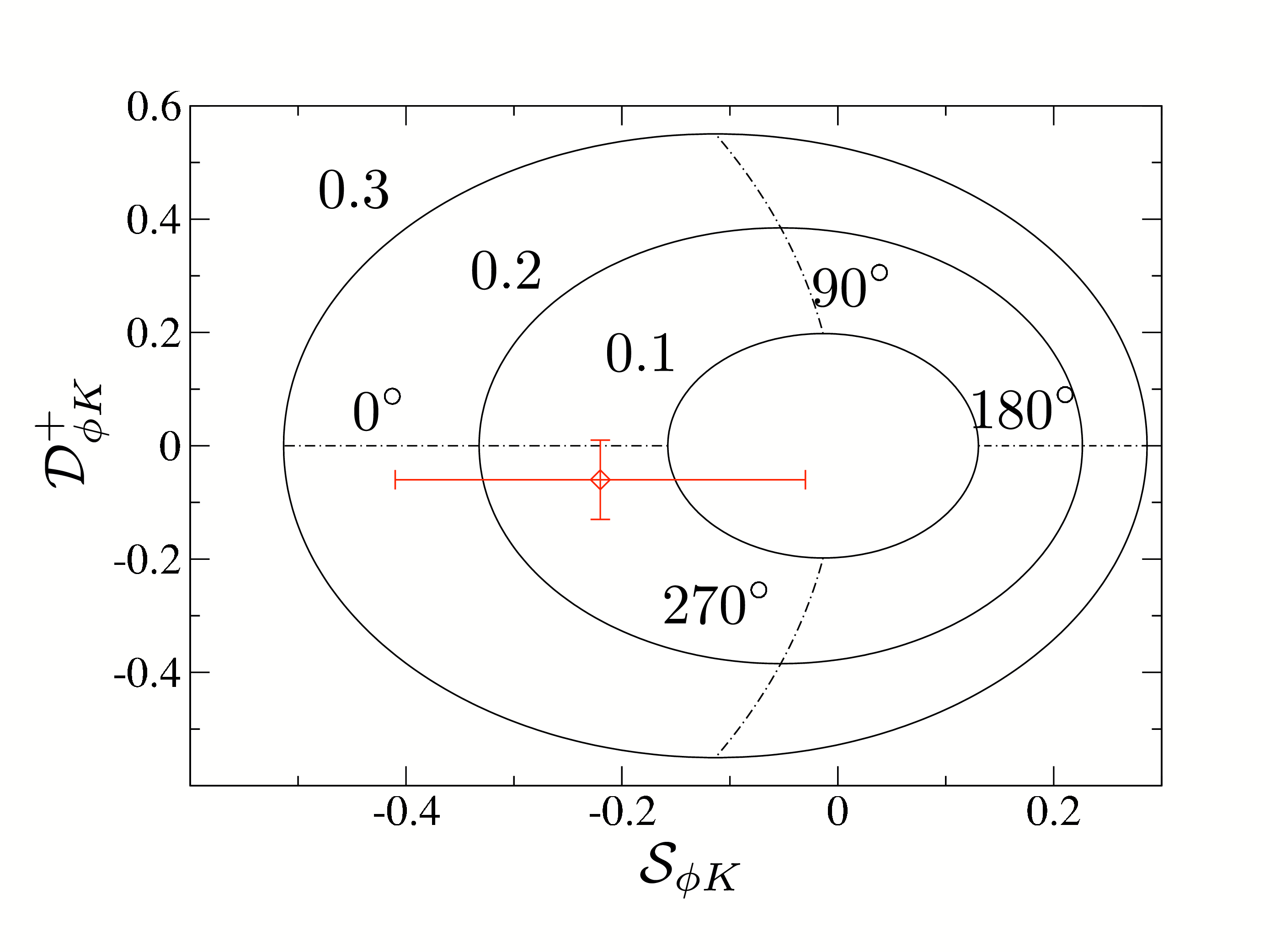}  &
   \includegraphics[width=7.3truecm]{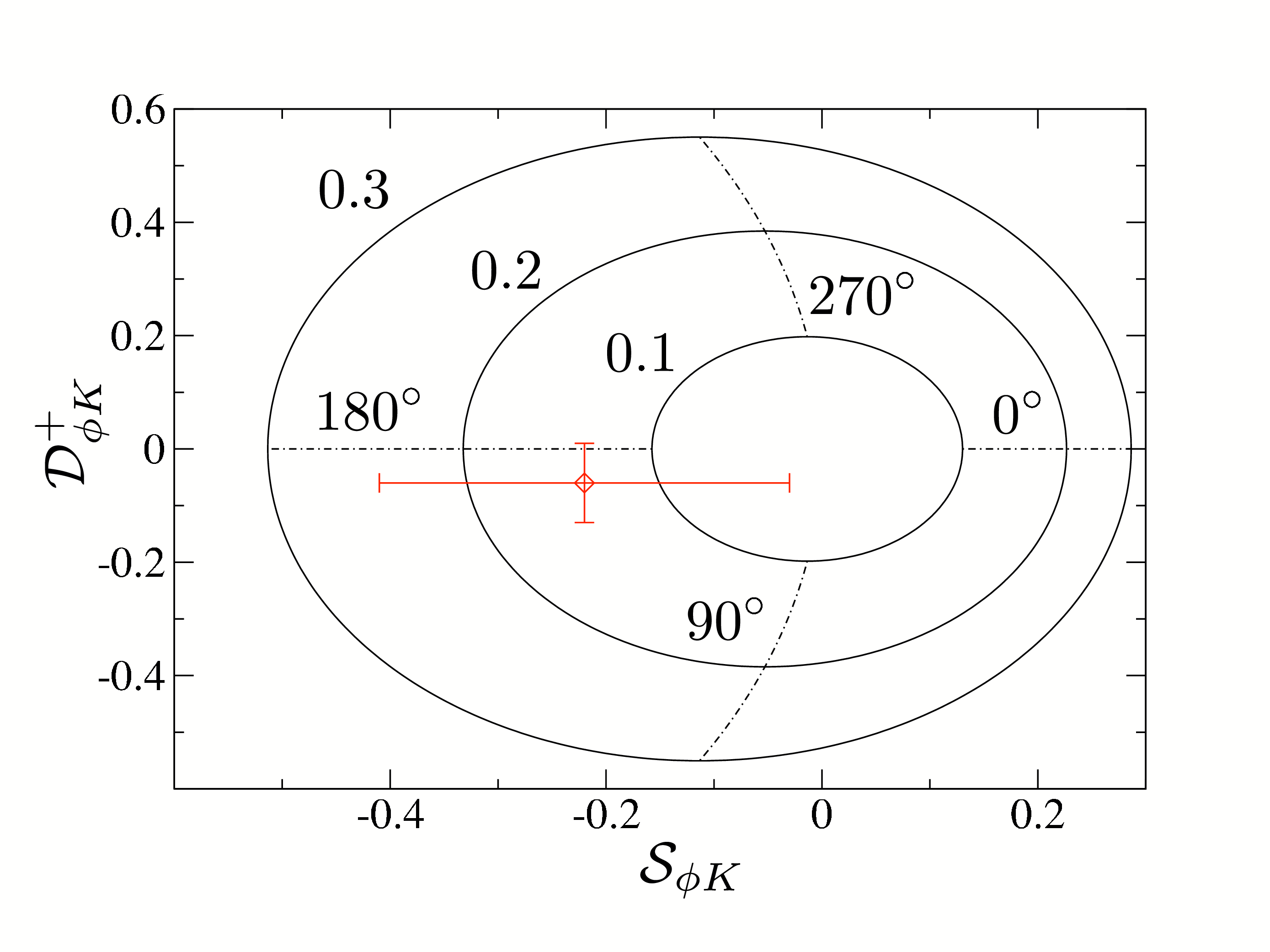}
 \end{tabular}}
 \vspace*{-0.4truecm}
\caption{The situation in the ${\cal S}_{\phi K}$--${\cal D}^+_{\phi K}$ plane for
NP contributions to the $B\to \phi K$ decay amplitudes in the $I=0$ isospin 
sector for NP phases $\phi_0=-90^\circ$ (a) and $\phi_0=+90^\circ$ (b). The 
diamonds with the error bars represent the averages of the current data, whereas 
the numbers correspond to the values of $\tilde\Delta_0$ and 
$\tilde v_0$.}\label{fig:Plot-BphiK}
\end{figure}

In Fig.~\ref{fig:BphiK-data}, we show the time evolution of the $B$-factory data
for the measurements of CP violation in $B_d\to\phi K_{\rm S}$, using the results 
reported at the LP~'03 \cite{LP03}, ICHEP~'04 \cite{ICHEP04} and LP~'05 \cite{LP05}
conferences. Because of (\ref{Obs-rel}), the corresponding observables have
to lie inside a circle with radius one around the origin, which is represented by the
dashed lines. The result announced by the Belle collaboration in
2003 led to quite some excitement in the community. Meanwhile, the Babar
\cite{BaBar-Bphi-K} and Belle \cite{Belle-Bphi-K} results are in good agreement 
with each other, yielding the following averages \cite{HFAG}:
\begin{equation}\label{BphiK-av}
{\cal A}_{\rm CP}^{\rm dir}(B_d\to \phi K_{\rm S})=-0.09\pm0.14, \quad
(\sin2\beta)_{\phi K_{\rm S}}=0.47\pm0.19. 
\end{equation}
If  we take (\ref{s2b-average}) into account, we obtain the following result for
the counterpart of (\ref{S-psi-K}):
\begin{equation}\label{S-phi-K}
{\cal S}_{\phi K}\equiv (\sin 2\beta)_{\phi K_{\rm S}}- (\sin 2\beta)_{\psi K_{\rm S}}
=-0.22\pm0.19.
\end{equation}
This number still appears to be somewhat on the lower side, thereby indicating potential 
NP contributions to $b\to s \bar s s$ processes.

Further insights into the origin and the isospin structure of NP contributions 
can be obtained through a combined analysis of the neutral and charged 
$B\to \phi K$ modes with the help of observables 
${\cal B}_{\phi K}$ and ${\cal D}^\pm_{\phi K}$ \cite{FM-BphiK}, which are 
defined in analogy to (\ref{BpsiK}) and (\ref{DpmPsiK}), respectively. The
current experimental results read as follows:
\begin{equation}\label{B-Dpm-PhiK-res}
\hspace*{-0.7truecm} {\cal B}_{\phi K} =0.00\pm0.08,\quad
{\cal D}^-_{\phi K}=-0.03\pm0.07, \quad
{\cal D}^+_{\phi K}=-0.06\pm0.07.
\end{equation}
As in the $B\to J/\psi K$ case, ${\cal B}_{\phi K}$ and ${\cal D}^-_{\phi K}$ probe
NP effects in the $I=1$ sector, which are expected to be dynamically suppressed,
whereas ${\cal D}^+_{\phi K}$ is sensitive to NP in the $I=0$ sector. The latter 
kind of NP could also manifest itself as a non-vanishing value of (\ref{S-phi-K}).

In order to illustrate these effects, let us consider again the case where NP enters 
only in the $I=0$ isospin sector. An important example is given by EW penguins, 
which have a significant impact on $B\to\phi K$ decays \cite{RF-EWP}. In analogy
to the discussion in Subsection~\ref{ssec:BpsiK}, we may then write
\begin{equation}\label{AphiK-NP}
A(B^0_d\to \phi K^0)=\tilde A_0\left[1+\tilde v_0e^{i(\tilde \Delta_0+\phi_0)}\right]=
A(B^+ \to \phi K^+),
\end{equation}
which implies ${\cal B}_{\phi K} ={\cal D}^-_{\phi K}=0$, in accordance with 
(\ref{B-Dpm-PhiK-res}). The notation corresponds to the one of (\ref{ApsiK-NP}).  
Using the factorization approach to deal with the QCD and EW penguin contributions, 
we obtain the following estimate in the SM, where the CP-violating NP phase
$\phi_0$ vanishes \cite{BFRS}:
\begin{equation}\label{v-SM-phiK}
\left.\tilde v_0e^{i\tilde \Delta_0}\right|_{\rm fact}^{\rm SM}\approx -0.2.
\end{equation}
In Figs.~\ref{fig:Plot-BphiK} (a) and (b), we show the situation in the 
${\cal S}_{\phi K}$--${\cal D}^+_{\phi K}$ plane for NP phases $\phi_0=-90^\circ$
and $\phi_0=+90^\circ$, respectively, and various values of $\tilde v_0$; each point
of the contours is parametrized by $\tilde\Delta_0\in[0^\circ,360^\circ]$. We observe 
that the central values of the current experimental data, which are represented by the 
diamonds with the error bars, can straightforwardly be accommodated in this scenario 
in the case of $\phi_0=+90^\circ$ for strong phases satisfying $\cos\tilde\Delta_0<0$, 
as in factorization. Moreover, as can also be seen in Fig.~\ref{fig:Plot-BphiK} (b),
the EW penguin contributions would then have to be suppressed with respect
to the SM estimate, which would be an interesting feature in view of the discussion of
the $B\to \pi K$ puzzle and the rare decay constraints in Section~\ref{sec:BpiK-puzzle}.

It will be interesting to follow the evolution of the $B$-factory data,
and to monitor also similar modes, such as $B^0_d\to \pi^0 K_{\rm S}$ 
\cite{PAPIII} and  $B^0_d\to \eta'K_{\rm S}$ \cite{loso}. For a compilation of 
the corresponding experimental results, see Ref.~\cite{HFAG}; recent 
theoretical papers dealing with these channels can be found in 
Refs.~\cite{BFRS,GGR,beneke,BFRS-5}. We will return to the CP 
asymmetries of the $B^0_d\to \pi^0 K_{\rm S}$ channel in 
Section~\ref{sec:BpiK-puzzle}.

\begin{figure}[t]
\centerline{
 \includegraphics[width=4.5truecm]{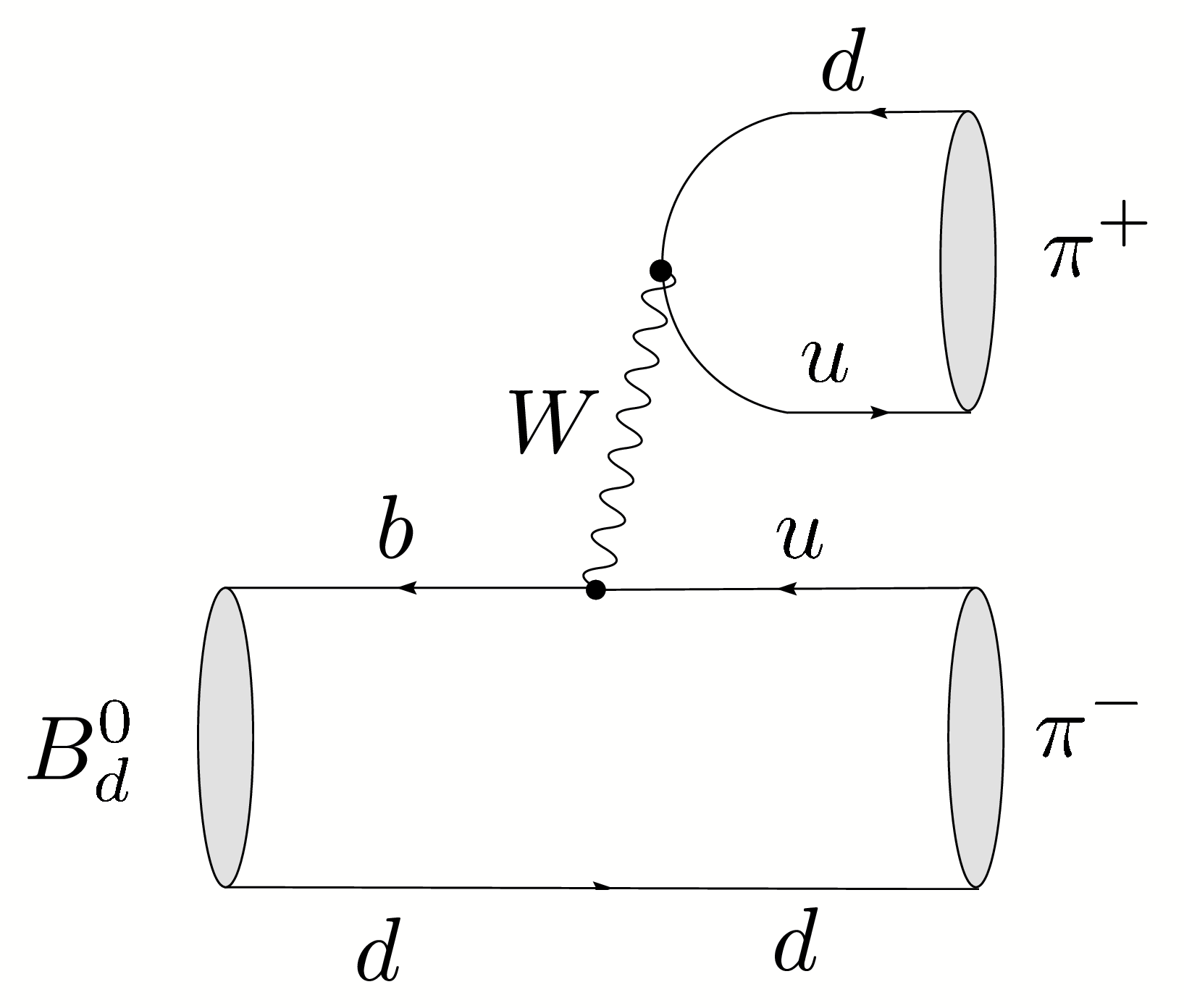}
 \hspace*{0.5truecm}
 \includegraphics[width=5.2truecm]{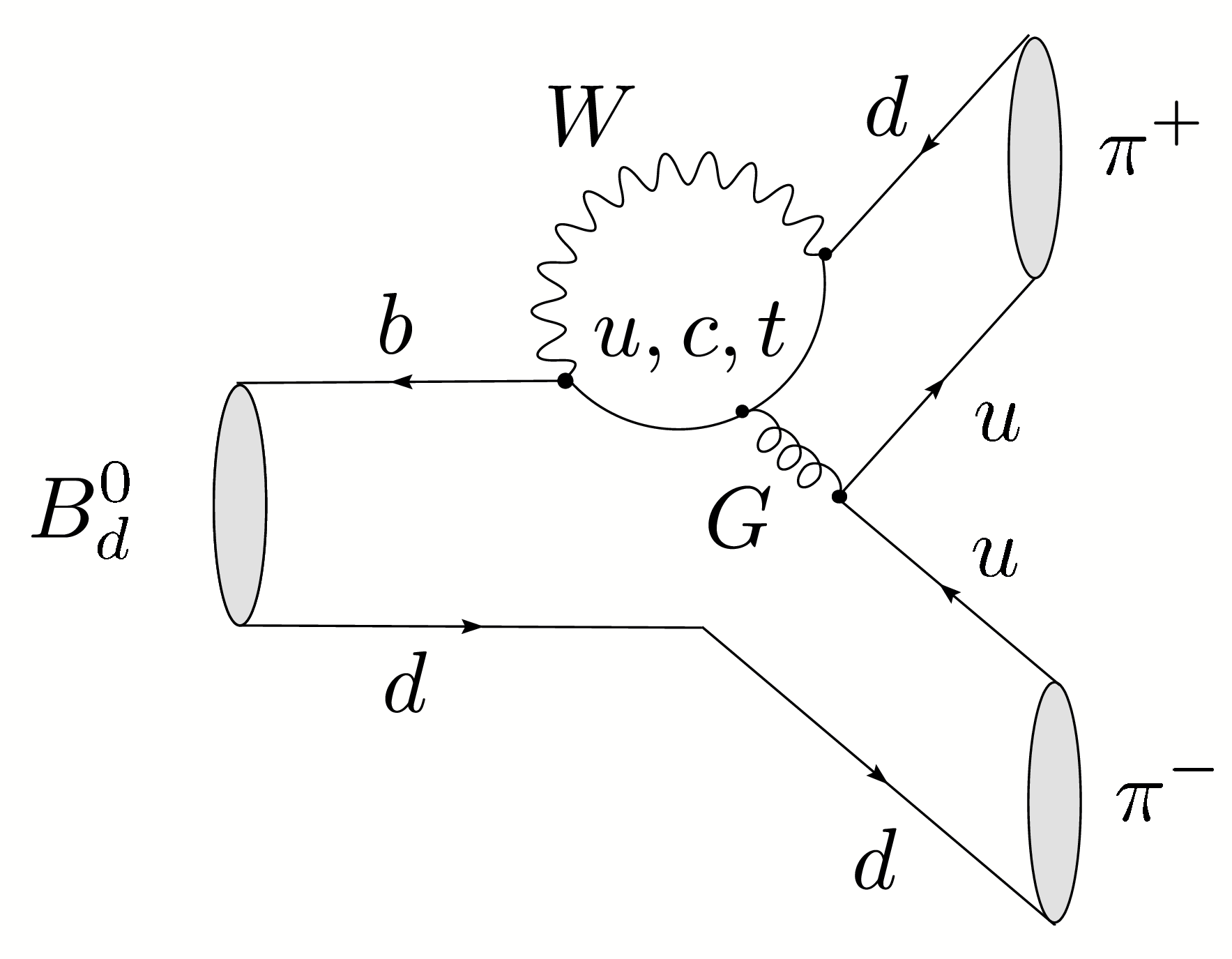}  
 }
 \vspace*{-0.3truecm}
\caption{Feynman diagrams contributing to $B^0_d\to \pi^+\pi^-$ 
decays.}\label{fig:Bpipi-diag}
\end{figure}

\subsection{$B^0_d\to \pi^+\pi^-$}\label{ssec:Bpi+pi-}
This decay is a transition into a CP eigenstate with eigenvalue $+1$, and 
originates from $\bar b\to\bar u u \bar d$ processes, as can be seen in 
Fig.~\ref{fig:Bpipi-diag}. In analogy to (\ref{Bd-ampl1}) and (\ref{B0phiK0-ampl}), 
its decay amplitude can be written as follows \cite{RF-BsKK}:
\begin{equation}
A(B_d^0\to\pi^+\pi^-)=
\lambda_u^{(d)}\left(A_{\rm T}^{u}+
A_{\rm P}^{u}\right)+\lambda_c^{(d)}A_{\rm P}^{c}+
\lambda_t^{(d)}A_{\rm P}^{t}.
\end{equation}
Using again (\ref{UT-rel}) to eliminate the CKM factor 
$\lambda_t^{(d)}=V_{td}V_{tb}^\ast$ and applying once more the 
Wolfenstein parametrization yields
\begin{equation}\label{Bpipi-ampl}
A(B_d^0\to\pi^+\pi^-)={\cal C}\left[e^{i\gamma}-de^{i\theta}\right],
\end{equation}
where the overall normalization ${\cal C}$ and
\begin{equation}\label{D-DEF}
d e^{i\theta}\equiv\frac{1}{R_b}
\left[\frac{A_{\rm P}^{c}-A_{\rm P}^{t}}{A_{\rm T}^{u}+
A_{\rm P}^{u}-A_{\rm P}^{t}}\right]
\end{equation}
are hadronic parameters. 
The formalism discussed in Subsection~\ref{ssec:CP-strat} then implies 
\begin{equation}\label{xi-Bdpipi}
\xi_{\pi^+\pi^-}^{(d)}=-e^{-i\phi_d}\left[\frac{e^{-i\gamma}-
d e^{i\theta}}{e^{+i\gamma}-d e^{i\theta}}\right].
\end{equation}
In contrast to the expressions (\ref{xi-BdpsiKS}) and (\ref{xi-phiKS}) 
for the $B_d^0\to J/\psi K_{\rm S}$ and $B^0_d\to\phi K_{\rm S}$ counterparts,
respectively, the hadronic parameter $d e^{i\theta}$, which suffers from large 
theoretical uncertainties, does {\it not} enter  (\ref{xi-Bdpipi}) 
in a doubly Cabibbo-suppressed way. This feature is at the basis of the
famous ``penguin problem'' in $B^0_d\to\pi^+\pi^-$, which was addressed
in many papers (see, for instance, \cite{GL}--\cite{GLSS}). If the penguin 
contributions to this channel were negligible, i.e.\ $d=0$, its CP asymmetries 
were simply given by
\begin{eqnarray}
{\cal A}_{\rm CP}^{\rm dir}(B_d\to\pi^+\pi^-)&=&0 \\
{\cal A}_{\rm CP}^{\rm mix}(B_d\to\pi^+\pi^-)&=&\sin(\phi_d+2\gamma)
\stackrel{\rm SM}{=}\sin(\underbrace{2\beta+2\gamma}_{2\pi-2\alpha})
=-\sin 2\alpha.
\end{eqnarray}
Consequently, ${\cal A}_{\rm CP}^{\rm mix}(B_d\to\pi^+\pi^-)$ would then
allow us to determine $\alpha$. However, in the general case, we obtain 
expressions with the help of (\ref{CPV-OBS}) and (\ref{xi-Bdpipi}) of the form
\begin{eqnarray}
{\cal A}_{\rm CP}^{\rm dir}(B_d\to \pi^+\pi^-)&=&
G_1(d,\theta;\gamma) \label{CP-Bpipi-dir-gen}\\
{\cal A}_{\rm CP}^{\rm mix}(B_d\to \pi^+\pi^-)&=&
G_2(d,\theta;\gamma,\phi_d);\label{CP-Bpipi-mix-gen}
\end{eqnarray}
for explicit formulae, see \cite{RF-BsKK}. We observe that actually the 
phases $\phi_d$ and $\gamma$ enter directly in the $B_d\to\pi^+\pi^-$ 
observables, and not $\alpha$. Consequently, since $\phi_d$ can be fixed 
through the mixing-induced CP violation in the ``golden'' mode 
$B_d\to J/\psi K_{\rm S}$, as we have seen in Subsection~\ref{ssec:BpsiK},
we may use $B_d\to\pi^+\pi^-$ to probe $\gamma$. 

The current measurements of the $B_d\to\pi^+\pi^-$ CP asymmetries 
are given as follows:
\begin{eqnarray}
{\cal A}_{\rm CP}^{\rm dir}(B_d\to\pi^+\pi^-)
&=&\left\{
\begin{array}{ll}
-0.09\pm0.15\pm0.04 & \mbox{(BaBar \cite{BaBar-Bpipi-05})}\\
-0.56\pm0.12\pm0.06 & \mbox{(Belle \cite{Belle-Bpipi-05})}
\end{array}
\right.\label{Adir-exp}\\
{\cal A}_{\rm CP}^{\rm mix}(B_d\to\pi^+\pi^-)
&=&\left\{
\begin{array}{ll}
+0.30\pm0.17\pm0.03& \mbox{(BaBar \cite{BaBar-Bpipi-05})}\\
+0.67\pm0.16\pm0.06 & \mbox{(Belle \cite{Belle-Bpipi-05}).}
\end{array}
\right.\label{Amix-exp}
\end{eqnarray}
The BaBar and Belle results are still not fully consistent with each other, 
although the experiments are now in better agreement. In \cite{HFAG}, 
the following averages were obtained:
\begin{eqnarray}
{\cal A}_{\rm CP}^{\rm dir}(B_d\to\pi^+\pi^-)&=&-0.37\pm0.10
\label{Bpipi-CP-averages}\\
{\cal A}_{\rm CP}^{\rm mix}(B_d\to\pi^+\pi^-)&=&+0.50\pm0.12.
\label{Bpipi-CP-averages2}
\end{eqnarray}
The central values of these averages are remarkably stable
in time. Direct CP violation at this level would require large penguin 
contributions with large CP-conserving strong phases, thereby indicating
large non-factorizable effects. 

This picture is in fact supported by the direct CP violation in $B^0_d\to\pi^-K^+$ 
modes that could be established by the $B$ factories in 
the summer of 2004 \cite{CP-B-dir}. Here the BaBar and Belle results agree 
nicely with each other, yielding the following average \cite{HFAG}:
\begin{equation}\label{AdirBdpimKp-exp}
{\cal A}_{\rm CP}^{\rm dir}(B_d\to\pi^\mp K^\pm)=0.115\pm 0.018.
\end{equation}
The diagrams contributing to $B^0_d\to\pi^-K^+$ can straightforwardly be 
obtained from those in Fig.~\ref{fig:Bpipi-diag} by just replacing the anti-down quark
emerging from the $W$ boson through an anti-strange quark. Consequently, the 
hadronic matrix elements entering $B^0_d\to\pi^+\pi^-$ and $B^0_d\to\pi^-K^+$ can be 
related to one another through the $SU(3)$ flavour symmetry of strong interactions
and the additional assumption that the penguin annihilation and exchange topologies
contributing to $B^0_d\to\pi^+\pi^-$, which have no counterpart in 
$B^0_d\to\pi^-K^+$ and involve the ``spectator" down quark in 
Fig.~\ref{fig:Bpipi-diag}, play actually a negligible r\^ole \cite{RF-Bpipi}. Following 
these lines, we obtain the following relation in the SM:
\begin{equation}\label{H-rel}
\hspace*{-1.7truecm}
H_{\rm BR}\equiv\underbrace{\frac{1}{\epsilon}
\left(\frac{f_K}{f_\pi}\right)^2\left[\frac{\mbox{BR}
(B_d\to\pi^+\pi^-)}{\mbox{BR}(B_d\to\pi^\mp K^\pm)}
\right]}_{\mbox{$7.5\pm 0.7$}} =
\underbrace{-\frac{1}{\epsilon}\left[\frac{{\cal A}_{\rm CP}^{\rm dir}(B_d\to\pi^\mp 
K^\pm)}{{\cal A}_{\rm CP}^{\rm dir}(B_d\to\pi^+\pi^-)}
\right]}_{\mbox{$6.7\pm 2.0$}} \equiv H_{{\cal A}_{\rm CP}^{\rm dir}},
\end{equation}
where 
\begin{equation}\label{eps-def}
\epsilon\equiv\frac{\lambda^2}{1-\lambda^2}=0.053, 
\end{equation}
and the ratio $f_K/f_\pi=160/131$ of the kaon and pion decay constants
defined through
\begin{equation}\label{decay-const-def}
\langle 0|\bar s \gamma_\alpha\gamma_5 u|K^+(k)\rangle=
i f_K k_\alpha, \quad
\langle 0|\bar d \gamma_\alpha\gamma_5 u|\pi^+(k)\rangle=
i f_\pi k_\alpha
\end{equation}
describes
factorizable $SU(3)$-breaking corrections. As usual, the CP-averaged 
branching ratios are defined as
\begin{equation}
\mbox{BR}\equiv\frac{1}{2}\left[\mbox{BR}(B\to f)+
\mbox{BR}(\bar B\to \bar f)\right].
\end{equation}
In (\ref{H-rel}), we have also given the
numerical values following from the data. Consequently, this relation 
is well satisfied within the experimental uncertainties, and does not
show any anomalous behaviour. It supports therefore the SM description
of the $B^0_d\to\pi^-K^+$, $B^0_d\to\pi^+\pi^-$ decay amplitudes,
and our working assumptions listed before (\ref{H-rel}). 

The quantities $H_{\rm BR}$ and $H_{{\cal A}_{\rm CP}^{\rm dir}}$ introduced
in this relation can be written as follows:
\begin{equation}\label{H-fct}
H_{\rm BR} = G_3(d,\theta;\gamma) =
H_{{\cal A}_{\rm CP}^{\rm dir}}.
\end{equation}
If we complement this expression with (\ref{CP-Bpipi-dir-gen}) and
(\ref{CP-Bpipi-mix-gen}), and use (see (\ref{phid-exp}))
\begin{equation}\label{phi-d-det}
\phi_d=(43.4\pm2.5)^\circ,
\end{equation}
we have sufficient information to determine $\gamma$, 
as well as $(d,\theta)$  \cite{RF-BsKK,RF-Bpipi,FleischerMatias}. In using
(\ref{phi-d-det}), we assume that the possible discrepancy with the SM
described by (\ref{S-psi-K}) is only due to NP in $B^0_d$--$\bar B^0_d$ mixing
and not to effects entering through the $B^0_d\to J/\psi K_{\rm S}$ decay amplitude.
As was recently shown in Ref.~\cite{BFRS-5}, the results following from $H_{\rm BR}$ 
and $H_{{\cal A}_{\rm CP}^{\rm dir}}$ give results that are in good agreement with 
one another. Since the avenue offered by $H_{{\cal A}_{\rm CP}^{\rm dir}}$
is cleaner than the one provided by $H_{\rm BR}$, it is preferable to use the former
quantity to determine $\gamma$, yielding the following result \cite{BFRS-5}:
\begin{equation}\label{gamma-det}
\gamma=(73.9^{+5.8}_{-6.5})^\circ.
\end{equation}
Here a second solution around $42^\circ$ was discarded, which can be exclueded 
through an analysis of the whole $B\to\pi\pi,\pi K$ system \cite{BFRS}. As was recently 
discussed  \cite{BFRS-5} (see also Refs.~\cite{RF-Bpipi,FleischerMatias}), even large 
non-factorizable $SU(3)$-breaking corrections have a remarkably small impact 
on the numerical result in (\ref{gamma-det}). The value of $\gamma$ in 
(\ref{gamma-det}) is higher than the results following from the CKM fits
\cite{CKMfitter,UTfit}. An even larger value in the ballpark of $80^\circ$ was 
recently extracted from the $B\to\pi\pi$ data with the help of SCET 
\cite{gam-SCET,SCET-Bdpi0K0}. Performing Dalitz analyses of the neutral
$D$-meson decays in $B^\pm\to D^{(*)} K^\pm$ and $B^\pm\to D K^{*\pm}$ 
transitions, the $B$ factories have obtained the following results for $\gamma$:
\begin{equation}\label{gam}
\gamma=\left\{
\begin{array}{ll}
(67\pm28\pm13\pm11)^\circ & \mbox{BaBar \cite{Babar-Dal}}\\
(68^{+14}_{-15}\pm13\pm11)^\circ & \mbox{Belle \cite{Belle-Dal},}
\end{array}
\right.
\end{equation}
which agree with (\ref{gamma-det}), although the errors are too large to
draw definite conclusions.

The interesting feature of the value of $\gamma$ in (\ref{gamma-det}) is 
that it should not receive significant NP contributions. If we complement
it with $|V_{ub}/V_{cb}|$ extracted from semi-leptonic tree-level 
$B$ decays, which are also very robust with respect to NP effects, we may 
determine the ``true'' UT, i.e.\  the reference UT introduced in 
Refs.~\cite{GNW,refut}. Using, as in Ref.~\cite{BFRS-5}, the average value 
$|V_{ub}/V_{cb}|= 0.102\pm  0.005$ (for a detailed discussion, see Ref.~\cite{UTfit}) 
yields
\begin{equation}\label{UT-true}
\alpha_{\rm true}=(80.3^{+6.6}_{-5.9})^\circ, \quad
\beta_{\rm true}=(25.8\pm1.3)^\circ,
\end{equation}
corresponding to $(\sin 2\beta)_{\rm true}=0.78\pm 0.03$, which is significantly
larger than (\ref{s2b-average}). This difference can be attributed to a 
non-vanishing value of the NP phase $\phi_d^{\rm NP}$ in (\ref{Dm-Phi-NP}), 
where $\phi_d^{\rm SM}$ corresponds to $2\beta_{\rm true}$. This
exercise yields $\phi^{\rm NP}_d=-(8.2\pm3.5)^\circ$ \cite{BFRS-5}, in 
excellent accordance with the discussion in Subsection~\ref{ssec:BpsiK}, and 
the recent study of  Ref.~\cite{UTfit-NP}.  Performing detailed analyses
of $B^0_d\to\rho^+\rho^-$ decays, the $B$ factories have extracted the 
following ranges of $\alpha$:
\begin{equation}\label{alph}
\alpha=\left\{
\begin{array}{ll}
(100\pm13)^\circ & \mbox{BaBar \cite{Babar-alph}}\\
(87\pm17)^\circ & \mbox{Belle \cite{Belle-alph},}
\end{array}
\right.
\end{equation}
which can be related to $\alpha_{\rm true}$ with the help of the simple
relation 
\begin{equation}
\alpha_{\rm true}=\alpha+ \phi_d^{\rm NP}/2. 
\end{equation}
Comparing (\ref{UT-true}) and (\ref{alph}), we observe that the latter 
measurements seem also to prefer a {\it negative} value of $\phi^{\rm NP}_d$,
in accordance with the discussion given above, although the current errors are 
of course not conclusive. Nevertheless, this pattern is interesting and should 
be monitored in the future as the quality of the data improves.

The decay $B^0_d\to\pi^+\pi^-$ plays also an important r\^ole in the next section, 
dealing with an analysis of the $B\to\pi K$ system.

\begin{figure}
   \centerline{
   \begin{tabular}{lc}
     {\small(a)} & \\
    &  \includegraphics[width=5.2truecm]{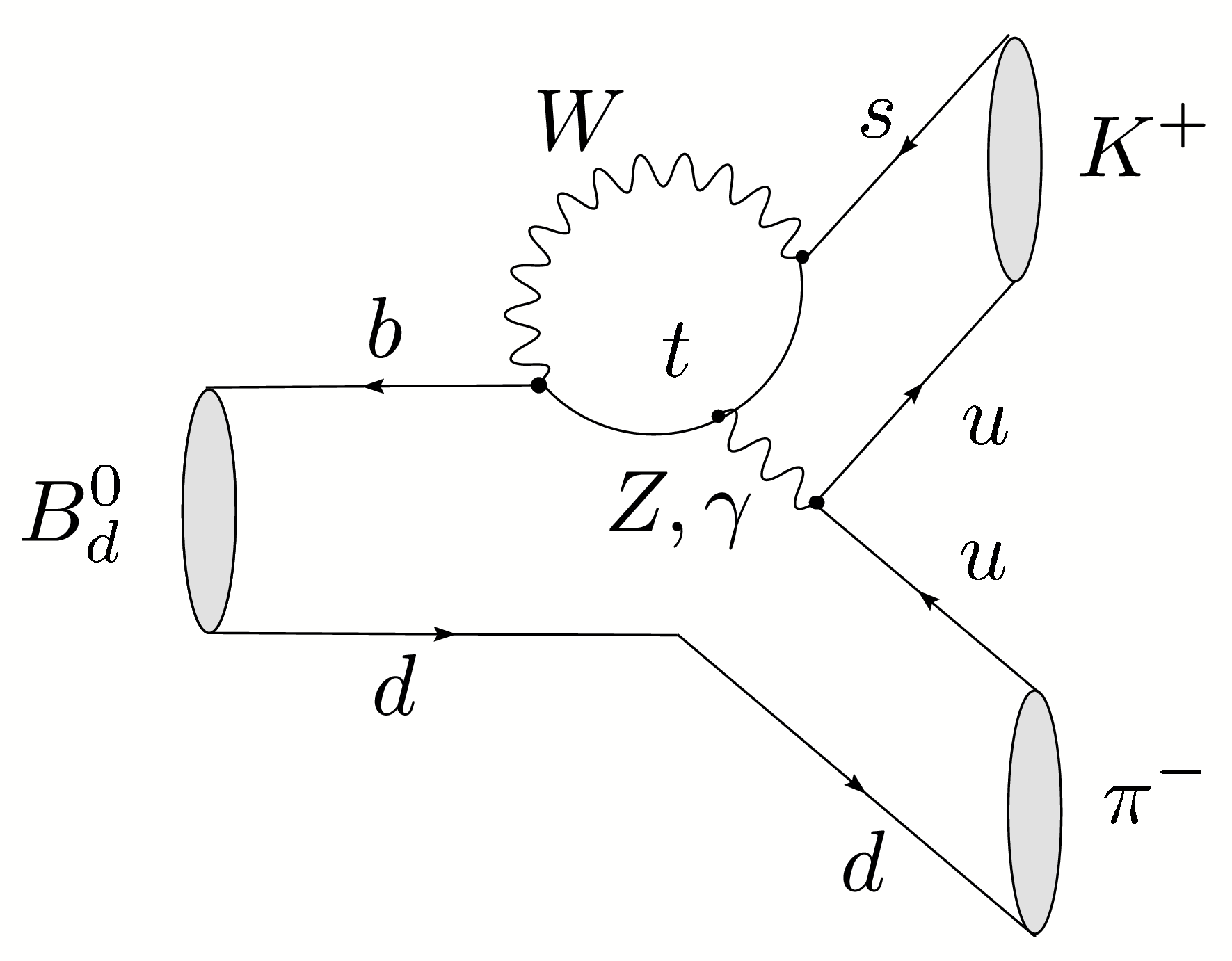}
     \hspace*{0.5truecm}
    \includegraphics[width=5.2truecm]{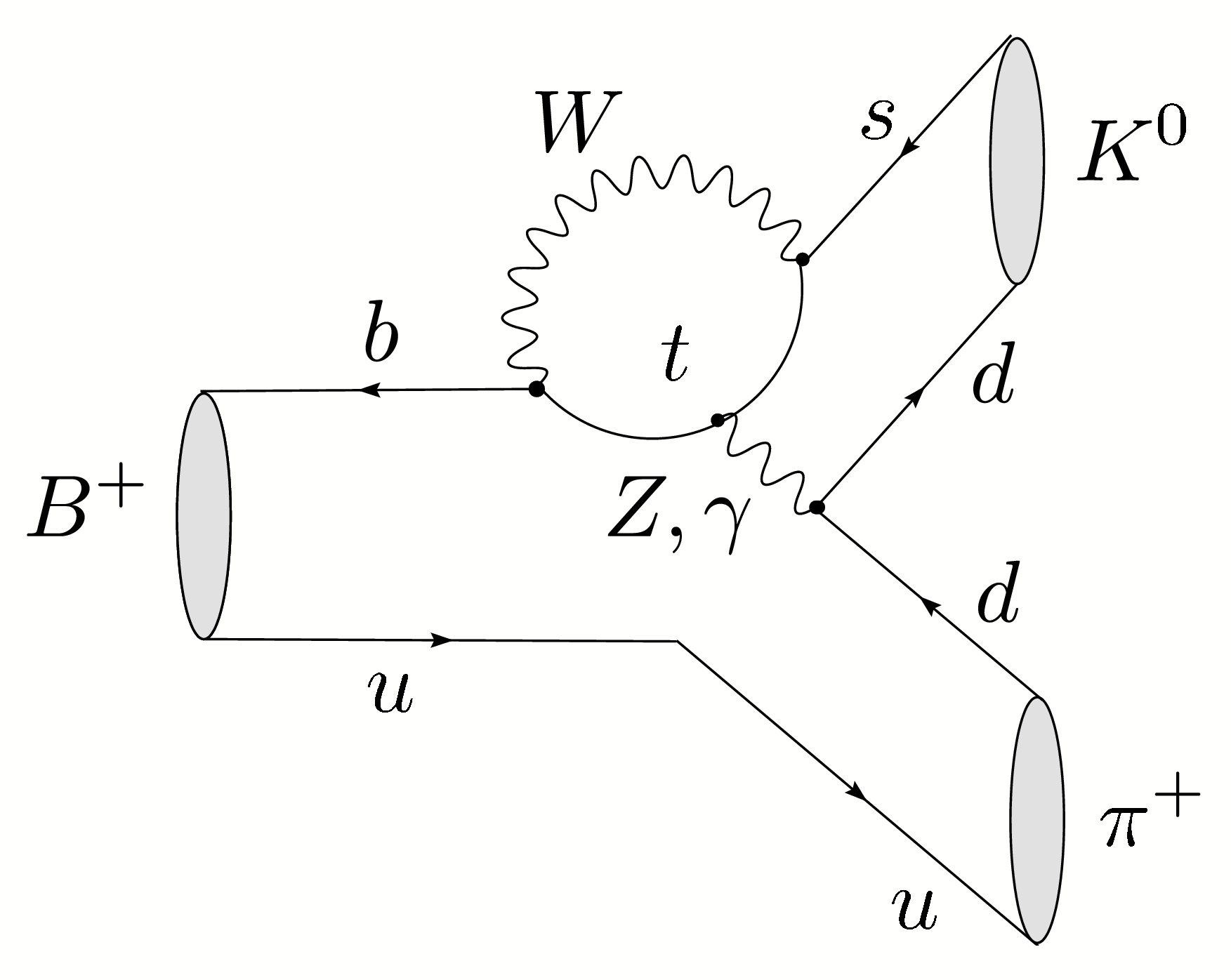} \\
     {\small(b)} & \\
    &     \includegraphics[width=5.2truecm]{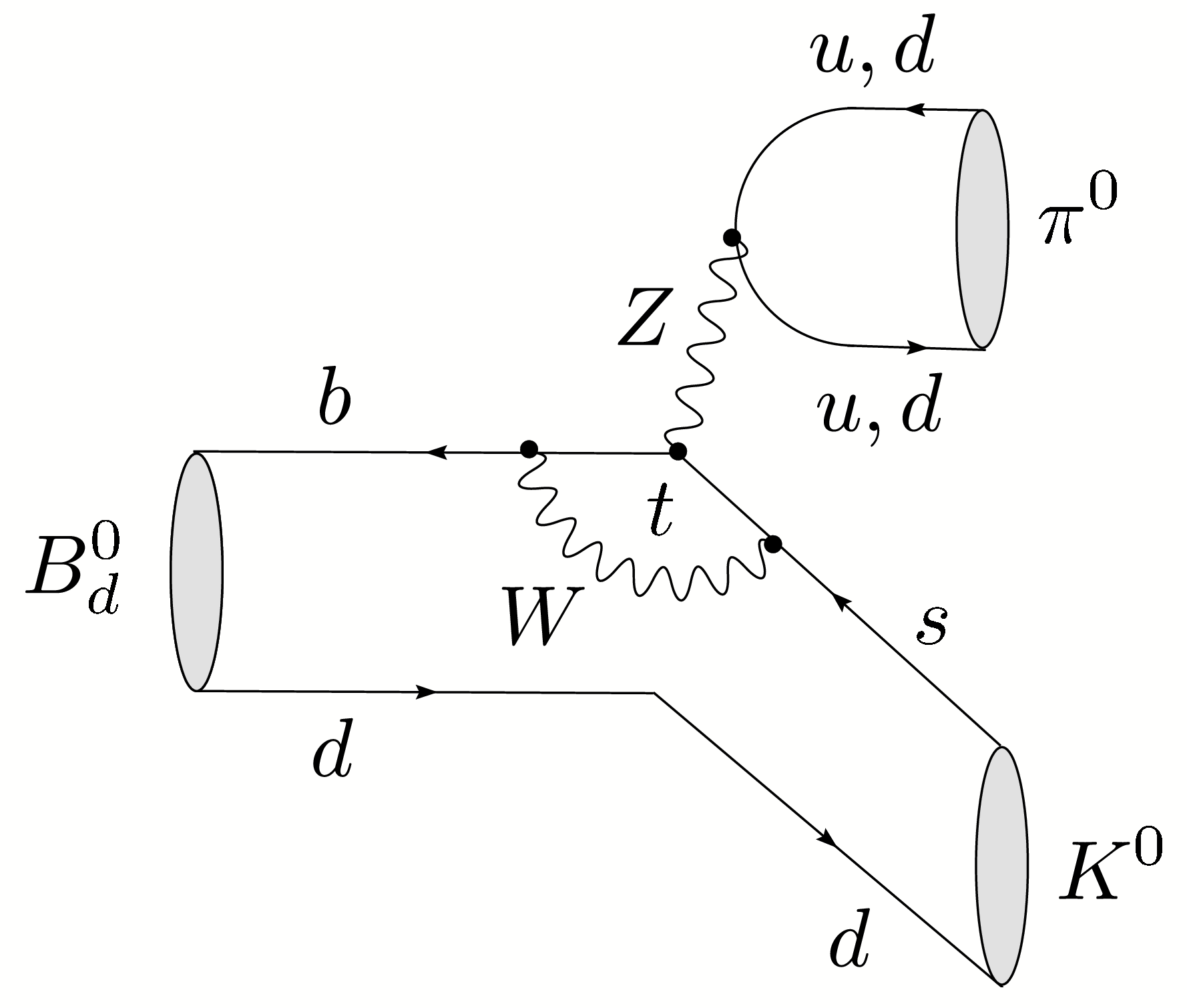} 
     \hspace*{0.5truecm}
    \includegraphics[width=5.2truecm]{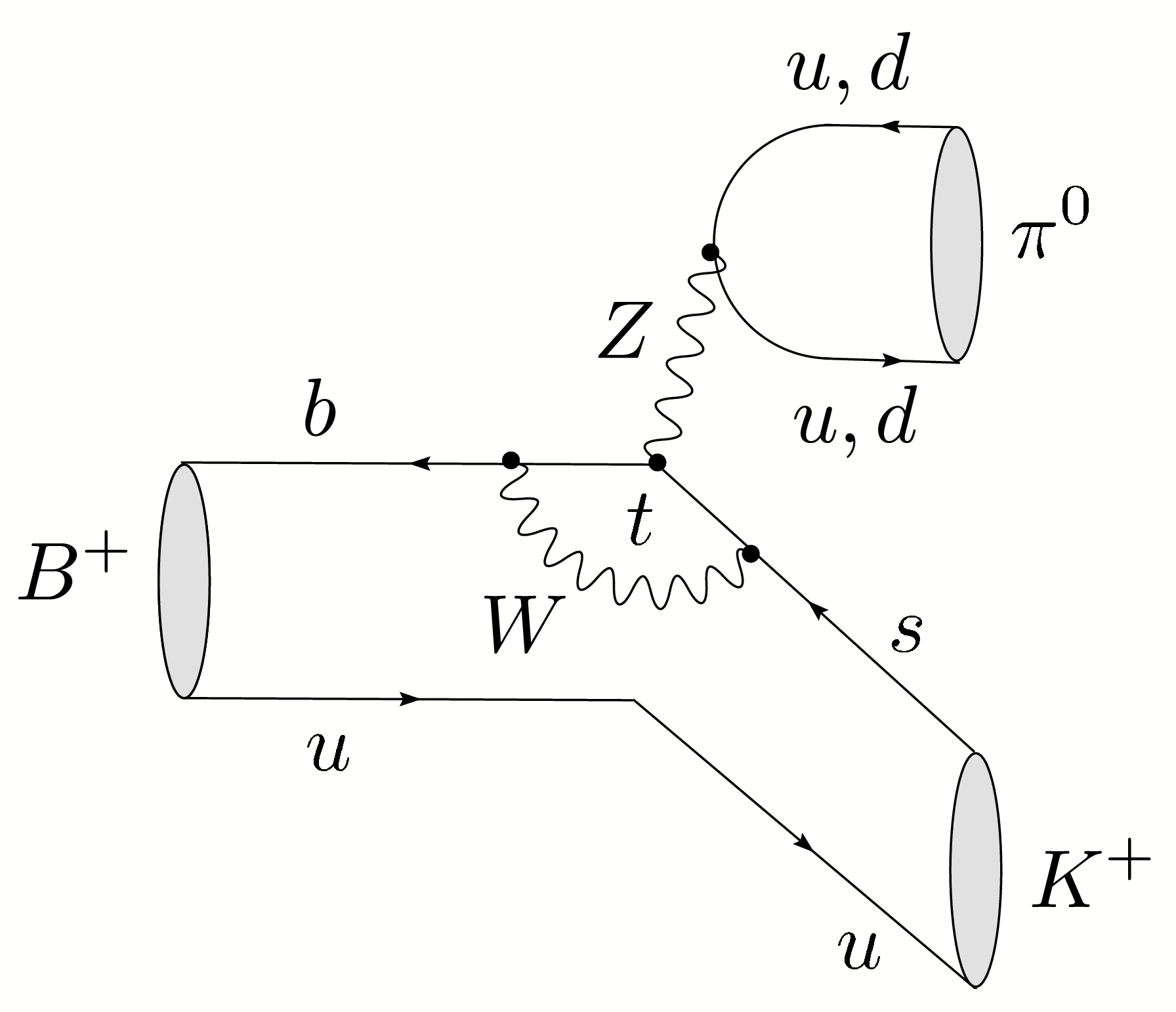} 
     \end{tabular}}
     \caption{Examples of the colour-suppressed (a) and colour-allowed (b) 
     EW penguin contributions to the $B\to\pi K$ system.}\label{fig:BpiK-EWP}
\end{figure}

\boldmath
\section{The $B\to\pi K$ Puzzle and its Relation to Rare $B$ and
$K$ Decays}\label{sec:BpiK-puzzle}
\unboldmath
\subsection{Preliminaries}\label{ssec:BpiK-prel}
We made already first contact with a $B\to\pi K$ decay in 
Subsection~\ref{ssec:Bpi+pi-}, the $B^0_d\to\pi^-K^+$ channel. It receives
contributions both from tree and from penguin topologies. Since this decay
originates from a $\bar b\to\bar s$ transition, the tree amplitude is suppressed
by a CKM factor $\lambda^2 R_b\sim 0.02$ with respect to the penguin
amplitude. Consequently, $B^0_d\to\pi^-K^+$ is governed by QCD penguins;
the tree topologies contribute only at the 20\% level to the decay amplitude.
The feature of the dominance of QCD penguins applies to all $B\to\pi K$ modes, 
which can be classified with respect to their EW penguin contributions 
as follows (see Fig.~\ref{fig:BpiK-EWP}):
\begin{itemize}
\item[(a)] In the $B^0_d\to\pi^-K^+$ and $B^+\to\pi^+K^0$ decays, EW penguins 
contribute in colour-suppressed form and are hence expected to play a minor r\^ole.
\item[(b)] In the $B^0_d\to\pi^0K^0$ and $B^+\to\pi^0K^+$ decays, EW penguins 
contribute in colour-allowed form and have therefore a significant impact on the decay 
amplitude, entering at the same order of magnitude as the tree contributions.
\end{itemize}
As we noted above, EW penguins offer an attractive avenue for NP to 
enter non-leptonic $B$ decays, which is also the case for the
$B\to\pi K$ system \cite{FM-BpiK-NP,trojan}. Indeed, the decays of class (b) 
show a puzzling pattern, which may point towards such a NP scenario.
This feature emerged already in 2000 \cite{BF00}, when the CLEO collaboration 
reported the observation of the $B^0_d\to\pi^0K^0$ channel with a surprisingly 
prominent rate \cite{CLEO00}, and is still present in the most recent BaBar and 
Belle data, thereby receiving a lot of attention in the literature (see, for instance, 
Refs.~\cite{Z-prime-BpiK} and \cite{BeNe}--\cite{WZ}).

In the following discussion, we focus on the systematic
strategy to explore the ``$B\to\pi K$ puzzle"  developed in Ref.~\cite{BFRS}; 
all numerical results refer to the most recent analysis presented in 
Ref.~\cite{BFRS-5}. The logical structure is very simple: the starting point is
given by the values of $\phi_d$ and $\gamma$ in (\ref{phi-d-det})
and (\ref{gamma-det}), respectively, and by the $B\to\pi\pi$ system, 
which allows us to extract a set of hadronic parameters from the data
with the help of the isospin symmetry of strong interactions. Then we make, in
analogy to the determination of $\gamma$ in Subsection~\ref{ssec:Bpi+pi-}, 
the following working hypotheses:
\begin{itemize}
\item[(i)] $SU(3)$ flavour symmetry of strong interactions (but taking factorizable 
$SU(3)$-breaking corrections into account),
\item[(ii)] neglect of penguin annihilation and exchange topologies,
\end{itemize}
which allow us to fix the hadronic $B\to\pi K$ parameters through their $B\to\pi\pi$
counterparts. Interestingly, we may gain confidence in these assumptions through 
internal consistency checks (an example is relation (\ref{H-rel})), which work nicely
within the experimental uncertainties. Having the hadronic $B\to\pi K$ parameters
at hand, we can predict the $B\to \pi K$ observables in the SM. The comparison
of the corresponding picture with the $B$-factory data will then guide us to NP
in the EW penguin sector, involving in particular a large CP-violating NP phase. In the
final step, we explore the interplay of this NP scenario with rare $K$ and $B$
decays.

\subsection{Extracting Hadronic Parameters from the $B\to\pi\pi$ 
System}\label{ssec:Bpipi-hadr}
In order to fully exploit the information that is provided by the whole $B\to\pi\pi$ 
system, we use -- in addition to the two CP-violating $B^0_d\to\pi^+\pi^-$
observables  -- the following ratios of CP-averaged branching ratios:
\begin{eqnarray}
R_{+-}^{\pi\pi}&\equiv&2\left[\frac{\mbox{BR}(B^+\to\pi^+\pi^0)
+\mbox{BR}(B^-\to\pi^-\pi^0)}{\mbox{BR}(B_d^0\to\pi^+\pi^-)
+\mbox{BR}(\bar B_d^0\to\pi^+\pi^-)}\right]
=2.04\pm0.28
\label{Rpm-def}\\
R_{00}^{\pi\pi}&\equiv&2\left[\frac{\mbox{BR}(B_d^0\to\pi^0\pi^0)+
\mbox{BR}(\bar B_d^0\to\pi^0\pi^0)}{\mbox{BR}(B_d^0\to\pi^+\pi^-)+
\mbox{BR}(\bar B_d^0\to\pi^+\pi^-)}\right]
=0.58\pm0.13.
\end{eqnarray}
The pattern of the experimental numbers in these expressions came as quite 
a surprise, as the central values calculated in QCDF gave 
$R_{+-}^{\pi\pi}=1.24$ and $R_{00}^{\pi\pi}=0.07$ \cite{BeNe}. As discussed in 
detail in \cite{BFRS}, this ``$B\to\pi\pi$ puzzle" can straightforwardly be accommodated 
in the SM through large non-factorizable hadronic interference effects, i.e.\ 
does not point towards NP. For recent SCET analyses, 
see Refs.~\cite{SCET-Bdpi0K0,BPRS,FeHu}. 

Using the isospin symmetry of strong interactions, we can write
\begin{equation}\label{Rpipi-gen}
R_{+-}^{\pi\pi}=F_1(d,\theta,x,\Delta;\gamma), \quad
R_{00}^{\pi\pi}=F_2(d,\theta,x,\Delta;\gamma),
\end{equation}
where $xe^{i\Delta}$ is another hadronic parameter, which was introduced
in \cite{BFRS}. Using now, in addition, the CP-violating observables in
(\ref{CP-Bpipi-dir-gen}) and (\ref{CP-Bpipi-mix-gen}), we arrive at the following 
set of haronic parameters:
\begin{equation}\label{Bpipi-par-det}
d=0.52^{+0.09}_{-0.09}, \quad
\theta=(146^{+7.0}_{-7.2})^\circ, \quad
x=0.96^{+0.13}_{-0.14}, \quad
\Delta=-(53^{+18}_{-26})^\circ.
\end{equation}
In the extraction of these quantites, also the EW penguin effects in the 
$B\to\pi\pi$ system are included \cite{BF98,GPY}, although these topologies have a 
tiny impact \cite{PAPIII}. Let us emphasize that the results for the hadronic 
parameters listed above, which are consistent with the picture emerging in the 
analyses of other authors (see, e.g., Refs.~\cite{CGRS,ALP-Bpipi}), 
are essentially clean and serve as a testing ground for 
calculations within QCD-related approaches. For instance, in 
recent QCDF \cite{busa} and PQCD \cite{kesa} analyses, the 
following numbers were obtained:
\begin{equation}
\left.d\right|_{\rm QCDF}=0.29\pm0.09, \quad
\left.\theta\right|_{\rm QCDF}=-\left(171.4\pm14.3\right)^\circ, 
\end{equation}
\begin{equation}
\left.d\right|_{\rm PQCD}=0.23^{+0.07}_{-0.05}, \quad
+139^\circ < \left.\theta\right|_{\rm PQCD} < +148^\circ,
\end{equation}
which depart significantly from the pattern in (\ref{Bpipi-par-det}) that is implied
by the data. 

Finally, we can predict the CP asymmetries of the decay $B_d\to\pi^0\pi^0$:
\begin{equation}\label{ACP-Bdpi0pi0-pred}
{\cal A}_{\rm CP}^{\rm dir}(B_d\to \pi^0\pi^0)=-0.30^{+0.48}_{-0.26}, \quad
{\cal A}_{\rm CP}^{\rm mix}(B_d\to \pi^0\pi^0)=-0.87^{+0.29}_{-0.19}.
\end{equation}
The current experimental value for the direct CP 
asymmetry is given as follows \cite{HFAG}:
\begin{equation}\label{ACP-Bdpi0pi0-exp}
{\cal A}_{\rm CP}^{\rm dir}(B_d\to \pi^0\pi^0)=-0.28^{+0.40}_{-0.39}.
\end{equation}
Consequently, no stringent test of the corresponding prediction 
in (\ref{ACP-Bdpi0pi0-pred}) is provided at this stage, although the 
indicated agreement is encouraging.

\subsection{Analysis of the $B\to\pi K$ System}\label{ssec:BpiK}
Let us begin the analysis of the $B\to\pi K$ system by having a closer
look at the modes of class (a) introduced above, $B_d\to\pi^\mp K^\pm$
and $B^\pm\to\pi^\pm K$, which are only marginally affected by
EW penguin contributions. We used the banching ratio and direct 
CP asymmetry of the former channel already in the $SU(3)$ relation (\ref{H-rel}),
which is nicely satisfied by the current data, and in the extraction of
$\gamma$ with the help of the CP-violating $B_d\to\pi^+\pi^-$ observables,
yielding the value in  (\ref{gamma-det}). The $B_d\to\pi^\mp K^\pm$
modes provide the CP-violating asymmetry
\begin{equation}\label{ACP-BppipK0}
\hspace*{-1.9truecm}{\cal A}_{\rm CP}^{\rm dir}(B^\pm\to\pi^\pm K)\equiv
\frac{\mbox{BR}(B^+\to\pi^+K^0)-
\mbox{BR}(B^-\to\pi^-\bar K^0)}{\mbox{BR}(B^+\to\pi^+K^0)+
\mbox{BR}(B^-\to\pi^-\bar K^0)} =
0.02 \pm 0.04,
\end{equation}
and enter in the following ratio \cite{FM}:
\begin{equation}\label{R-def}
\hspace*{-0.7truecm}R\equiv\left[\frac{\mbox{BR}(B_d^0\to\pi^- K^+)+
\mbox{BR}(\bar B_d^0\to\pi^+ K^-)}{\mbox{BR}(B^+\to\pi^+ K^0)+
\mbox{BR}(B^-\to\pi^- \bar K^0)}
\right]\frac{\tau_{B^+}}{\tau_{B^0_d}} =
0.86\pm0.06;
\end{equation}
the numerical values refer again to the most recent compilation in \cite{HFAG}. 
The $B^+\to\pi^+ K^0$ channel involves another hadronic parameter,
$\rho_{\rm c}e^{i\theta_{\rm c}}$, which cannot be determined through
the $B\to\pi\pi$ data \cite{BF98,defan,neubert}:
\begin{equation}\label{B+pi+K0}
A(B^+\to\pi^+K^0)=-P'\left[1+\rho_{\rm c}e^{i\theta_{\rm c}}e^{i\gamma}
\right];
\end{equation}
the overall normalization $P'$ cancels in (\ref{ACP-BppipK0}) and
(\ref{R-def}). Usually, it is assumed that the parameter $\rho_{\rm c}e^{i\theta_{\rm c}}$  
can be neglected. In this case, the direct CP asymmetry in (\ref{ACP-BppipK0}) 
vanishes, and $R$ can be calculated through the $B\to\pi\pi$ data with the help 
of the assumptions specified in Subsection~\ref{ssec:BpiK-prel}:
\begin{equation}\label{R-pred-0}
R|_{\rm SM}=0.963^{+0.019}_{-0.022}.
\end{equation}

This numerical result is $1.6 \sigma$ larger than the experimental value
in (\ref{R-def}). As was discussed in detail in \cite{BFRS-up}, 
the experimental range for the direct CP asymmetry in (\ref{ACP-BppipK0})  
and the first direct signals for the $B^\pm\to K^\pm K$ decays favour a 
value of $\theta_{\rm c}$ around $0^\circ$. This feature allows us to essentially 
resolve the small discrepancy concerning $R$ for values of $\rho_{\rm c}$ around 
0.05. The remaining small numerical difference between the calculated value of
$R$ and the experimental result, if confirmed by future data, could be due to
(small) colour-suppressed EW penguins, which enter $R$ as well \cite{BFRS}.
As was recently discussed in Ref.~\cite{BFRS-5}, even large non-factorizable
$SU(3)$-breaking effects would have a small impact on the predicted value 
of $R$. In view of these results, it would not be a surprise to see an increase 
of the experimental value of $R$ in the future.

\begin{figure}
\begin{center}
\includegraphics[width=10cm]{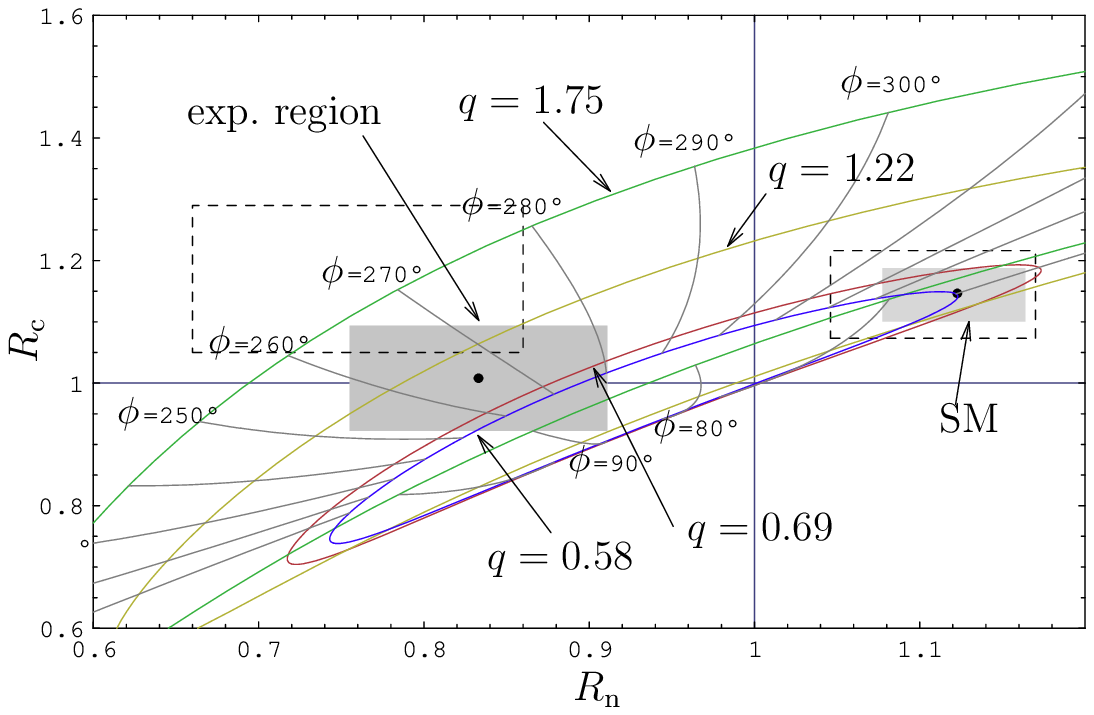}
\end{center}
\vspace*{-0.5truecm}
\caption{The current situation in the $R_{\rm n}$--$R_{\rm c}$ plane: the shaded 
areas indicate the experimental and SM $1 \sigma$ ranges, while the lines show the
theory predictions for the central values of the hadronic parameters
and various values of $q$ with $\phi\in[0^\circ,360^\circ]$.}\label{fig:RnRc}
\end{figure}

Let us now turn to the $B^+\to\pi^0K^+$ and $B^0_d\to\pi^0K^0$ channels,
which are the $B\to\pi K$ modes with significant contributions from EW
penguin topologies. The key observables for the exploration of these modes 
are the following ratios of their CP-averaged branching ratios \cite{BF00,BF98}:
\begin{equation}\label{Rc-def}
R_{\rm c}\equiv2\left[\frac{\mbox{BR}(B^+\to\pi^0K^+)+
\mbox{BR}(B^-\to\pi^0K^-)}{\mbox{BR}(B^+\to\pi^+ K^0)+
\mbox{BR}(B^-\to\pi^- \bar K^0)}\right] =
1.01\pm0.09
\end{equation}
\begin{equation}\label{Rn-def}
R_{\rm n}\equiv\frac{1}{2}\left[
\frac{\mbox{BR}(B_d^0\to\pi^- K^+)+
\mbox{BR}(\bar B_d^0\to\pi^+ K^-)}{\mbox{BR}(B_d^0\to\pi^0K^0)+
\mbox{BR}(\bar B_d^0\to\pi^0\bar K^0)}\right] =
0.83\pm0.08,
\end{equation}
where the overall normalization factors of the decay amplitudes cancel, 
as in (\ref{R-def}). In order to describe the EW penguin effects, both a parameter 
$q$, which measures the strength of the EW penguins with respect to 
tree-like topologies, and a CP-violating phase $\phi$ are introduced. In the SM, this 
phase vanishes, and $q$ can be calculated with the help of the $SU(3)$ flavour 
symmetry, yielding a value of $0.69 \times 0.086/|V_{ub}/V_{cb}|= 0.58$ \cite{NR}.
Following the strategy described above yields the following SM predictions: 
\begin{equation}\label{RncSM}
R_{\rm c}|_{\rm SM}=1.15 \pm 0.05, \quad R_{\rm n}|_{\rm SM}=1.12 \pm 0.05,
\end{equation}
where in particular the value of $R_{\rm n}$ does not agree with the experimental
number, which is a manifestation of the $B\to\pi K$ puzzle. As was recently
discussed in Ref.~\cite{BFRS-5}, the internal consistency checks of the
working assumptions listed in Subsection~\ref{ssec:BpiK-prel} are
currently satisfied at the level of $25\%$, and can be systematically improved 
through better data. A detailed study of the numerical predictions in 
(\ref{RncSM}) (and those given below) shows that their sensitivity on 
non-factorizable $SU(3)$-breaking effects of this order of magnitude is
surprisingly small. Consequently, it is very exciting to speculate that NP
effects in the EW penguin sector, which are described effectively through
$(q,\phi)$, are at the origin of the $B\to\pi K$ puzzle. 
Following Ref.~\cite{BFRS}, we show the situation in the $R_{\rm n}$--$R_{\rm c}$ 
plane in Fig.~\ref{fig:RnRc}, where -- for the convenience of the reader --
also the experimental range and the SM predictions at the time of the 
original analysis of Ref.~\cite{BFRS} are indicated through the dashed rectangles. 
We observe that although
the central values of $R_{\rm n}$ and $R_{\rm c}$ have slightly moved towards
each other, the puzzle is as prominent as ever. The experimental
region can now be reached without an enhancement of $q$, but
a large CP-violating phase $\phi$ of the order of $-90^\circ$ is
still required:
\begin{equation}
\label{q-phi}
q=0.99\,^{+0.66}_{-0.70} ,\quad \phi=-(94\,^{+16}_{-17} )^\circ.
\end{equation}
Interestingly, $\phi$ of the order of $+90^\circ$ can now also bring us rather 
close to the experimental range of $R_{\rm n}$ and $R_{\rm c}$. 

An interesting probe of the NP phase $\phi$ is also provided
by the CP violation in the decay $B^0_d\to\pi^0 K_{\rm S}$. Within the SM,
the corresponding observables are expected to satisfy the following 
relations \cite{PAPIII}:
\begin{equation}\label{Bdpi0K0-rel}
{\cal A}_{\rm CP}^{\rm dir}(B_d\!\to\!\pi^0 K_{\rm S})\approx 0, \quad
{\cal A}_{\rm CP}^{\rm mix}(B_d\!\to\!\pi^0 K_{\rm S})\approx
{\cal A}_{\rm CP}^{\rm mix}(B_d\!\to\!\psi K_{\rm S}).
\end{equation}
The most recent  Belle \cite{Belle-Bphi-K} and BaBar \cite{BaBar-pi0KS}
measurements of these quantities are in agreement with each other, and 
lead to the following averages \cite{HFAG}:
\begin{eqnarray}
{\cal A}_{\rm CP}^{\rm dir}(B_d\!\to\!\pi^0 K_{\rm S})&=&-0.02\pm0.13\\
{\cal A}_{\rm CP}^{\rm mix}(B_d\!\to\!\pi^0 K_{\rm S})&=&-0.31\pm0.26
\equiv -(\sin2\beta)_{\pi^0K_{\rm S}}.
\end{eqnarray}
Taking (\ref{s2b-average}) into account yields
\begin{equation}\label{DS}
\Delta S \equiv (\sin2\beta)_{\pi^0K_{\rm S}}-
(\sin2\beta)_{\psi K_{\rm S}} =
-0.38\pm 0.26,
\end{equation}
which may indicate a sizeable deviation of the
experimentally measured value of $(\sin2\beta)_{\pi^0K_{\rm S}}$ from
$(\sin2\beta)_{\psi K_{\rm S}}$, and is therefore one of the recent hot topics. 
Since the strategy developed in Ref.~\cite{BFRS} allows us also to predict the 
CP-violating observables of the $B^0_d\to\pi^0 K_{\rm S}$ channel both within the 
SM and within our scenario of NP, it allows us to address this issue, yielding
\begin{equation}\label{pi0KS-SM}
{\cal A}_{\rm CP}^{\rm dir}(B_d\!\to\!\pi^0 K_{\rm S})|_{\rm SM}=0.06^{+0.09}_{-0.10},
\qquad
\Delta S\vert_{\rm SM}= 0.13\pm0.05, 
\end{equation}
\begin{equation}\label{pi0KS-NP}
{\cal A}_{\rm CP}^{\rm dir}(B_d\!\to\!\pi^0 K_{\rm S})|_{\rm NP}=0.01\,^{+0.14}_{-0.18}, 
\qquad \Delta S\vert_{\rm NP}=0.27\,^{+0.05}_{-0.09},
\end{equation}
where the NP results refer to the EW penguin parameters in (\ref{q-phi}). Consequently,
$\Delta S$ is found to be {\it positive} in the SM. In the literature, values of
$\Delta S\vert_{\rm SM}\sim0.04$--$0.08$ can be found, which were obtained
-- in contrast to (\ref{pi0KS-SM})  -- with the help of dynamical approaches such as QCDF \cite{beneke} and SCET \cite{SCET-Bdpi0K0}. Moreover, 
bounds were derived with the help of the $SU(3)$ flavour symmetry 
\cite{SU3-bounds}. Looking at (\ref{pi0KS-NP}), we see that the modified  
parameters $(q,\phi)$ in (\ref{q-phi}) imply an enhancement of $\Delta S$ with 
respect to the SM case. Consequently, the best values of $(q,\phi)$ that are 
favoured by the measurements of $R_{\rm n,c}$ make the potential 
${\cal A}_{\rm CP}^{\rm mix}(B_d\!\to\!\pi^0 K_{\rm S})$ discrepancy 
even larger than in the SM.

\begin{figure}
\begin{center}
\includegraphics[width=10cm]{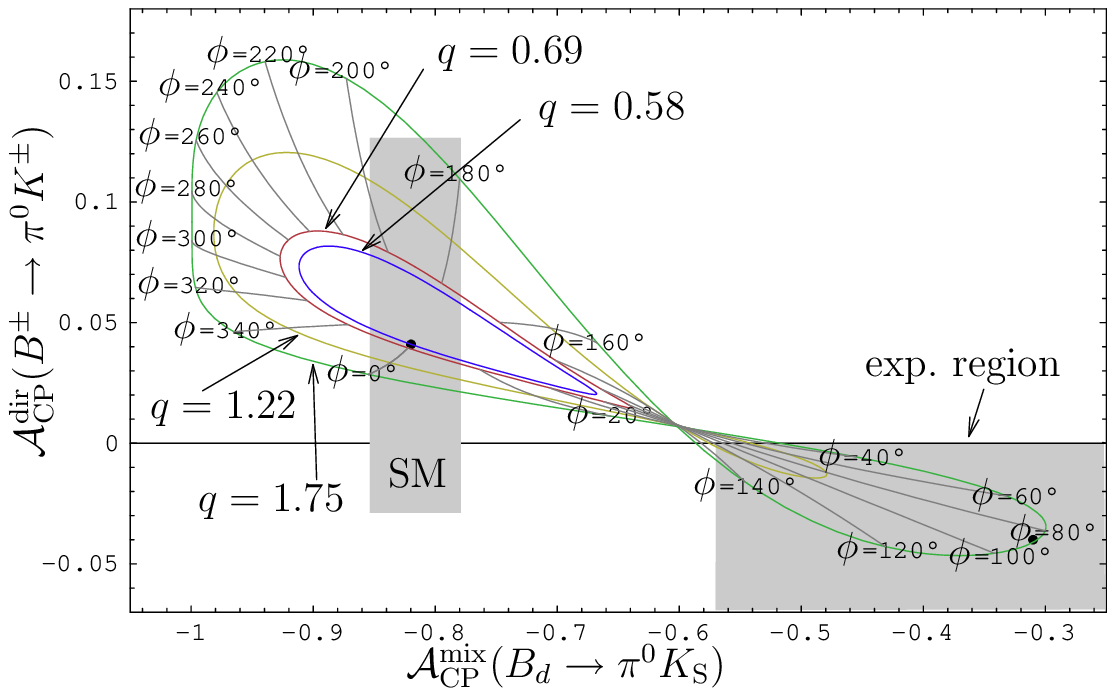}
\end{center}
\vspace*{-0.5truecm}
\caption{The situation in the 
${\cal A}_{\rm CP}^{\rm mix}(B_d\to\pi^0K_{\rm S})$--${\cal A}_{\rm CP}^{\rm dir}
(B^\pm\to\pi^0K^\pm)$ plane:  the shaded regions represent the experimental
and SM $1 \sigma$ ranges, while the lines show the
theory predictions for the central values of the hadronic parameters
and various values of $q$ with 
$\phi\in[0^\circ,360^\circ]$.\label{fig:Adirpi0KS-Amixpi0K+}}
\end{figure}

There is one CP asymmetry of the $B\to\pi K$ system left,  
which is measured as 
\begin{equation}\label{AdirBppi0KP-exp}
{\cal A}_{\rm CP}^{\rm dir}(B^\pm\to\pi^0K^\pm) =
-0.04\pm 0.04.
\end{equation}
In the limit of vanishing colour-suppressed tree and EW penguin topologies,
it is expected to be equal to the direct CP asymmetry of the $B_d\to\pi^\mp K^\pm$
modes. Since the experimental value of the latter asymmetry in 
(\ref{AdirBdpimKp-exp}) does not agree with (\ref{AdirBppi0KP-exp}), the 
direct CP violation in $B^\pm\to\pi^0K^\pm$ has also received
a lot of attention. The lifted colour suppression described by the large value of 
$x$ in (\ref{Bpipi-par-det}) could, in principle, be responsible for a non-vanishing
difference between (\ref{AdirBdpimKp-exp}) and (\ref{AdirBppi0KP-exp}),
\begin{equation}
\Delta A \equiv {\cal A}_{\rm CP}^{\rm dir}(B^\pm\to\pi^0K^\pm)
               -{\cal A}_{\rm CP}^{\rm dir}(B_d\to\pi^\mp K^\pm)\,\stackrel{{\rm exp}}{=} 
                -0.16\pm0.04.  \label{DeltaA}
\end{equation}
However, applying once again the strategy described above yields
\begin{equation}\label{AdirBppi0KP-SM}
{\cal A}_{\rm CP}^{\rm dir}(B^\pm\to\pi^0K^\pm)|_{\rm SM} 
= 0.04\,^{+0.09}_{-0.07},
\end{equation}
so that the SM still prefers a positive value of this CP asymmetry; 
the NP scenario characterized by (\ref{q-phi}) corresponds to 
\begin{equation}\label{AdirBppi0KP-NP}
{\cal A}_{\rm CP}^{\rm dir}(B^\pm\to\pi^0K^\pm)|_{\rm NP}
= 0.09\,^{+0.20}_{-0.16}.
\end{equation}

In view of the large uncertainties, no stringent test is provided at this point.
Nevertheless, it is tempting to play a bit with the CP asymmetries of the
$B^\pm\to\pi^0K^\pm$ and $B_d\to\pi^0K_{\rm S}$ decays. In 
Fig.~\ref{fig:Adirpi0KS-Amixpi0K+}, we show the situation in the 
${\cal A}_{\rm CP}^{\rm mix}(B_d\to\pi^0K_{\rm S})$--${\cal A}_{\rm CP}^{\rm dir}
(B^\pm\to\pi^0K^\pm)$ plane for various values of $q$ with $\phi\in[0^\circ,360^\circ]$.
We see that these observables seem to show a preference for positive values of 
$\phi$ around $+90^\circ$. As we noted above, in this case, we can also get 
rather close to the experimental region in the $R_{\rm n}$--$R_{\rm c}$ plane.
It is now interesting to return to the discussion of the NP effects in the
$B\to\phi K$ system given in Subsection~\ref{ssec:BphiK}. In our scenario of NP 
in the EW penguin sector, we have just to identify the CP-violating phase $\phi_0$ 
in (\ref{AphiK-NP}) with the NP phase $\phi$ \cite{BFRS}. Unfortunately, we 
cannot determine the hadronic $B\to\phi K$ parameters $\tilde v_0$ and 
$\tilde\Delta_0$ through the $B\to\pi\pi$ data as in the case of the $B\to\pi K$ 
system. However, if we take 
into account that $\tilde\Delta_0=180^\circ$ in factorization and look at 
Fig.~\ref{fig:Plot-BphiK}, we see again that the case of $\phi\sim+90^\circ$ would 
be favoured by the data for ${\cal S}_{\phi K}$. Alternatively, in the case of 
$\phi\sim-90^\circ$, $\tilde\Delta_0\sim 0^\circ$ would be required to 
accommodate a negative value of ${\cal S}_{\phi K}$, which appears unlikely. Interestingly, a similar comment applies to the $B\to J/\psi K$ observables 
shown in Fig.~\ref{fig:Plot-BpsiK}, although here a dramatic enhancement of the 
EW penguin parameter $v_0$ relative to the SM estimate would be simultaneously
needed to reach the central experimental values, in contract to the reduction of 
$\tilde v_0$ in the $B\to\phi K$ case. In view of rare decay constraints, the behaviour 
of the $B\to \phi K$ parameter $\tilde v_0$ appears much more likely, 
thereby supporting the assumption after (\ref{phi-d-det}).

\subsection{The Interplay with Rare $K$ and $B$ Decays and
Future Scenarios}\label{ssec:rareKB}
In order to explore the implications of the $B\to\pi K$ puzzle for rare 
$K$ and $B$ decays, we
assume that the NP enters the EW penguin sector through 
$Z^0$ penguins with a new CP-violating phase. This scenario was already
considered in the literature, where model-independent analyses and 
studies within SUSY can be found \cite{Z-pen-analyses,BuHi}.
In the strategy discussed here, the short-distance function $C$ characterizing 
the $Z^0$ penguins is determined through the $B\to\pi K$ data \cite{BFRS-I}. 
Performing a renormalization-group analysis yields
\begin{equation}\label{RG}
C(\bar q)= 2.35~ \bar q e^{i\phi} -0.82 \quad\mbox{with}\quad 
\bar q= q \left[\frac{|V_{ub}/V_{cb}|}{0.086}\right].
\end{equation}
Evaluating then the relevant box-diagram contributions in the SM 
and using (\ref{RG}), the short-distance functions
\begin{equation}\label{X-C-rel}
X=2.35~ \bar q e^{i\phi} -0.09 \quad \mbox{and} \quad 
Y=2.35~ \bar q e^{i\phi} -0.64
\end{equation}
can also be calculated, which govern the rare $K$, $B$ decays with $\nu\bar\nu$ 
and $\ell^+\ell^-$ in the final states, respectively. In the SM, we have 
$C=0.79$, $X=1.53$ and $Y=0.98$, with {\it vanishing} CP-violating phases. 
An analysis along these lines shows that the value of $(q,\phi)$ in (\ref{q-phi}), 
which is preferred by the $B\to\pi K$ observables $R_{\rm n,c}$, requires the 
following lower bounds for $X$ and $Y$ \cite{BFRS-5}:
\begin{equation}\label{XY1}
|X|_{\rm min}\approx 
|Y|_{\rm min}\approx 2.2,
\end{equation}
which appear to violate the $95\%$ probability upper bounds
\begin{equation}\label{XY2}
X\le 1.95, \quad Y\le 1.43
\end{equation}
that were recently obtained within the context of MFV \cite{Bobeth:2005ck}. 
Although we have to deal with CP-violating NP phases in our scenario,
which goes therefore beyond the MFV framework, a closer look at 
$B\to X_s \ell^+\ell^-$ shows that the upper 
bound on $|Y|$ in (\ref{XY2}) is difficult to avoid if NP enters only through
EW penguins  and the operator basis is the same as in the SM. A possible
solution to the clash between (\ref{XY1}) and (\ref{XY2}) would be given
by more complicated NP scenarios \cite{BFRS-5}. However, unless a specific 
model is chosen, the predictive power is then significantly reduced. For the
exploration of the NP effects in rare decays, we will therefore not follow
this avenue.

\begin{table}
\vspace{0.4cm}
\begin{center}
\begin{tabular}{|c||c|c|c|c|c|}
\hline
  Quantity & SM & Scen A & Scen B &  Scen C & Experiment
 \\ \hline 
$R_{\rm n}$  & 1.12 &$0.88$ & 1.03 &  1 & $0.83 \pm 0.08$ \\\hline
$R_{\rm c}$  & 1.15 &$0.96$ & 1.13 & 1  & $1.01 \pm 0.09$ \\\hline
${\cal A}_{\rm CP}^{\rm dir}(B^\pm\!\to\!\pi^0 K^\pm) $ &
  0.04 & $0.07$  \rule{0em}{1.05em}& 0.06 & 0.02  &  $-0.04 \pm 0.04$ \\ \hline
${\cal A}_{\rm CP}^{\rm dir}(B_d\!\to\!\pi^0 K_{\rm S})$ & 
  0.06 & $0.04$  \rule{0em}{1.05em}& 0.03  & 0.09 & $-0.02 \pm 0.13$ \\ \hline 
${\cal A}_{\rm CP}^{\rm mix}(B_d\!\to\!\pi^0 K_{\rm S})$ & 
  $-0.82$ & $-0.89$\rule{0em}{1.05em}& $-0.91$ & $-0.70$ &  $-0.31 \pm 0.26$ \\ \hline
$\Delta S$ & 0.13& 0.21& 0.22& 0.01& $-0.38\pm0.26$ \\ \hline
$\Delta A$ & $-0.07$& $-0.04$& $-0.05$& $-0.09$& $-0.16\pm0.04$ \\ \hline
\end{tabular}
\caption{\label{Scentab1} The $B\to\pi K$ observables for the 
 three scenarios introduced in the text. }
\end{center}
\end{table}

\begin{table}
\begin{center}
\begin{tabular}{|c||c|c|c|c|c|}
\hline
  Decay & \quad SM \quad &   Scen A &   Scen B &   Scen C &
  \parbox{2.3cm}{\rule{0em}{1em}Exp. bound \\(90\% {\rm C.L.})}
 \\ \hline
$\mbox{BR}(K^+ \to \pi^+ \nu \bar\nu)/10^{-11}$  &   
 $ 9.3$ & $2.7 $ &  $8.3 $ & $8.4 $ &  $(14.7^{+13.0}_{-8.9}) $\rule{0em}{1.05em} \\ \hline
$\mbox{BR}(K_{\rm L} \to \pi^0 \nu \bar \nu)/10^{-11}$  &
 $ 4.4$ &  $ 11.6$ &  $27.9$ & $7.2$ &  $ < 2.9 \times10^{4} $ \\ \hline
$\mbox{BR}(K_{\rm L} \to \pi^0  e^+ e^-)/10^{-11}$ &  
 $ 3.6$ & $4.6$  &    $7.1$ &  $4.9 $&   $<28$ \\ \hline
$\mbox{BR}(B \to X_s \nu \bar\nu)/10^{-5}$  &  
 $3.6$  &  $ 2.8 $&   $4.8$ &  $3.3 $ &  $<64$ \\ \hline
$\mbox{BR}(B_s \to \mu^+ \mu^-)/10^{-9}$  & 
 $3.9$ & $9.2$ & $ 9.1$ &  $7.0 $&  $<1.5\times 10^{2}$ \rule{0em}{1.05em}\\ \hline 
$\mbox{BR}(K_{\rm L} \to \mu^+ \mu^-)_{\rm SD}/10^{-9}$ &  
 $ 0.9$ & $0.9$  &    $0.001$ &  $0.6 $&   $<2.5$ \\ 
\hline
\end{tabular}
\caption{\label{Scentab2} Rare decay branching ratios for the three scenarios 
introduced in the text. We will have a closer look at the $B_s\to\mu^+\mu^-$ 
channel in Subsection~\ref{ssec:Bmumu}.}
\end{center}
\end{table}

Using an only slightly more generous bound on $|Y|$ by imposing 
$\left|Y \right| \leq 1.5$ and taking only those values of (\ref{q-phi}) 
that satisfy the constraint $\left|Y \right|=1.5$ yields
\begin{equation}
\label{q-phi-RD}
q= 0.48 \pm 0.07 ,\quad \phi=-(93 \pm 17 )^\circ,
\end{equation}
corresponding to a modest {\it suppression} of $q$ relative to its
updated SM value of $0.58$. It is interesting to investigate the impact
of various modifications of $(q,\phi)$, which allow us to satisfy the bounds 
in (\ref{XY2}), for the $B\to\pi K$ observables and rare decays. To this
end, three scenarios for the possible future evolution of the measurements
of $R_{\rm n}$ and $R_{\rm c}$ were introduced in \cite{BFRS-5}:
\begin{itemize}
\item {\it Scenario A:} $q=0.48$, $\phi = -93^{\circ}$, which is in accordance with 
the currrent rare decay bounds and the $B \to \pi K$ data (see (\ref{q-phi-RD})).
\item {\it Scenario B:} $q=0.66$, $\phi=-50^{\circ}$, which yields an increase
of $R_{\rm n}$ to 1.03, and some interesting effects in rare decays. This could,
for example, happen if radiative corrections to the  $B_d^0\to\pi^- K^+$ branching 
ratio enhance $R_{\rm n}$ \cite{Baracchini:2005wp}, though this alone would 
probably account for only about $5\%$.
\item {\it Scenario C:} here it is assumed that $R_{\rm n}=R_{\rm c}=1$, which
corresponds to $q=0.54$ and $\phi=61^{\circ}$. The {\it positive} sign of 
$\phi$ distinguishes this scenario strongly from the others.
\end{itemize}
The patterns of the observables of the $B\to\pi K$ and rare decays corresponding
to these scenarios are collected in Tables \ref{Scentab1} and \ref{Scentab2},
respectively. We observe that the $K \to \pi \nu \bar \nu$ modes, which are
theoretically very clean (for a recent review, see Ref.~\cite{BSU}), offer a particularly
interesting probe for the different scenarios. Concerning the observables of the 
$B \to \pi K$ system, ${\cal A}_{\rm CP}^{\rm mix}(B_d\!\to\!\pi^0 K_{\rm S})$ 
is very interesting: this CP asymmetry is found to be very large in Scenarios A and B, 
where the NP phase $\phi$ is negative. On the other hand, the positive sign of 
$\phi$ in Scenario C brings ${\cal A}_{\rm CP}^{\rm mix}(B_d\!\to\!\pi^0 K_{\rm S})$ 
closer to the data, in agreement with the features discussed in
Subsection~\ref{ssec:BpiK}. A similar comment applies to the
direct CP asymmetry of $B^\pm\to\pi^0K^\pm$. 

In view of the large uncertainties, unfortunately no definite conclusions on the
presence of NP can be drawn at this stage. However, the possible anomalies
in the $B\to\pi K$ system complemented with the one in $B\to\phi K$ may actually
indicate the effects of a modified EW penguin sector with a large CP-violating
NP phase. As we just saw, rare $K$ and $B$ decays have an impressive power
to reveal such a kind of NP. Let us finally stress that the analysis of the $B\to\pi\pi$
modes, which signals large non-factorizable effects, and the determination of the 
UT angle $\gamma$ described above are not affected by such NP effects. It will 
be interesting to monitor the evolution of the corresponding data with the help
of the strategy discussed above.

\boldmath
\section{A New Territory: $b\to d$ Penguins}\label{sec:bd-pengs}
\unboldmath
\subsection{Preliminaries}
Another hot topic which emerged recently is the exploration of 
$b\to d$ penguin processes. The non-leptonic decays belonging
to this category, which are mediated by $b\to d \bar s s$ quark transitions 
(see the classification in Subsection~\ref{ssec:non-lept}), are now coming 
within experimental reach at the $B$ factories. A similar comment applies 
to the radiative decays originating from $b\to d\gamma$ processes, whereas
$b\to d\ell^+\ell^-$ modes are still far from being accessible. The $B$ factories
are therefore just entering a new territory, which is still essentially unexplored.
Let us now have a closer look at the corresponding processes.

\subsection{A Prominent Example: $B^0_d\to K^0\bar K^0$}
The Feynman diagrams contributing to this decay can straightforwardly
be obtained from those for $B^0_d\to\phi K^0$ shown in 
Fig.~\ref{fig:BphiK-diag} by replacing the anti-strange quark emerging from the 
$W$ boson through an anti-down quark. The $B^0_d\to K^0\bar K^0$ 
decay is described by the low-energy effective Hamiltonian in (\ref{e4}) with $r=d$, 
where the current--current operators may only contribute  through penguin-like 
contractions, corresponding to the penguin topologies with internal up- and
charm-quark exchanges. The dominant r\^ole is played by QCD penguins; 
since EW penguins contribute only in colour-suppressed form, they have a minor 
impact on $B^0_d\to K^0\bar K^0$, in contrast to the case of $B^0_d\to\phi K^0$,
where they may also contribute in colour-allowed form. 

If apply the notation 
introduced in Section~\ref{sec:bench}, make again use of the
unitarity of the CKM matrix and apply the Wolfenstein parametrization, 
we may write the $B^0_d\to K^0\bar K^0$ amplitude as follows:
\begin{equation}\label{ampl-BdKK-lamt}
A(B^0_d\to K^0\bar K^0)=\lambda^3A(\tilde A_{\rm P}^t-\tilde A_{\rm P}^c)
\left[1-\rho_{K\!K} e^{i\theta_{K\!K}}e^{i\gamma}\right],
\end{equation}
where 
\begin{equation}\label{rho-KK-def}
\rho_{K\!K} e^{i\theta_{K\!K}}\equiv R_b
\left[\frac{\tilde A_{\rm P}^t-\tilde A_{\rm P}^u}{\tilde A_{\rm P}^t-\tilde A_{\rm P}^c}\right].
\end{equation}
This expression allows us to calculate the CP-violating asymmetries with 
the help of the formulae given in Subsection~\ref{ssec:CP-strat},
taking the following form:
\begin{eqnarray}
{\cal A}_{\rm CP}^{\rm dir}(B_d\to K^0\bar K^0)&=&
D_1(\rho_{K\!K},\theta_{K\!K};\gamma) \label{CP-BKK-dir-gen}\\
{\cal A}_{\rm CP}^{\rm mix}(B_d\to K^0\bar K^0)&=&
D_2(\rho_{K\!K},\theta_{K\!K};\gamma,\phi_d).\label{CP-BKK-mix-gen}
\end{eqnarray}

Let us assume, for a moment, that the penguin contributions are dominated 
by top-quark exchanges. In this case, (\ref{rho-KK-def}) simplifies as
\begin{equation}
\rho_{K\!K} e^{i\theta_{K\!K}} \to R_b.
\end{equation}
Since the CP-conserving strong phase $\theta_{K\!K}$ vanishes in this limit,
the direct CP violation in $B^0_d\to K^0\bar K^0$ vanishes, too. Moreover, 
if we take into account that $\phi_d=2\beta$ in the SM and use trigonometrical
relations which can be derived for the UT, we find that also the mixing-induced
CP asymmetry would be zero. These features suggest an interesting test
of the $b\to d$ flavour sector of the SM (see, for instance, \cite{quinn}). 
However, contributions from penguins with internal up- and charm-quark 
exchanges are expected to yield sizeable CP asymmetries in 
$B_d^0\to K^0\bar K^0$ even within the SM, so that the interpretation of these 
effects is much more complicated \cite{RF-BdKK}; these contributions 
contain also possible long-distance rescattering effects \cite{BFM},
which are often referred to as ``GIM" and ``charming" penguins and received
recently a lot of attention \cite{charming}.

\begin{figure}
\vspace*{0.3truecm}
\begin{center}
\includegraphics[width=10.0cm]{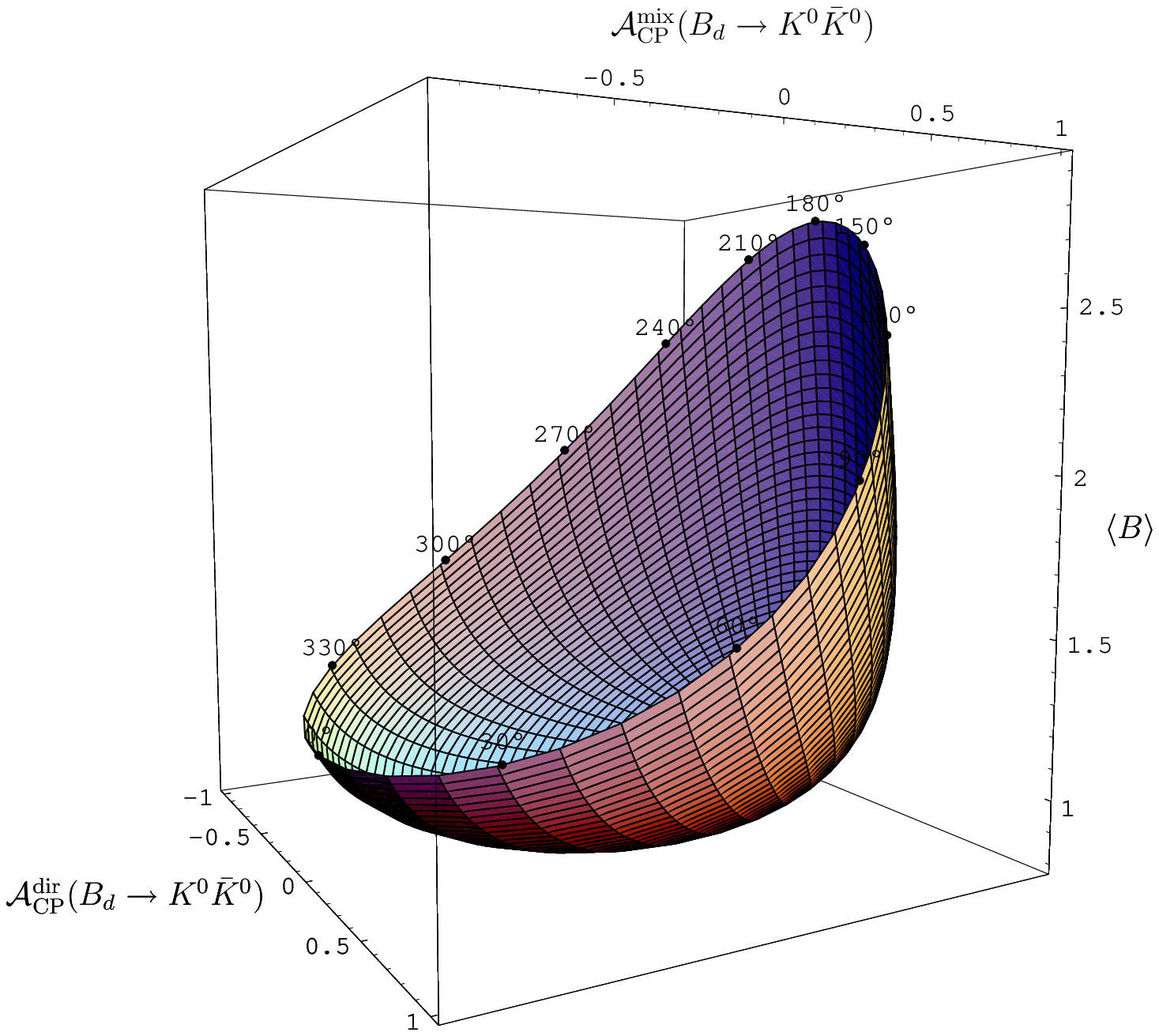}
\end{center}
\vspace*{-0.6truecm}
\caption{Illustration of the surface in the 
${\cal A}_{\rm CP}^{\rm dir}$--${\cal A}_{\rm CP}^{\rm mix}$--$\langle B \rangle$
observable space characterizing the $B^0_d\to K^0\bar K^0$ decay in the SM. 
The intersecting lines on the surface correspond to constant 
values of $\rho_{K\!K}$ and $\theta_{K\!K}$; the numbers on the fringe indicate 
the value of $\theta_{K\!K}$, while the fringe itself is defined by 
$\rho_{K\!K}=1$.}\label{fig:SM-surface}
\end{figure}

Despite this problem, interesting insights can be obtained through the
$B^0_d\to K^0\bar K^0$ observables \cite{FR1}.
By the time the CP-violating asymmetries in (\ref{CP-BKK-dir-gen}) and 
(\ref{CP-BKK-mix-gen}) can be measured, also the angle $\gamma$ of
the UT will be reliably known, in addition to the $B^0_d$--$\bar B^0_d$
mixing phase $\phi_d$. The experimental values of the CP asymmetries
can then be converted into $\rho_{K\!K}$ and $\theta_{K\!K}$, in analogy
to the $B\to\pi\pi$ discussion in Subsection~\ref{ssec:Bpipi-hadr}. Although 
these quantities are interesting to obtain insights into the $B\to\pi K$
parameter $\rho_{\rm c}e^{i\theta_{\rm c}}$ (see (\ref{B+pi+K0}))
through $SU(3)$ arguments, and can be compared  with theoretical predictions,  
for instance, those of QCDF, PQCD or SCET, they do not 
provide -- by themselves -- a test of the SM description of the 
FCNC processes mediating the decay $B^0_d\to K^0\bar K^0$. However, so far, 
we have not yet used the information offered by the CP-averaged branching 
ratio of this channel. It takes the following form:
\begin{equation}\label{BR-BKK-expr}
\mbox{BR}(B_d\to K^0\bar K^0)=\frac{\tau_{B_d}}{16\pi M_{B_d}}
\times \Phi_{KK} \times
|\lambda^3 A \, \tilde A_{\rm P}^{tc}|^2 \langle B \rangle,
\end{equation}
where $\Phi_{KK}$ denotes a two-body phase-space factor, 
$\tilde A_{\rm P}^{tc}\equiv \tilde A_{\rm P}^t-\tilde A_{\rm P}^c$, and
\begin{equation}\label{B-DEF}
\langle B \rangle\equiv 1-2\rho_{K\!K}\cos\theta_{K\!K}
\cos\gamma+\rho_{K\!K}^2.
\end{equation}
If we now use $\phi_d$ and the SM value of $\gamma$, we may characterize
the decay $B^0_d\to K^0\bar K^0$ -- within the SM -- through a surface in 
the observable space of ${\cal A}_{\rm CP}^{\rm dir}$, 
${\cal A}_{\rm CP}^{\rm mix}$ and $\langle B \rangle$. In 
Fig.~\ref{fig:SM-surface}, we show this surface, where each point 
corresponds to a given value of $\rho_{K\!K}$ and $\theta_{K\!K}$. It should 
be emphasized that this surface is {\it theoretically clean} since it 
relies only on the general SM parametrization of $B^0_d\to K^0\bar K^0$. 
Consequently, should future measurements give a value in observable space 
that should {\it not} lie on the SM surface, we would have immediate evidence 
for NP contributions to $\bar b\to \bar d s \bar s$ processes. 

Looking at Fig.~\ref{fig:SM-surface}, we see that $\langle B \rangle$ takes
an absolute minimum. Indeed, if we keep $\rho_{K\!K}$ and $\theta_{K\!K}$
as free parameters in (\ref{B-DEF}), we find
\begin{equation}\label{B-bound}
\langle B \rangle\geq \sin^2\gamma,
\end{equation}
which yields a strong lower bound because of the favourably large value of
$\gamma$. Whereas the direct and mixing-induced CP asymmetries can
be extracted from a time-dependent rate asymmetry (see (\ref{time-dep-CP})), 
the determination of $\langle B \rangle$ requires further information to
fix the overall normalization factor involving the penguin amplitude
$\tilde A_{\rm P}^{tc}$. The strategy developed in Ref.~\cite{BFRS} offers the 
following two avenues, using data for
\begin{itemize}
\item[i)] $B\to\pi\pi$ decays, i.e.\ $b\to d$ transitions, implying the following
lower bound:
\begin{equation}\label{BdKK-bound1}
\mbox{BR}(B_d\to K^0\bar K^0)_{\rm min}= 
\Xi^K_\pi\times\left(1.39\,^{+1.54}_{-0.95}\right) \times 10^{-6},
\end{equation}
\item[ii)] $B\to\pi K$ decays, i.e.\ $b\to s$ transitions, which are complemented 
by the $B\to\pi\pi$ system to determine a small correction, implying the following
lower bound:
\begin{equation}\label{BdKK-bound2}
\mbox{BR}(B_d\to K^0\bar K^0)_{\rm min}= 
\Xi^K_\pi\times\left(1.36\,^{+0.18}_{-0.21}\right) \times 10^{-6}.
\end{equation}
\end{itemize}
Here factorizable $SU(3)$-breaking corrections are included, 
as is made explicit through
\begin{equation}\label{Xi-K-pi}
\Xi^K_\pi=\left[\frac{f_0^K}{0.331}\frac{0.258}{f_0^\pi}\right]^2,
\end{equation}
where the numerical values for the $B\to K,\pi$ form factors $f_0^{K,\pi}$ 
refer to a recent light-cone sum-rule analysis \cite{Ball}. At the time of the 
derivation of these bounds, the $B$ factories reported an experimental {\it upper} 
bound of $\mbox{BR}(B_d\to K^0\bar K^0)<1.5\times 10^{-6}$ (90\% C.L.). Consequently, the theoretical {\it lower} bounds given above suggested that
the observation of this channel should just be ahead of us. Subsequently, the
first signals were indeed announced, in accordance with (\ref{BdKK-bound1}) and 
(\ref{BdKK-bound2}):
\begin{equation}\label{BdK0K0-data}
\hspace*{-0.5truecm}
\mbox{BR}(B_d\to K^0\bar K^0)=\left\{
\begin{array}{ll}
(1.19^{+0.40}_{-0.35}\pm0.13) \times 10^{-6} & \mbox{(BaBar 
\cite{BaBar-BKK}),}\\
 (0.8\pm0.3\pm0.1) \times 10^{-6} & \mbox{(Belle \cite{Belle-BKK}).}
 \end{array}\right.
\end{equation}
The SM description of $B^0_d\to K^0\bar K^0$ has thus successfully passed its 
first test. However, the experimental errors are still very large, and the next crucial 
step -- a measurement of the CP asymmetries -- is still missing. Using QCDF, 
an analysis of NP effects in this channel was recently performed in the minimal 
supersymmetric standard model \cite{giri-moh}. For further aspects of 
$B^0_d\to K^0\bar K^0$, the reader is referred to Ref.~\cite{FR1}.

\subsection{Radiative $b\to d$ Penguin Decays: $\bar B\to\rho\gamma$}
Another important tool to explore $b\to d$ penguins is
provided by $\bar B\to\rho\gamma$ modes.  In the SM, these decays 
are described by a Hamiltonian with the following 
structure \cite{B-LH98}:
\begin{equation}\label{Ham-bdgam}
{\cal H}_{\rm eff}^{b\to d\gamma}=\frac{G_{\rm F}}{\sqrt{2}}
\sum_{j=u,c} \! V_{jd}^\ast V_{jb}\left[\sum_{k=1}^{2}C_k Q_k^{jd}\!+\!
\sum_{k=3}^{8}C_k Q_k^{d}\right].
\end{equation}
Here the $Q_{1,2}^{jd}$ denote the current--current operators, whereas the 
$Q_{3\ldots 6}^{d}$ are the QCD penguin operators, which govern the
decay $\bar B^0_d\to  K^0\bar K^0$ together with the
penguin-like contractions of $Q_{1,2}^{cd}$ and $Q_{1,2}^{ud}$. In contrast 
to these four-quark operators,
\begin{equation}
Q_{7,8}^{d}=\frac{1}{8\pi^2}m_b\bar d_i \sigma^{\mu\nu}(1+\gamma_5)
\left\{e b_i F_{\mu\nu} ,\, g_{\rm s}T^a_{ij}b_j G^a_{\mu\nu} \right\}
\end{equation}
are electro- and chromomagnetic penguin operators. 
The most important contributions to $\bar B\to\rho\gamma$
originate from $Q_{1,2}^{jd}$ and $Q_{7,8}^{d}$, 
whereas the QCD penguin operators play only a minor r\^ole, in contrast 
to $\bar B^0_d\to K^0\bar K^0$. If we use again the
unitarity of the CKM matrix and apply the Wolfenstein parametrization,
we may write
\begin{equation}\label{Ampl-Brhogam}
A(\bar B \to \rho\gamma)=c_\rho \lambda^3 A {\cal P}_{tc}^{\rho\gamma}
\left[1-\rho_{\rho\gamma}e^{i\theta_{\rho\gamma}}e^{-i\gamma}\right],
\end{equation}
where $c_\rho=1/\sqrt{2}$ and 1 for $\rho=\rho^0$ and $\rho^\pm$,
respectively, ${\cal P}_{tc}^{\rho\gamma}\equiv
{\cal P}_t^{\rho\gamma}-{\cal P}_c^{\rho\gamma}$, and
\begin{equation}
\rho_{\rho\gamma}e^{i\theta_{\rho\gamma}}\equiv R_b\left[
\frac{{\cal P}_t^{\rho\gamma}-
{\cal P}_u^{\rho\gamma}}{{\cal P}_t^{\rho\gamma}-
{\cal P}_c^{\rho\gamma}}\right].
\end{equation}
Here we follow our previous notation, i.e.\ the ${\cal P}_j^{\rho\gamma}$ 
are strong amplitudes with the following interpretation: 
${\cal P}_u^{\rho\gamma}$ and ${\cal P}_c^{\rho\gamma}$ refer to the matrix 
elements of $\sum_{k=1}^{2}C_k Q_k^{ud}$ and $\sum_{k=1}^{2}C_k Q_k^{cd}$, 
respectively, whereas ${\cal P}_t^{\rho\gamma}$ corresponds to 
$-\sum_{k=3}^{8}C_k Q_k^{d}$. Consequently, ${\cal P}_u^{\rho\gamma}$
and ${\cal P}_c^{\rho\gamma}$ describe the penguin topologies with
internal up- and charm-quark exchanges, respectively, whereas 
${\cal P}_t^{\rho\gamma}$ corresponds to the penguins with the top
quark running in the loop. Let us note that 
(\ref{Ampl-Brhogam}) refers to a given photon helicity. However, 
the $b$ quarks couple predominantly to left-handed photons in 
the SM, so that the right-handed amplitude is usually neglected \cite{GP}; 
we shall return to this point below. Comparing (\ref{Ampl-Brhogam}) with 
(\ref{ampl-BdKK-lamt}), we observe that the structure of both amplitudes is 
the same. In analogy to $\rho_{K\!K} e^{i\theta_{K\!K}}$, 
$\rho_{\rho\gamma}e^{i\theta_{\rho\gamma}}$ may also be affected by 
long-distance effects, which represent a key uncertainty of 
$\bar B\to\rho\gamma$ decays \cite{LHC-Book,GP}. 

If we replace all down quarks in (\ref{Ham-bdgam}) by strange quarks, we obtain the 
Hamiltonian for $b\to s\gamma$ processes, which are already well established 
experimentally \cite{HFAG}:
\begin{eqnarray}
\mbox{BR}(B^\pm\to K^{\ast\pm}\gamma)&=&(40.3\pm2.6)\times 
10^{-6}\label{BR-charged}\\
\mbox{BR}(B_d^0\to K^{\ast0}\gamma)&=&(40.1\pm2.0)\times 
10^{-6}.\label{BR-neutral}
\end{eqnarray}
In analogy to (\ref{Ampl-Brhogam}), we may write
\begin{equation}\label{Ampl-BKastgam}
A(\bar B \!\to\! K^\ast \!\gamma)\!=-\!
\frac{\lambda^3 \! A {\cal P}_{tc}^{K^\ast\!\gamma}}{\sqrt{\epsilon}} \!
\left[1\!+\!\epsilon\rho_{K\!^\ast\!\gamma}e^{i\theta_{K\!^\ast\!\gamma}}
e^{-i\!\gamma}\right]\!,
\end{equation}
where $\epsilon$ was introduced in (\ref{eps-def}). Thanks to the smallness
of $\epsilon$, the parameter 
$\rho_{K\!^\ast\gamma}e^{i\theta_{K\!^\ast\gamma}}$ 
plays an essentially negligible r\^ole for the $\bar B \to K^\ast \gamma$ 
transitions.

Let us have a look at the charged decays $B^\pm \to \rho^{\pm} \gamma$ 
and $B^\pm \to K^{\ast\pm} \gamma$ first. If we consider their 
CP-averaged branching ratios, we obtain
\begin{equation}\label{rare-ratio}
\frac{\mbox{BR}(B^\pm \to \rho^{\pm} 
\gamma)}{\mbox{BR}(B^\pm \to K^{\ast\pm} \gamma)}=\epsilon
\left[\frac{\Phi_{\rho\gamma}}{\Phi_{K\!^\ast\gamma}}\right]
\left|\frac{{\cal P}_{tc}^{\rho\gamma}}{{\cal P}_{tc}^{K\!^\ast\gamma}}
\right|^2 H^{\rho\gamma}_{K\!^\ast\gamma},
\end{equation}
where $\Phi_{\rho\gamma}$ and $\Phi_{K\!^\ast\gamma}$ denote phase-space 
factors, and 
\begin{equation}
H^{\rho\gamma}_{K\!^\ast\gamma}\equiv
\frac{1-2\rho_{\rho\gamma}\cos\theta_{\rho\gamma}\cos\gamma+
\rho_{\rho\gamma}^2}{1+2\epsilon\rho_{K\!^\ast\gamma}
\cos\theta_{K\!^\ast\gamma}
\cos\gamma+\epsilon^2\rho_{K\!^\ast\gamma}^2}.
\end{equation}
Since $B^\pm \to \rho^{\pm} \gamma$ and $B^\pm \to K^{\ast\pm} \gamma$ 
are related through the interchange of all down and strange quarks, 
the $U$-spin flavour symmetry of strong interactions allows us to relate 
the corresponding hadronic amplitudes to each other; the $U$-spin
symmetry is an $SU(2)$ subgroup of the full $SU(3)_{\rm F}$ flavour-symmetry
group, which relates down and strange quarks in the same manner as the 
conventional strong isospin symmetry relates down and up quarks. Following 
these lines, we obtain
\begin{equation}\label{U-spin1}
|{\cal P}_{tc}^{\rho\gamma}|=|{\cal P}_{tc}^{K\!^\ast\gamma}|
\end{equation}
\begin{equation}\label{U-spin2}
\rho_{\rho\gamma}e^{i\theta_{\rho\gamma}}=
\rho_{K\!^\ast\gamma}e^{i\theta_{K\!^\ast\gamma}}\equiv
\rho e^{i\theta}.
\end{equation}
Although we may determine the ratio of the penguin amplitudes 
$|{\cal P}_{tc}|$ in (\ref{rare-ratio}) with the help of  (\ref{U-spin1}) -- up to 
$SU(3)$-breaking effects to be discussed below -- we are still left
with the dependence on $\rho$ and $\theta$. However, keeping $\rho$ 
and $\theta$ as free parameters, it can be shown that
$H^{\rho\gamma}_{K\!^\ast\gamma}$ satisfies the following relation \cite{FR2}:
\begin{equation}\label{H-bound}
H^{\rho\gamma}_{K\!^\ast\gamma}\geq \left[1-2\epsilon
\cos^2\gamma+{\cal O}(\epsilon^2)\right]\sin^2\gamma,
\end{equation}
where the term linear in $\epsilon$ gives a shift of about $1.9\%$. 

Concerning possible $SU(3)$-breaking effects to (\ref{U-spin2}), they 
may only enter this tiny correction and are negligible for our analysis. 
On the other hand, the $SU(3)$-breaking corrections to (\ref{U-spin1}) 
have a sizeable impact. Following \cite{ALP-rare,BoBu}, we write
\begin{equation}\label{SU3-break-rare}
\left[\frac{\Phi_{\rho\gamma}}{\Phi_{K\!^\ast\gamma}}\right]
\left|\frac{{\cal P}_{tc}^{\rho\gamma}}{{\cal P}_{tc}^{K\!^\ast\gamma}}
\right|^2=\left[\frac{M_B^2-M_\rho^2}{M_B^2-M_{K^\ast}^2}\right]^3
\zeta^2,
\end{equation}
where $\zeta=F_\rho/F_{K^\ast}$ is the $SU(3)$-breaking ratio of the
$B^\pm\to\rho^\pm\gamma$ and $B^\pm\to K^{\ast\pm}\gamma$ form factors; a 
light-cone sum-rule analysis gives $\zeta^{-1}=1.31\pm0.13$ \cite{Ball-Braun}.
Consequently, (\ref{H-bound}) and (\ref{SU3-break-rare}) allow us to convert 
the measured $B^\pm\to K^{\ast\pm}\gamma$ branching ratio (\ref{BR-charged}) 
into a {\it lower} SM bound for 
$\mbox{BR}(B^\pm\to\rho^\pm\gamma)$ with the help of (\ref{rare-ratio}) \cite{FR2}:
\begin{equation}\label{Brhogam-char}
\mbox{BR}(B^\pm\to \rho^\pm\gamma)_{\rm min}=\left(1.02\,^{+0.27}_{-0.23}
\right)\times10^{-6}.
\end{equation}

A similar kind of reasoning holds also for the $U$-spin 
pairs $B^\pm\to K^\pm K, \pi^\pm K$ and $B^\pm\to K^\pm K^\ast, \pi^\pm K^\ast$,
where the following lower bounds can be derived \cite{FR2}:
\begin{eqnarray}
\mbox{BR}(B^\pm\!\to\! K^\pm K)_{\rm min} \!\!&=&\!\! \Xi^K_\pi\!\times\!
\left(1.69\,^{+0.21}_{-0.24}\right)\!\times\! 10^{-6}\label{BKK-char}\\
\mbox{BR}(B^\pm\!\to\! K^\pm K^\ast)_{\rm min} \!\!&=&\!\! \Xi^K_\pi\!\times\!
\left(0.68\,^{+0.11}_{-0.13}
\right)\!\times \! 10^{-6},\label{BpiKast}
\end{eqnarray}
with $\Xi^K_\pi$ given in (\ref{Xi-K-pi}). Thanks to the most recent
$B$-factory data, we have now also evidence for $B^\pm\to K^\pm K$
decays:
\begin{equation}
\mbox{BR}(B^\pm\!\to\! K^\pm K)=
\left\{\begin{array}{ll}
(1.5\pm0.5\pm0.1)\times 10^{-6} & \mbox{(BaBar \cite{BaBar-BKK})}\\
(1.0\pm0.4\pm0.1)\times 10^{-6} & \mbox{(Belle \cite{Belle-BKK}),}
\end{array} \right.
\end{equation}
whereas the upper limit of $5.3\times 10^{-6}$ for $B^\pm\to K^\pm K^\ast$
still leaves a lot of space. Obviously, we may also consider the
$B^\pm\to K^{\ast\pm} K, \rho^\pm K$ system \cite{FR2}. However,
since currently only the upper bound 
$\mbox{BR}(B^\pm\to \rho^\pm K)<48\times 10^{-6}$ is available, 
we cannot yet give a number for the lower bound on 
$\mbox{BR}(B^\pm\to K^{\ast\pm} K)$. Experimental analyses of
these modes are strongly encouraged.

Let us now turn to $\bar B^0_d\to\rho^0\gamma$, which receives 
contributions from exchange and penguin annihilation topologies that are 
not present in 
$\bar B^0_d\to \bar K^{\ast0}\gamma$; in the case of $B^\pm\to\rho^\pm\gamma$ 
and $B^\pm\to K^{\ast\pm}\gamma$, which are related by the $U$-spin symmetry, 
there is a one-to-one correspondence of topologies. Making the 
plausible assumption that the topologies involving the spectator quarks play 
a minor r\^ole, and taking the factor of $c_{\rho^0}=1/\sqrt{2}$ in 
(\ref{Ampl-Brhogam}) into account, the counterpart of (\ref{Brhogam-char}) 
is given by 
\begin{equation}\label{Brhogam-neut}
\mbox{BR}(B_d\to \rho^0\gamma)_{\rm min}=\left(0.51\,^{+0.13}_{-0.11}
\right)\times10^{-6}.
\end{equation}

At the time of the derivation of the {\it lower} bounds for the
$B\to\rho\gamma$ branching ratios given above, the following experimental {\it upper}
bounds ($90\%$ C.L.) were available:
\begin{equation}\label{Brho-gam-char-EXP}
\mbox{BR}(B^\pm\to \rho^\pm\gamma)<\left\{\begin{array}{ll}
1.8\times 10^{-6} & \mbox{(BaBar \cite{Babar-Brhogamma-bound})}\\
2.2\times 10^{-6} & \mbox{(Belle \cite{Belle-Brhogamma-bound})}\\
\end{array} \right.
\end{equation}
\begin{equation}\label{Brho-gam-neut-EXP}
\mbox{BR}(B_d\to \rho^0\gamma)<\left\{\begin{array}{ll}
0.4\times 10^{-6} & \mbox{(BaBar \cite{Babar-Brhogamma-bound})}\\
0.8\times 10^{-6} & \mbox{(Belle \cite{Belle-Brhogamma-bound}).}
\end{array} \right.
\end{equation}
Consequently, it was expected that the $\bar B\to\rho\gamma$ modes should 
soon be discovered at the $B$ factories \cite{FR2}. Indeed, the Belle 
collaboration reported recently the first observation of $b\to d\gamma$ processes
\cite{Belle-bdgam-obs}:
\begin{eqnarray}
\mbox{BR}(B^\pm\to \rho^\pm\gamma)&=&\left(0.55^{+0.43+0.12}_{-0.37-0.11}\right)
\times 10^{-6}
\label{Belle-Brhogam-p}\\
\mbox{BR}(B_d\to \rho^0\gamma)&=&\left(1.17^{+0.35+0.09}_{-0.31-0.08}\right)
\times 10^{-6}
\label{Belle-Brhogam-n}\\
\mbox{BR}(B\to(\rho,\omega)\gamma)&=&
\left(1.34^{+0.34+0.14}_{-0.31-0.10}\right)\times 10^{-6},
\end{eqnarray}
which was one of the hot topics of the 2005 summer conferences \cite{Belle-press}.
These measurements still suffer from large uncertainties, and the pattern of the 
central values of (\ref{Belle-Brhogam-p}) and (\ref{Belle-Brhogam-n}) would be in
conflict with the expectation following from the isospin symmetry. It will be interesting
to follow the evolution of the data. The next important conceptual step would be the measurement of the corresponding CP-violating observables, though this is still
in the distant future.

An alternative avenue to confront the data for the $B\to \rho\gamma$
branching ratios with the SM is provided by converting them into information
on the side $R_t$ of the UT. To this end, the authors of Refs.~\cite{ALP-rare,BoBu}
use also (\ref{SU3-break-rare}), and calculate the CP-conserving (complex) 
parameter $\delta a$ entering 
$\rho_{\rho\gamma}e^{i\theta_{\rho\gamma}}=R_b\left[1+\delta a\right]$
in the QCDF approach. The corresponding result, which favours a small impact 
of $\delta a$, takes leading and next-to-leading order QCD corrections into 
account and holds to leading order in the heavy-quark limit \cite{BoBu}. 
In view of the remarks about possible long-distance effects made above and the 
$B$-factory data for the $B\to\pi\pi$ system, which indicate large corrections 
to the QCDF picture for non-leptonic $B$ decays into two light pseudoscalar 
mesons (see Subsection~\ref{ssec:Bpipi-hadr}), it is, however, not obvious that 
the impact of $\delta a$ is actually small. The advantage of the bound
following from (\ref{H-bound}) is that it is  -- by construction -- {\it not} affected 
by $\rho_{\rho\gamma}e^{i\theta_{\rho\gamma}}$ at all.

\subsection{General Lower Bounds for $b\to d$ Penguin Processes}
Interestingly, the bounds discussed above are actually 
realizations of a general, model-independent bound that can be derived
in the SM for $b\to d$ penguin processes \cite{FR2}. If we consider such
a decay, $\bar B \to \bar f_d$, we may -- in analogy to (\ref{ampl-BdKK-lamt}) 
and (\ref{Ampl-Brhogam}) -- write
\begin{equation}
A(\bar B \to \bar f_d)= A^{(0)}_d
\left[1-\rho_de^{i\theta_d}e^{-i\gamma}\right],
\end{equation}
so that the CP-averaged amplitude square is given as follows:
\begin{equation}
\langle|A(B \to f_d)|^2\rangle=|A^{(0)}_d|^2
\left[1-2\rho_d\cos\theta_d\cos\gamma+\rho_d^2\right].
\end{equation}
In general, $\rho_d$ and $\theta_d$ depend on the point in phase space
considered. Consequently, the expression
\begin{equation}
\mbox{BR}(B \to f_d)=\tau_B\left[\sum_{\rm Pol}
\int \!\! d \, {\rm PS} \, \langle|A(B \to f_d)|^2\rangle \right]
\end{equation}
for the CP-averaged branching ratio, where the sum runs over possible
polarization configurations of $f_d$, does {\it not} factorize into 
$|A^{(0)}_d|^2$ and $[1-2\rho_d\cos\theta_d\cos\gamma+\rho_d^2]$ as 
in the case of the two-body decays considered above. However, if we 
keep $\rho_d$ and $\theta_d$ as free, ``unknown'' parameters at any 
given point in phase space, we obtain
\begin{equation}
\langle|A(B \to f_d)|^2\rangle\geq|A^{(0)}_d|^2 \sin^2\gamma,
\end{equation}
which implies
\begin{equation}
\mbox{BR}(B \to f_d)\geq\tau_B\left[\sum_{\rm Pol}
\int \!\! d \, {\rm PS} \, |A^{(0)}_d|^2 \right]\sin^2\gamma.
\end{equation}

In order to deal with the term in square brackets, we use a $b\to s$ 
penguin decay $\bar B \to \bar f_s$, which is the counterpart of $\bar B \to \bar f_d$ 
in that the corresponding CP-conserving strong amplitudes can be related
to one another through the $SU(3)$ flavour symmetry. In analogy to 
(\ref{Ampl-BKastgam}), we may then write
\begin{equation}
A(\bar B \to \bar f_s)= - \frac{A^{(0)}_s}{\sqrt{\epsilon}}
\left[1+\epsilon\rho_s e^{i\theta_s}e^{-i\gamma}\right].
\end{equation}
If we neglect the term proportional to $\epsilon$ in the square bracket, 
we arrive at
\begin{equation}\label{general-bound}
\frac{\mbox{BR}(B \to f_d)}{\mbox{BR}(B \to f_s)}
\geq \epsilon \left[\frac{\sum_{\rm Pol}\int \! d \, {\rm PS} \, 
|A^{(0)}_d|^2 }{\sum_{\rm Pol}\int \! d \, {\rm PS} \, |A^{(0)}_s|^2 }
\right]\sin^2\gamma.
\end{equation}
Apart from the tiny $\epsilon$ correction, which gave a shift of about
$1.9\%$ in (\ref{H-bound}), (\ref{general-bound}) is valid
exactly in the SM. If we now apply the $SU(3)$ flavour symmetry, we obtain
\begin{equation}\label{SU3-limit}
\frac{\sum_{\rm Pol}\int \! d \, {\rm PS} \, 
|A^{(0)}_d|^2 }{\sum_{\rm Pol}\int \! d \, {\rm PS} \, |A^{(0)}_s|^2 }
\stackrel{SU(3)_{\rm F}}{\longrightarrow} 1.
\end{equation}
Since $\sin^2\gamma$ is favourably large in the SM and the decay
$\bar B \to \bar f_s$ will be measured before its $b\to d$ 
counterpart  -- simply because of the CKM enhancement -- 
(\ref{general-bound}) provides strong lower bounds for 
$\mbox{BR}(B \to f_d)$. 

It is instructive to return briefly to $B\to\rho\gamma$. If we look at 
(\ref{general-bound}), we observe immediately that the assumption that 
these modes are governed by a single photon helicity is no longer 
required. Consequently, (\ref{Brhogam-char}) and (\ref{Brhogam-neut}) 
are actually very robust with respect to this issue, which may only affect 
the $SU(3)$-breaking corrections to a small extend. This feature is interesting
in view of the recent discussion in \cite{GGLP}, where the photon polarization
in $B\to \rho\gamma$ and $B\to K^\ast \gamma$ decays was critically analyzed. 

We can now also derive a bound for the  
$B^\pm\to K^{\ast\pm}K^{\ast}, \rho^\pm K^\ast$ system, where 
we have to sum in (\ref{general-bound}) over three polarization configurations
of the vector mesons. The analysis of the $SU(3)$-breaking corrections is
more involved than in the case of the decays considered above, and the 
emerging lower bound of 
$\mbox{BR}(B^\pm\to K^{\ast\pm} K^\ast)_{\rm min}\sim0.6\times 10^{-6}$
is still very far from the experimental upper bound of $71\times 10^{-6}$.
Interestingly, the theoretical lower bound would be reduced by $\sim 0.6$ in 
the strict $SU(3)$ limit, i.e.\ would be more conservative \cite{FR2}. A similar 
comment applies to (\ref{BdKK-bound1}), (\ref{BdKK-bound2}) and  
(\ref{BKK-char}), (\ref{BpiKast}). On the other hand, the 
$B\to\rho\gamma$ bounds in (\ref{Brhogam-char}) and 
(\ref{Brhogam-neut}) would be enhanced by $\sim 1.7$ in this case.
However, here the theoretical situation is more favourable since we 
have not to rely on the factorization hypothesis to deal with the 
$SU(3)$-breaking effects as in the case of the non-leptonic decays. 

Let us finally come to another application of (\ref{general-bound}), which 
is offered by decays of the kind $\bar B\to \pi \ell^+\ell^-$ and 
$\bar B\to \rho \ell^+\ell^-$. It is
well known that the $\rho_d$ terms complicate the interpretation of
the corresponding data considerably \cite{LHC-Book}; the bound offers
SM tests that are not affected by these contributions. The 
structure of the $b\to d \ell^+\ell^-$ Hamiltonian is similar to 
(\ref{Ham-bdgam}), but involves the additional operators
\begin{equation}
Q_{9,10}=\frac{\alpha}{2\pi}(\bar\ell\ell)_{\rm V\!,\,A}
(\bar d_i b_i)_{\rm V-A}.
\end{equation}
The $b \to s \ell^+\ell^-$ modes $\bar B\to K \ell^+\ell^-$ 
and $\bar B\to K^\ast \ell^+\ell^-$ were already observed at the $B$ 
factories, with branching ratios at the $0.6\times 10^{-6}$ and 
$1.4\times 10^{-6}$ levels \cite{HFAG}, respectively, and received considerable
theoretical attention (see, e.g., \cite{BKll}). For the application
of (\ref{general-bound}), the charged decay combinations
$B^\pm\to \pi^\pm \ell^+\ell^-, K^\pm \ell^+\ell^-$ and
$B^\pm\to \rho^\pm \ell^+\ell^-, K^{\ast\pm} \ell^+\ell^-$ are suited
best since the corresponding decay pairs are related to each other 
through the $U$-spin symmetry \cite{HM}. The numbers given above
suggest
\begin{equation}\label{Bpi-ellell-bounds}
\mbox{BR}(B^\pm\to \pi^\pm \ell^+\ell^-), \quad
\mbox{BR}(B^\pm\to \rho^\pm \ell^+\ell^-)
\mathrel{\hbox{\rlap{\hbox{\lower4pt\hbox{$\sim$}}}\hbox{$>$}}}
10^{-8},
\end{equation}
thereby leaving the exploration of these $b\to d$ penguin decays for the more 
distant future. Detailed studies of the associated $SU(3)$-breaking corrections 
are engouraged. By the time the $B^\pm\to \pi^\pm \ell^+\ell^-$, 
$\rho^\pm \ell^+\ell^-$ modes will come within experimental reach, we will
hopefully have a good picture of these effects. 

It will be interesting to confront all of these bounds with experimental data.
In the case of the non-leptonic $B_d\to K^0\bar K^0$, $B^\pm\to K^\pm K$ modes
and their radiative $B\to\rho \gamma$ counterparts, they have already provided a
first successful test of the SM description of the corresponding FCNC processes,
although the uncertainties are still very large in view of the fact that 
we are just at the beginning
of the experimental exploration of these channels. A couple of other non-leptonic
decays of this kind may just be around the corner. It would be exciting if some
bounds were significantly violated through destructive interference between
SM and NP contributions. Since the different decay classes are governed by
different operators, we could actually encounter surprises!

\boldmath
\section{A Key Target of $B$-Decay Studies in the LHC Era:
$B_s$ Mesons}\label{sec:LHC}
\unboldmath
\subsection{Preliminaries}\label{ssec:Bs-prelim}
First insights into the $B_s$ system could already be obtained through the 
LEP experiments (CERN) and SLD (SLAC) \cite{LEPBOSC}. 
Since the currently operating $e^+e^-$ $B$ factories run at the $\Upsilon(4S)$ 
resonance, which decays only into $B_{u,d}$ but not into $B_s$ mesons, 
the $B_s$ system cannot be explored by the BaBar and Belle experiments. On the 
other hand, plenty of $B_s$ mesons will be produced at hadron colliders.
After important steps at the Tevatron, the physics potential of the $B_s$-meson
system can then be fully exploited at the LHC, in particular by the LHCb
experiment \cite{LHC-Book,schneider}. 

In the SM, the $B^0_s$--$\bar B^0_s$ oscillations are expected to be much faster
than their $B_d$-meson counterparts, and could so far not be observed.
Using the data of the LEP experiments, SLD and the Tevatron, only 
lower bounds on $\Delta M_s$ could be obtained. The most recent world
average reads as follows \cite{oldeman}:
\begin{equation}\label{DMs-bound}
\Delta M_s > 16.6 \, \mbox{ps}^{-1} \, \mbox{(90\% C.L.)}.
\end{equation}
The mass difference $\Delta M_s$ plays an important r\^ole in the CKM fits
discussed in Subsection~\ref{ssec:UT}. Let us now have a closer look at this 
topic. Following the discussion given in Section~\ref{sec:B}, the mass
difference of the $B_q$ mass eigenstates satisfies the following relation in the SM:
\begin{equation}\label{DMq-simple}
\Delta M_q\propto M_{B_q}\hat B_{B_q}f_{B_q}^2 |V_{tq}^\ast V_{tb}|^2,
\end{equation}
where $M_{B_q}\equiv [M_{\rm H}^{(q)}+M_{\rm L}^{(q)}]/2$, and the
factor of $\hat B_{B_q}f_{B_q}^2$ involving a ``bag'' parameter and the
$B_q$ decay constant defined in analogy to (\ref{decay-const-def})
arises from the parametrization of the
hadronic matrix element of the $(\bar b q)_{\rm V-A}(\bar bq)_{\rm V-A}$
operator of the low-energy effective Hamiltonian describing $B^0_q$--$\bar B^0_q$
mixing. Looking at (\ref{DMq-simple}), we see that knowledge of these
non-perturbative hadronic parameters, which typically comes from 
lattice \cite{CKM-book,lattice} or QCD sum-rule calculations \cite{SR-calc}, allows us
to determine $|V_{td}|$, which can then be converted into the UT side
$R_t$ with the help of (\ref{Rb-Rt-def}), as $|V_{cb}|= A \lambda^2$ can
be determined  through semi-leptonic $B$ decays \cite{CKM-book}.
On the other hand, the
Wolfenstein expansion allows us also to derive the relation
\begin{equation}\label{Rt-simple-rel}
R_t\equiv\frac{1}{\lambda}\left|\frac{V_{td}}{V_{cb}}\right|=
\frac{1}{\lambda}\left|\frac{V_{td}}{V_{ts}}\right|
\left[1+{\cal O}(\lambda^2)\right].
\end{equation}
Consequently, we may -- up to corrections entering at the $\lambda^2$ 
level -- determine $R_t$ through
\begin{equation}\label{RT2-DM}
\hspace*{-0.8truecm}\frac{\Delta M_d}{\Delta M_s}=
\left[\frac{M_{B_d}}{M_{B_s}}\right]
\left[\frac{\hat B_{B_d}}{\hat B_{B_s}}\right]
\left[\frac{f_{B_d}}{f_{B_s}}\right]^2
\left|\frac{V_{td}}{V_{ts}}\right|^2 
\, \Rightarrow \,
\left|\frac{V_{td}}{V_{ts}}\right|=\xi \sqrt{\left[\frac{M_{B_s}}{M_{B_d}}\right]
\left[\frac{\Delta M_d}{\Delta M_s}\right]},
\end{equation}
where
\begin{equation}\label{xi-SU3}
\xi\equiv\frac{\sqrt{\hat B_s}f_{B_s}}{\sqrt{\hat B_d}f_{B_d}}
\end{equation}
equals 1 in the strict $SU(3)$ limit. The evaluation of the $SU(3)$-breaking 
corrections entering $\xi$ is an 
important aspect of lattice QCD; recent studies give \cite{lattice}
\begin{equation}\label{xi-lat}
\xi=1.23\pm0.06.
\end{equation}
In comparison with the determination of $R_t$ through the absolute value of
$\Delta M_d$, the advantage of (\ref{RT2-DM}) is that the hadronic parameters
enter only through $SU(3)$-breaking corrections. Moreover, the CKM factor
$A$, the short-distance QCD corrections, and the Inami--Lim function $S_0(x_t)$  
cancel in this expression. Thanks to the latter feature, the determination 
of $R_t$ with the help of (\ref{RT2-DM}) is not only valid in the SM, but also in
the NP scenarios with MFV, in contrast to the extraction using only the information
about $\Delta M_d$ \cite{buras-MFV}. As can be
see in Fig.~\ref{fig:UTfits}, the main implication of the experimental lower
bound for $\Delta M_s$ is $\gamma
\mathrel{\hbox{\rlap{\hbox{\lower4pt\hbox{$\sim$}}}\hbox{$<$}}}
90^\circ$. 

In Subsection~\ref{ssec:Bmix} we saw that the width difference $\Delta\Gamma_d$
is negligibly small, whereas its $B_s$ counterpart is expected to be sizeable. As was
recently reviewed in Ref.~\cite{lenz}, the current theoretical status of these
quantities is given as follows:
\begin{equation}\label{DGam-numbers}
\frac{|\Delta\Gamma_d|}{\Gamma_d}=(3\pm1.2)\times 10^{-3}, \quad
\frac{|\Delta\Gamma_s|}{\Gamma_s}=0.12\pm0.05.
\end{equation}
The width difference $\Delta\Gamma_s$ may provide interesting studies of CP 
violation through ``untagged'' $B_s$ rates \cite{dun}--\cite{DFN}, which are defined as 
\begin{equation}
\langle\Gamma(B_s(t)\to f)\rangle
\equiv\Gamma(B^0_s(t)\to f)+\Gamma(\bar B^0_s(t)\to f),
\end{equation}
and are characterized by the feature that we do not distinguish between
initially, i.e.\ at time $t=0$, present $B^0_s$ or $\bar B^0_s$ mesons. 
If we consider a final state $f$ to which both a $B^0_s$ and a $\bar B^0_s$ 
may decay, and use the expressions in (\ref{rates}), we find
\begin{equation}\label{untagged-rate}
\hspace*{-0.7truecm}\langle\Gamma(B_s(t)\to f)\rangle
\propto \left[\cosh(\Delta\Gamma_st/2)-{\cal A}_{\Delta\Gamma}(B_s\to f)
\sinh(\Delta\Gamma_st/2)\right]e^{-\Gamma_s t},
\end{equation}
where ${\cal A}_{\Delta\Gamma}(B_s\to f)\propto \mbox{Re}\,\xi_f^{(s)}$ was 
introduced in (\ref{ADGam}). We observe that the rapidly oscillating 
$\Delta M_st$ terms cancel, and that we may obtain information about the 
phase structure of the observable $\xi_f^{(s)}$, thereby providing valuable
insights into CP violation. Following these lines, for instance,  the 
untagged observables offered by the angular distribution of the 
$B_s\to K^{*+}K^{*-}, K^{*0}\bar K^{*0}$ decay products allow
a determination of $\gamma$, provided $\Delta\Gamma_s$ is 
actually sizeable \cite{FD-CP}. Although $B$-decay experiments at hadron 
colliders should be able to resolve the $B^0_s$--$\bar B^0_s$ oscillations, 
untagged $B_s$-decay rates are interesting in terms of efficiency, 
acceptance and purity. Recently, the first results for $\Delta\Gamma_s$ 
were reported from the Tevatron, using the $B^0_s\to J/\psi\phi$ channel \cite{DDF}:
\begin{equation}
\frac{|\Delta\Gamma_s|}{\Gamma_s}=\left\{
\begin{array}{ll}
0.65^{+0.25}_{-0.33}\pm0.01 & \mbox{(CDF \cite{CDF-DG})}\\
0.24^{+0.28+0.03}_{-0.38-0.04} & \mbox{(D0 \cite{D0-DG})}.
\end{array}
\right.
\end{equation}
It will be interesting to follow the evolution of the data for this quantity. 

Finally, let us emphasize that the $B^0_s$--$\bar B^0_s$ mixing phase
takes a tiny value in the SM, $\phi_s=-2\delta\gamma=-2\lambda^2\eta\sim
-2^\circ$, whereas a large value of $\phi_d\sim 43^\circ$ was measured.
This feature has interesting implications for the pattern of the CP-violating
effects in certain $B_s$ decays, including the ``golden" channel
$B^0_s\to J/\psi \phi$.

\subsection{$B^0_s\to J/\psi \phi$}\label{ssec:BsPsiPhi}
As can be seen in Fig.~\ref{fig:BpsiK-diag}, the decay $B^0_s\to J/\psi \phi$
is simply related to $B^0_d\to J/\psi K_{\rm S}$ through a
replacement of the down spectator quark by a strange quark. Consequently,
the structure of the $B^0_s\to J/\psi\phi$ decay amplitude is  
completely analogous to that of (\ref{BdpsiK-ampl2}). On the other hand, the 
final state of  $B^0_s\to J/\psi\phi$ consists of two vector mesons, and is hence
an admixture of different CP eigenstates, which can, however, be disentangled 
through an angular analysis of the $B^0_s\to J/\psi [\to\ell^+\ell^-]\phi[\to K^+K^-]$
decay products \cite{DDF,DDLR}. The corresponding angular distribution 
exhibits tiny direct CP violation, and allows the extraction of
\begin{equation}\label{sinphis}
\sin\phi_s+{\cal O}(\overline{\lambda}^3)=\sin\phi_s+{\cal O}(10^{-3})
\end{equation}
through mixing-induced CP violation.
Since we have $\phi_s={\cal O}(10^{-2})$ in the SM, the 
determination of this phase from (\ref{sinphis}) is affected by
hadronic uncertainties of ${\cal O}(10\%)$, which may become an issue 
for the LHC era. These uncertainties can be controlled with
the help of flavour-symmetry arguments through the 
$B^0_d\to J/\psi \rho^0$ decay \cite{RF-ang}.

Thanks to its nice experimental signature, $B^0_s\to J/\psi\phi$ is very accessible
at hadron colliders, and can be fully exploited at the LHC.
Needless to note, the big hope is that large CP violation
will be found in this channel. Since the CP-violating effects in 
$B^0_s\to J/\psi\phi$ are tiny in the SM, such an observation 
would give us an unambiguous
signal for NP \cite{DFN,NiSi,Branco}. As the situation for NP entering 
through the decay amplitude is similar to $B\to J/\psi K$, we would get 
evidence for CP-violating NP contributions to $B^0_s$--$\bar B^0_s$ mixing, 
and could extract the corresponding sizeable value of $\phi_s$ \cite{DFN}.
Such a scenario may generically arise in the presence of NP with 
$\Lambda_{\rm NP}\sim\mbox{TeV}$ \cite{RF-Phys-Rep}, as well as 
in specific models; for examples, see Refs.~\cite{JN,BKK,Z-prime}.

\begin{figure}[t]
\centerline{
 \includegraphics[width=5.6truecm]{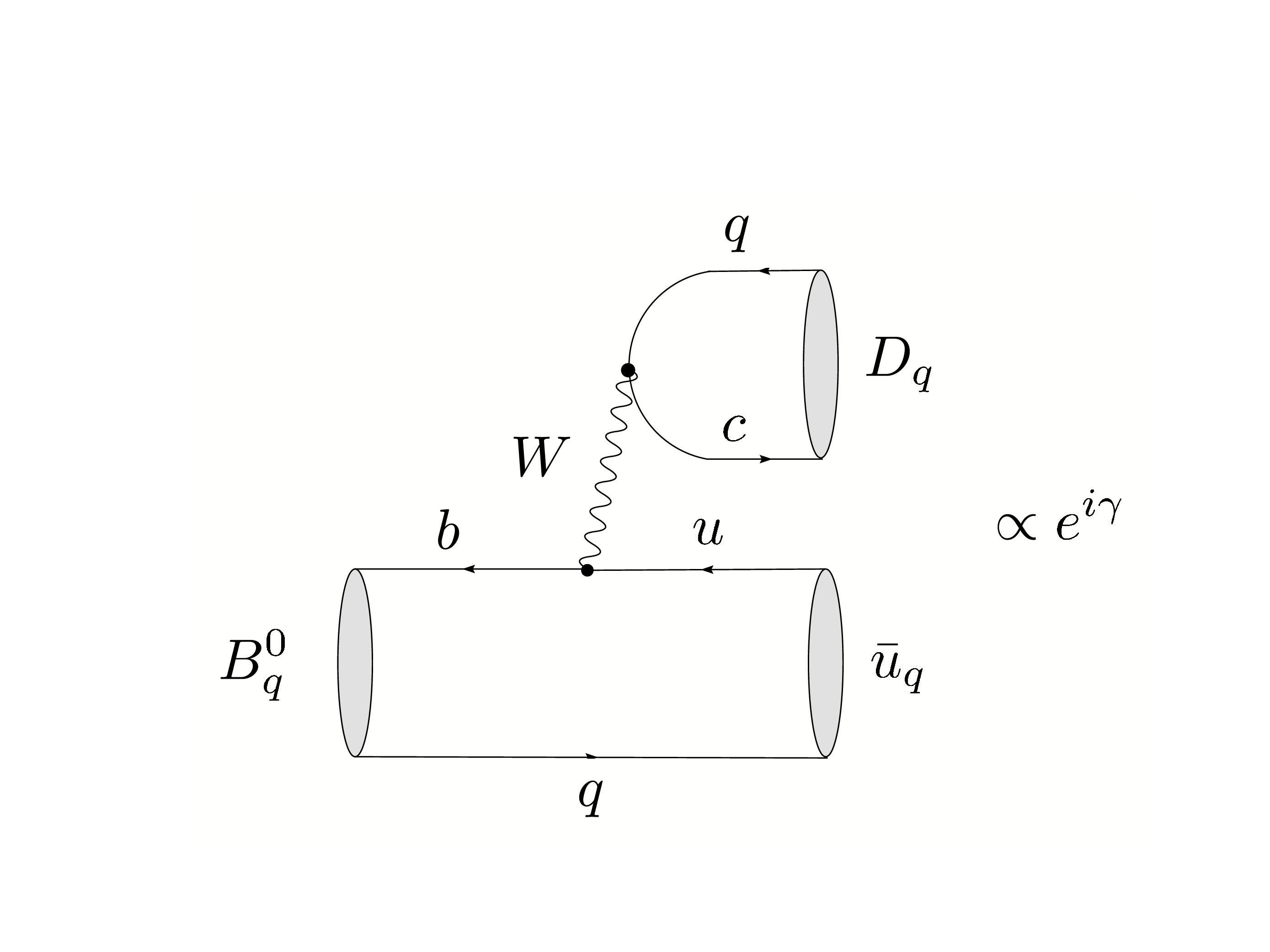}
 \hspace*{0.5truecm}
 \includegraphics[width=5.8truecm]{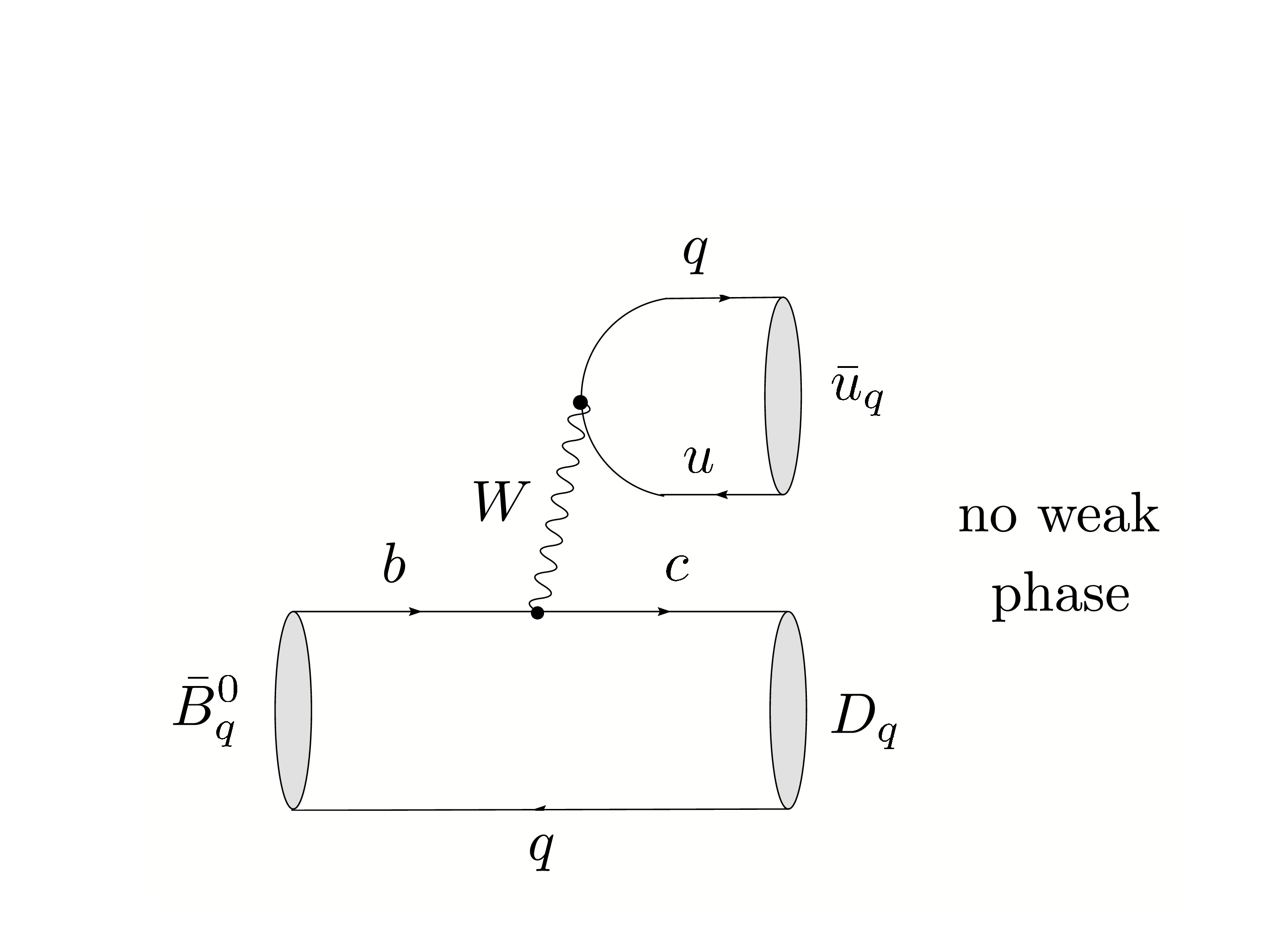}  
 }
 \vspace*{-0.3truecm}
\caption{Feynman diagrams contributing to $B^0_q\to D_q\bar u_q$
and $\bar B^0_q\to D_q \bar u_q$ 
decays.}\label{fig:BqDquq}
\end{figure}

\subsection{$B_s\to D_s^\pm K^\mp$ and $B_d\to D^\pm \pi^\mp$}\label{ssec:BsDsK}
\begin{figure}
\centerline{
 \includegraphics[width=3.2truecm]{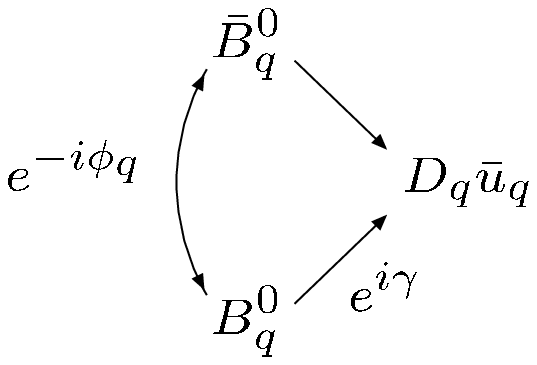} }
 \vspace*{-0.3truecm}
\caption{Interference effects between $B^0_q\to D_q\bar u_q$
and $\bar B^0_q\to D_q\bar u_q$ 
decays.}\label{fig:BqDquq-int}
\end{figure}

The decays $B_s\to D_s^\pm K^\mp$ \cite{BsDsK} and $B_d\to D^\pm \pi^\mp$
\cite{BdDpi} can be 
treated on the same theoretical basis, and provide new strategies to determine 
$\gamma$ \cite{RF-gam-ca}. Following this paper, we write these modes, which 
are pure ``tree" decays according to the classification of 
Subsection~\ref{ssec:non-lept}, generically as $B_q\to D_q \bar u_q$. 
As can be seen from the Feynman diagrams in Fig.~\ref{fig:BqDquq}, their
characteristic feature is that both a $B^0_q$ and a $\bar B^0_q$ meson may decay 
into the same final state $D_q \bar u_q$. Consequently,  as illustrated in 
Fig.~\ref{fig:BqDquq-int}, interference effects between $B^0_q$--$\bar B^0_q$ 
mixing and decay processes arise, which allow us to probe the weak phase 
$\phi_q+\gamma$ through measurements of the corresponding time-dependent
decay rates. 

In the case of $q=s$, i.e.\ $D_s\in\{D_s^+, D_s^{\ast+}, ...\}$ and 
$u_s\in\{K^+, K^{\ast+}, ...\}$, these interference effects are governed 
by a hadronic parameter $X_s e^{i\delta_s}\propto R_b\approx0.4$, where
$R_b\propto |V_{ub}/V_{cb}|$ is the usual UT side, and hence are large. 
On the other hand, for $q=d$, i.e.\ $D_d\in\{D^+, D^{\ast+}, ...\}$ 
and $u_d\in\{\pi^+, \rho^+, ...\}$, the interference effects are described 
by $X_d e^{i\delta_d}\propto -\lambda^2R_b\approx-0.02$, and hence are tiny. 
In the following, we shall only consider $B_q\to D_q \overline{u}_q$ modes, 
where at least one of the $D_q$, $\bar u_q$ states is a pseudoscalar 
meson; otherwise a complicated angular analysis has to be performed.

The time-dependent rate asymmetries of these decays take the same form
as (\ref{time-dep-CP}).  It is well known that they allow a {\it theoretically 
clean} determination of $\phi_q+\gamma$, where the ``conventional'' 
approach works as follows \cite{BsDsK,BdDpi}: 
if we measure the observables 
$C(B_q\to D_q\bar u_q)\equiv C_q$ 
and $C(B_q\to \bar D_q u_q)\equiv \overline{C}_q$ provided by the
$\cos(\Delta M_qt)$ pieces, we may determine the following quantities:
\begin{equation}\label{Cpm-def}
\hspace*{-0.9truecm}
\langle C_q\rangle_+\equiv
\frac{1}{2}\left[\overline{C}_q+ C_q\right]=0, \quad
\langle C_q\rangle_-\equiv
\frac{1}{2}\left[\overline{C}_q-C_q\right]=\frac{1-X_q^2}{1+X_q^2},
\end{equation}
where $\langle C_q\rangle_-$ allows us to extract $X_q$. However, to this
end we have to resolve terms entering at the $X_q^2$ level. In the case 
of $q=s$, we have $X_s={\cal O}(R_b)$, implying $X_s^2={\cal O}(0.16)$, so 
that this should actually be possible, though challenging. On the other hand, 
$X_d={\cal O}(-\lambda^2R_b)$ is doubly Cabibbo-suppressed. Although it 
should be possible to resolve terms of ${\cal O}(X_d)$, this will be 
impossible for the vanishingly small $X_d^2={\cal O}(0.0004)$ 
terms, so that other approaches to fix $X_d$ are required
\cite{BdDpi}. For the extraction of $\phi_q+\gamma$, the 
mixing-induced observables $S(B_q\to D_q\bar u_q)\equiv S_q$ and 
$S(B_q\to \bar D_q u_q)\equiv \overline{S}_q$ associated with the
$\sin(\Delta M_qt)$ terms of the time-dependent rate asymmetry must be 
measured. In analogy to (\ref{Cpm-def}), it is convenient to
introduce observable combinations $\langle S_q\rangle_\pm$. Assuming 
that $X_q$ is known, we may consider the quantities
\begin{eqnarray}
s_+&\equiv& (-1)^L
\left[\frac{1+X_q^2}{2 X_q}\right]\langle S_q\rangle_+
=+\cos\delta_q\sin(\phi_q+\gamma)\\
s_-&\equiv&(-1)^L
\left[\frac{1+X_q^2}{2X_q}\right]\langle S_q\rangle_-
=-\sin\delta_q\cos(\phi_q+\gamma),
\end{eqnarray}
which yield
\begin{equation}\label{conv-extr}
\sin^2(\phi_q+\gamma)=\frac{1}{2}\left[(1+s_+^2-s_-^2) \pm
\sqrt{(1+s_+^2-s_-^2)^2-4s_+^2}\right],
\end{equation}
implying an eightfold solution for $\phi_q+\gamma$. If we fix the sign of
$\cos\delta_q$ through factorization,  still a fourfold discrete ambiguity is left,
which is limiting the power for the search of NP significantly.  
Note that this assumption allows us also 
to fix the sign of $\sin(\phi_q+\gamma)$ through $\langle S_q\rangle_+$. 
To this end, the factor $(-1)^L$, where $L$ is the $D_q\bar u_q$ 
angular momentum, has to be properly taken into account. 
This is a crucial issue for the extraction of 
the sign of $\sin(\phi_d+\gamma)$ from $B_d\to D^{\ast\pm}\pi^\mp$ decays.

Let us now discuss new strategies to explore CP violation through 
$B_q\to D_q \bar u_q$ modes, following Ref.~\cite{RF-gam-ca}. 
If $\Delta\Gamma_s$ is sizeable, the ``untagged'' 
rates introduced in (\ref{untagged-rate}) allow us to measure 
${\cal A}_{\rm \Delta\Gamma}(B_s\to D_s\bar u_s)
\equiv {\cal A}_{\rm \Delta\Gamma_s}$ and 
${\cal A}_{\rm \Delta\Gamma}(B_s\to \bar D_s u_s)\equiv 
\overline{{\cal A}}_{\rm \Delta\Gamma_s}$. Introducing, in analogy 
to (\ref{Cpm-def}), observable combinations 
$\langle{\cal A}_{\rm \Delta\Gamma_s}\rangle_\pm$, we may derive the relations
\begin{equation}\label{untagged-extr}
\tan(\phi_s+\gamma)=
-\left[\frac{\langle S_s\rangle_+}{\langle{\cal A}_{\rm \Delta\Gamma_s}
\rangle_+}\right]
=+\left[\frac{\langle{\cal A}_{\rm \Delta\Gamma_s}
\rangle_-}{\langle S_s\rangle_-}\right],
\end{equation}
which allow an {\it unambiguous} extraction of $\phi_s+\gamma$ if we fix
the sign of $\cos\delta_q$ through factorization. 
Another important advantage 
of (\ref{untagged-extr}) is that we do {\it not} have to rely on 
${\cal O}(X_s^2)$ terms, as $\langle S_s\rangle_\pm$ and 
$\langle {\cal A}_{\rm \Delta\Gamma_s}\rangle_\pm$ are proportional to $X_s$.
On the other hand, a sizeable value of $\Delta\Gamma_s$ is of course
needed.

If we keep the hadronic quantities $X_q$ and $\delta_q$  
as ``unknown'', free parameters in the expressions for the
$\langle S_q\rangle_\pm$, we may obtain bounds on $\phi_q+\gamma$ from
\begin{equation}
|\sin(\phi_q+\gamma)|\geq|\langle S_q\rangle_+|, \quad
|\cos(\phi_q+\gamma)|\geq|\langle S_q\rangle_-|.
\end{equation}
If $X_q$ is known, stronger constraints are implied by 
\begin{equation}\label{bounds}
|\sin(\phi_q+\gamma)|\geq|s_+|, \quad
|\cos(\phi_q+\gamma)|\geq|s_-|.
\end{equation}
Once $s_+$ and $s_-$ are known, we may of course determine
$\phi_q+\gamma$ through the ``conventional'' approach, using 
(\ref{conv-extr}). However, the bounds following from 
(\ref{bounds}) provide essentially the same information 
and are much simpler to 
implement. Moreover, as discussed in detail in Ref.~\cite{RF-gam-ca}
for several examples within the SM, the bounds following from the $B_s$ and 
$B_d$ modes may be highly complementary, thereby providing particularly 
narrow, theoretically clean ranges for $\gamma$. 

Let us now further exploit the complementarity between the 
$B_s^0\to D_s^{(\ast)+}K^-$ and $B_d^0\to D^{(\ast)+}\pi^-$ processes.
Looking at the corresponding decay topologies, we see that
these channels are related to each other through an interchange of 
all down and strange quarks. Consequently, applying again the $U$-spin 
symmetry implies $a_s=a_d$ and $\delta_s=\delta_d$, where $a_s\equiv X_s/R_b$ 
and $a_d\equiv -X_d/(\lambda^2 R_b)$ are the ratios of the hadronic matrix elements 
entering $X_s$ and $X_d$, respectively. There are various possibilities 
to implement these relations \cite{RF-gam-ca}. A particularly simple
picture arises if we assume that $a_s=a_d$ {\it and} $\delta_s=\delta_d$, 
which yields
\begin{equation}
\tan\gamma=-\left[\frac{\sin\phi_d-S
\sin\phi_s}{\cos\phi_d-S\cos\phi_s}
\right]\stackrel{\phi_s=0^\circ}{=}
-\left[\frac{\sin\phi_d}{\cos\phi_d-S}\right].
\end{equation}
Here we have introduced
\begin{equation}
S\equiv-R\left[\frac{\langle S_d\rangle_+}{\langle S_s\rangle_+}\right]
\end{equation}
with
\begin{equation}
R\equiv\left(\frac{1-\lambda^2}{\lambda^2}\right)
\left[\frac{1}{1+X_s^2}\right],
\end{equation}
where $R$ can be fixed with the help of untagged $B_s$ rates through
\begin{equation}
R=\left(\frac{f_K}{f_\pi}\right)^2 \left[
\frac{\Gamma(\bar B^0_s \to D_s^{(\ast)+}\pi^-)+
\Gamma(B^0_s\to D_s^{(\ast)-}\pi^+)}{\langle\Gamma(B_s\to D_s^{(\ast)+}K^-)
\rangle+\langle\Gamma(B_s\to D_s^{(\ast)-}K^+)\rangle}\right].
\end{equation}
Alternatively, we can {\it only} assume that $\delta_s=\delta_d$ {\it or} 
that $a_s=a_d$ \cite{RF-gam-ca}. An important feature of this strategy
is that it allow us to extract an {\it unambiguous} value of $\gamma$, 
which is crucial for the search of NP; first studies for LHCb are very promising 
in this respect \cite{wilkinson-CKM}.
Another advantage with respect to the ``conventional'' approach is that 
$X_q^2$ terms have not to be resolved experimentally. In 
particular, $X_d$ does {\it not} have to be fixed, and $X_s$ may only enter 
through a $1+X_s^2$ correction, which can straightforwardly be determined 
through untagged $B_s$ rate measurements. In the most refined implementation 
of this strategy, the measurement of $X_d/X_s$ would only be interesting for 
the inclusion of $U$-spin-breaking corrections in $a_d/a_s$. Moreover, we may 
obtain interesting insights into hadron dynamics and $U$-spin breaking. 

The colour-suppressed counterparts
of the $B_q\to D_q \bar u_q$ modes are also interesting
for the exploration of CP violation. 
In the case of the $B_d\to D K_{\rm S(L)}$, $B_s\to D \eta^{(')}, D \phi$, ...\
modes, the interference effects between $B^0_q$--$\bar B^0_q$ mixing
and decay processes are governed by $x_{f_s}e^{i\delta_{f_s}}\propto R_b$.
If we consider the CP eigenstates $D_\pm$ of the neutral $D$-meson system, 
we obtain additional interference effects at the amplitude level, which involve 
$\gamma$, and may introduce the following ``untagged'' rate asymmetry 
\cite{RF-BdDpi0}:
\begin{equation}
\Gamma_{+-}^{f_s}\equiv
\frac{\langle\Gamma(B_q\to D_+ f_s)\rangle-\langle
\Gamma(B_q\to D_- f_s)\rangle}{\langle\Gamma(B_q\to D_+ f_s)\rangle
+\langle\Gamma(B_q\to D_- f_s)\rangle},
\end{equation}
which allows us to constrain $\gamma$ through the relation
\begin{equation}
|\cos\gamma|\geq |\Gamma_{+-}^{f_s}|. 
\end{equation}
Moreover, if we complement
$\Gamma_{+-}^{f_s}$ with 
\begin{equation}
\langle S_{f_s}\rangle_\pm\equiv \frac{1}{2}\left[S_+^{f_s}\pm S_-^{f_s}\right],
\end{equation}
where $S_\pm^{f_s}\equiv {\cal A}_{\rm CP}^{\rm mix}(B_q\to D_\pm f_s)$,
we may derive the following simple but {\it exact} relation:
\begin{equation}
\tan\gamma\cos\phi_q=
\left[\frac{\eta_{f_s} \langle S_{f_s}
\rangle_+}{\Gamma_{+-}^{f_s}}\right]+\left[\eta_{f_s}\langle S_{f_s}\rangle_--
\sin\phi_q\right],
\end{equation}
with $\eta_{f_s}\equiv(-1)^L\eta_{\rm CP}^{f_s}$. This expression allows 
a conceptually simple, theoretically clean and essentially unambiguous 
determination of $\gamma$ \cite{RF-BdDpi0}. Since the interference effects are 
governed by the tiny parameter $x_{f_d}e^{i\delta_{f_d}}\propto -\lambda^2R_b$
in the case of $B_s\to D_\pm K_{\rm S(L)}$, 
$B_d\to D_\pm \pi^0, D_\pm \rho^0, ...$, these modes are not as interesting 
for the extraction of $\gamma$. However, they provide the relation
\begin{equation}
\eta_{f_d}\langle S_{f_d}\rangle_-=\sin\phi_q + {\cal O}(x_{f_d}^2)
=\sin\phi_q + {\cal O}(4\times 10^{-4}),
\end{equation}
allowing very interesting determinations of $\phi_q$ with theoretical 
accuracies one order of magnitude higher than those of
the conventional  $B^0_d\to J/\psi K_{\rm S}$ and $B^0_s\to J/\psi \phi$
approaches~\cite{RF-BdDpi0}. As we pointed out in Subsection~\ref{ssec:BpsiK},
these measurements would be very interesting in view of the new world
average of $(\sin2\beta)_{\psi K_{\rm S}}$.

\subsection{$B^0_s\to K^+K^-$ and $B^0_d\to\pi^+\pi^-$}
\begin{figure}
   \centering
   \includegraphics[width=9.4truecm]{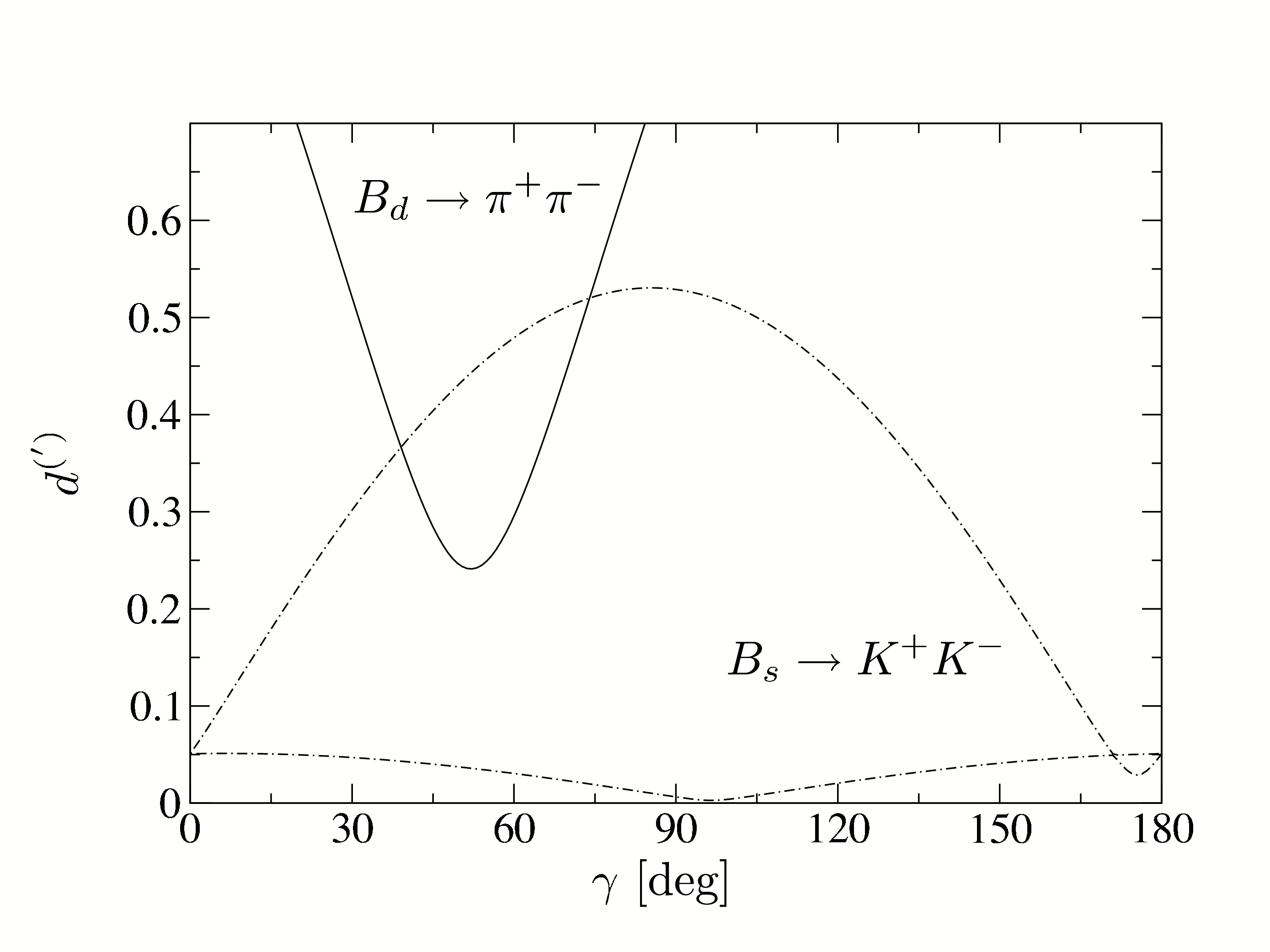} 
   \vspace*{-0.6truecm}
   \caption{The contours in the $\gamma$--$d^{(')}$ plane for an example with
   $d=d'=0.52$, $\theta=\theta'=146^\circ$, $\phi_d=43.4^\circ$, $\phi_s=-2^\circ$,
   $\gamma=74^\circ$, which corresponds to the CP asymmetries
   ${\cal A}_{\rm CP}^{\rm dir}(B_d\to\pi^+\pi^-)=-0.37$ and 
   ${\cal A}_{\rm CP}^{\rm mix}(B_d\to\pi^+\pi^-)=+0.50$
   (see Subsections~\ref{ssec:Bpi+pi-} and \ref{ssec:Bpipi-hadr}), as well as
   ${\cal A}_{\rm CP}^{\rm dir}(B_s\to K^+K^-)=+0.12$ and
   ${\cal A}_{\rm CP}^{\rm mix}(B_s\to K^+K^-)=-0.19$.}\label{fig:Bs-Bd-contours}
\end{figure}

The decay $B^0_s\to K^+K^-$ is a $\bar b \to \bar s$ transition, and
involves tree and penguin amplitudes, as the $B^0_d\to\pi^+\pi^-$ mode 
\cite{RF-BsKK}. However, because of the different CKM structure, the latter 
topologies play actually the dominant r\^ole in the $B^0_s\to K^+K^-$ channel. 
In analogy to (\ref{Bpipi-ampl}), we may write
\begin{equation}\label{BsKK-ampl}
A(B_s^0\to K^+K^-)=\sqrt{\epsilon} \,\, {\cal C}'
\left[e^{i\gamma}+\frac{1}{\epsilon}\,d'e^{i\theta'}\right],
\end{equation}
where $\epsilon$ was introduced in (\ref{eps-def}), and 
the CP-conserving hadronic parameters ${\cal C}'$ and $d'e^{i\theta'}$ 
correspond to ${\cal C}$ and $de^{i\theta}$, respectively. The corresponding
observables take then the following generic form:
\begin{eqnarray}
{\cal A}_{\rm CP}^{\rm dir}(B_s\to K^+K^-)&=&
G_1'(d',\theta';\gamma) \label{CP-BsKK-dir-gen}\\
{\cal A}_{\rm CP}^{\rm mix}(B_s\to K^+K^-)&=&
G_2'(d',\theta';\gamma,\phi_s),\label{CP-BsKK-mix-gen}
\end{eqnarray}
in analogy to the expressions for the CP-violating $B^0_d\to\pi^+\pi^-$
asymmetries in (\ref{CP-Bpipi-dir-gen}) and (\ref{CP-Bpipi-mix-gen}). 
Since $\phi_d=(43.4\pm 2.5)^\circ$ is already known (see
Subsection~\ref{ssec:BpsiK}) and $\phi_s$ is negligibly small
in the SM -- or can be determined through $B^0_s\to J/\psi \phi$ should CP-violating
NP contributions to $B^0_s$--$\bar B^0_s$ mixing make it sizeable -- 
we may convert the measured values of 
${\cal A}_{\rm CP}^{\rm dir}(B_d\to \pi^+\pi^-)$, 
${\cal A}_{\rm CP}^{\rm mix}(B_d\to \pi^+\pi^-)$ and
${\cal A}_{\rm CP}^{\rm dir}(B_s\to K^+K^-)$, 
${\cal A}_{\rm CP}^{\rm mix}(B_s\to K^+K^-)$ into {\it theoretically clean}
contours in the $\gamma$--$d$ and $\gamma$--$d'$ planes, respectively.
In Fig.~\ref{fig:Bs-Bd-contours}, we show these contours for an example,
which corresponds to the central values of (\ref{Bpipi-CP-averages}) and 
(\ref{Bpipi-CP-averages2}) with the hadronic parameters $(d,\theta)$
in (\ref{Bpipi-par-det}). 

As can be seen in Fig.~\ref{fig:Bpipi-diag}, the decay 
$B^0_d\to\pi^+\pi^-$ is actually related to $B^0_s\to K^+K^-$ through the interchange 
of {\it all} down and strange quarks. Consequently, each decay topology contributing
to $B^0_d\to\pi^+\pi^-$ has a counterpart in $B^0_s\to K^+K^-$, and
the corresponding hadronic parameters can be related to each other
with the help of the $U$-spin flavour symmetry of strong interactions,
implying the following relations \cite{RF-BsKK}:
\begin{equation}\label{U-spin-rel}
d'=d, \quad \theta'=\theta.
\end{equation}
Applying the former, we may extract $\gamma$ and $d$ through the 
intersections of the theoretically clean $\gamma$--$d$ and $\gamma$--$d'$ 
contours. As discussed in Ref.~\cite{RF-BsKK}, it is also possible to resolve 
straightforwardly the 
twofold ambiguity for $(\gamma,d)$ arising in Fig.~\ref{fig:Bs-Bd-contours},
thereby leaving us with the ``true" solution of $\gamma=74^\circ$ in
this example. Moreover, we may determine $\theta$ and $\theta'$, which
allow an interesting internal consistency check of the second $U$-spin relation 
in (\ref{U-spin-rel}). An alternative avenue is provided if we eliminate $d$ and 
$d'$ through the CP-violating $B_d\to\pi^+\pi^-$ and $B_s\to K^+K^-$
observables, respectively, and extract then these parameters and $\gamma$ 
through the $U$-spin relation $\theta'=\theta$.

This strategy is very promising from an experimental point of view
for LHCb, where an accuracy for $\gamma$ of a few degrees
can be achieved \cite{LHC-Book,schneider,LHCb-analyses}. As far as 
possible $U$-spin-breaking 
corrections to $d'=d$ are concerned, they enter the determination of $\gamma$ 
through a relative shift of the $\gamma$--$d$ and $\gamma$--$d'$ contours; 
their impact on the extracted value of $\gamma$ therefore depends on the form 
of these curves, which is fixed through the measured observables. In the examples discussed in Refs.~\cite{RF-Phys-Rep,RF-BsKK}, as well as in the one shown in 
Fig.~\ref{fig:Bs-Bd-contours}, the extracted value of $\gamma$ would be very 
stable under such effects. Let us also note that the $U$-spin
relations in (\ref{U-spin-rel}) are particularly robust since they involve only
ratios of hadronic amplitudes, where all $SU(3)$-breaking decay constants
and form factors cancel in factorization and also chirally enhanced terms
would not lead to  $U$-spin-breaking corrections \cite{RF-BsKK}. 
On the other hand, the ratio $|{\cal C}'/{\cal C}|$, which equals 1 in the strict 
$U$-spin limit and enters the $U$-spin relation
\begin{equation}
\frac{{\cal A}_{\rm CP}^{\rm mix}
(B_s\to K^+K^-)}{{\cal A}_{\rm CP}^{\rm dir}(B_d\to\pi^+\pi^-)}=
-\left|\frac{{\cal C}'}{{\cal C}}\right|^2
\left[\frac{\mbox{BR}(B_d\to\pi^+\pi^-)}{\mbox{BR}(B_s\to K^+K^-)}\right]
\frac{\tau_{B_s}}{\tau_{B_d}},
\end{equation}
is affected by $U$-spin-breaking effects within factorization. An 
estimate of the corresponding form factors was recently performed
in Ref.~\cite{KMM} with the help of QCD sum rules, which is an important 
ingredient for a SM prediction of the CP-averaged $B_s\to K^+K^-$ branching 
ratio \cite{BFRS,BFRS-up}, yielding a value in accordance with the first results 
reported by the CDF collaboration \cite{CDF-BsKK}. For other recent analyses
of the $B_s\to K^+K^-$ decay, see Refs.~\cite{safir,BLMV}

In addition to the $B_s\to K^+K^-$, $B_d\to\pi^+\pi^-$ and
$B_s\to D_s^\pm K^\mp$, $B_d\to D^\pm \pi^\mp$ strategies discussed
above, also other $U$-spin methods for the extraction of $\gamma$ were 
proposed, using 
$B_{s(d)}\to J/\psi K_{\rm S}$ or $B_{d(s)}\to D_{d(s)}^+D_{d(s)}^-$ 
\cite{RF-BdsPsiK}, $B_{d(s)}\to K^{0(*)}\bar K^{0(*)}$ \cite{RF-Phys-Rep,RF-ang}, 
$B_{(s)}\to \pi K$ \cite{GR-BspiK}, or $B_{s(d)}\to J/\psi \eta$
modes \cite{skands}. In a very recent paper \cite{SoSu}, also two-body decays 
of charged $B$ mesons were considered.

\begin{figure}[t]
\centerline{
 \includegraphics[width=4.6truecm]{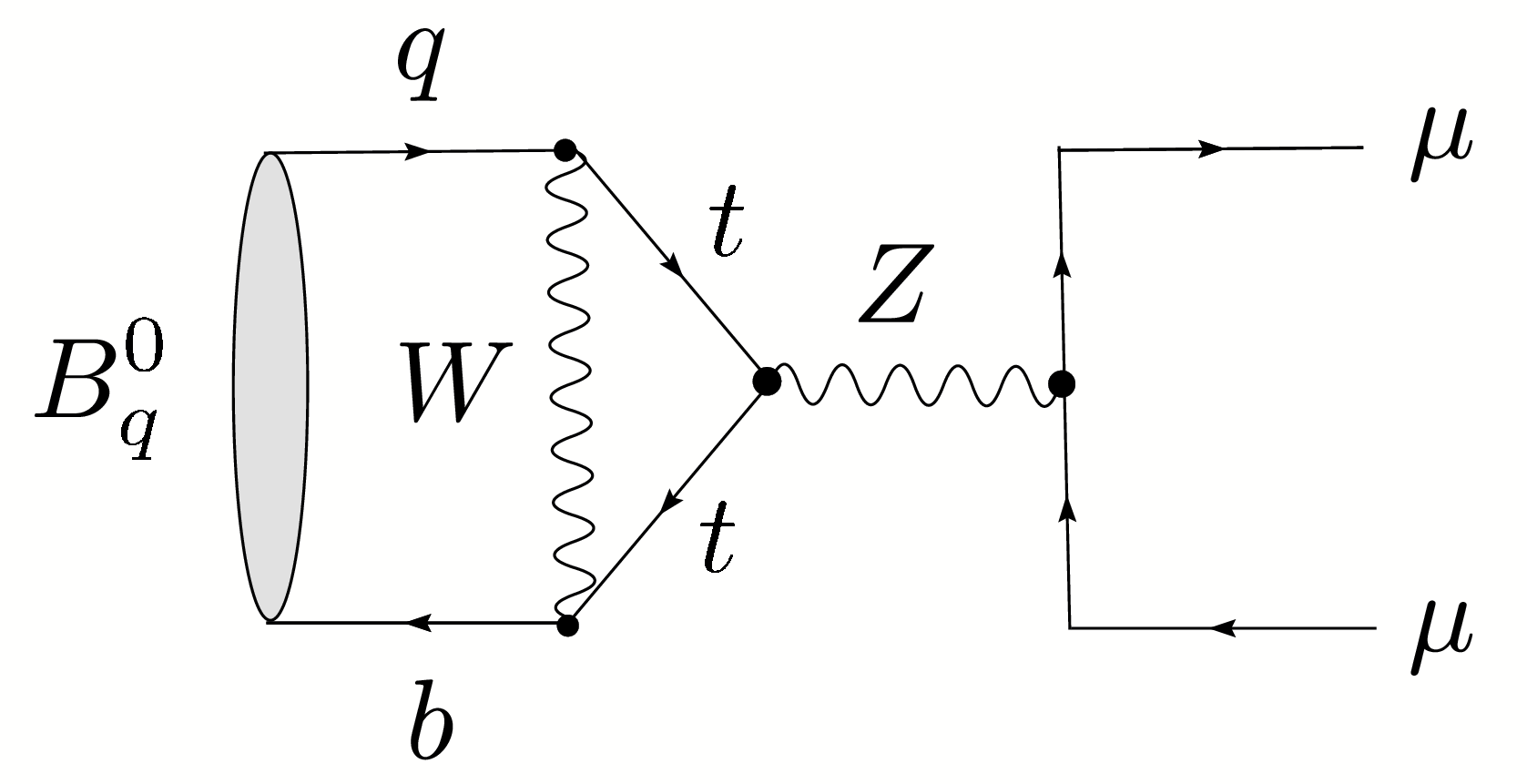}
 \hspace*{0.5truecm}
 \includegraphics[width=4.3truecm]{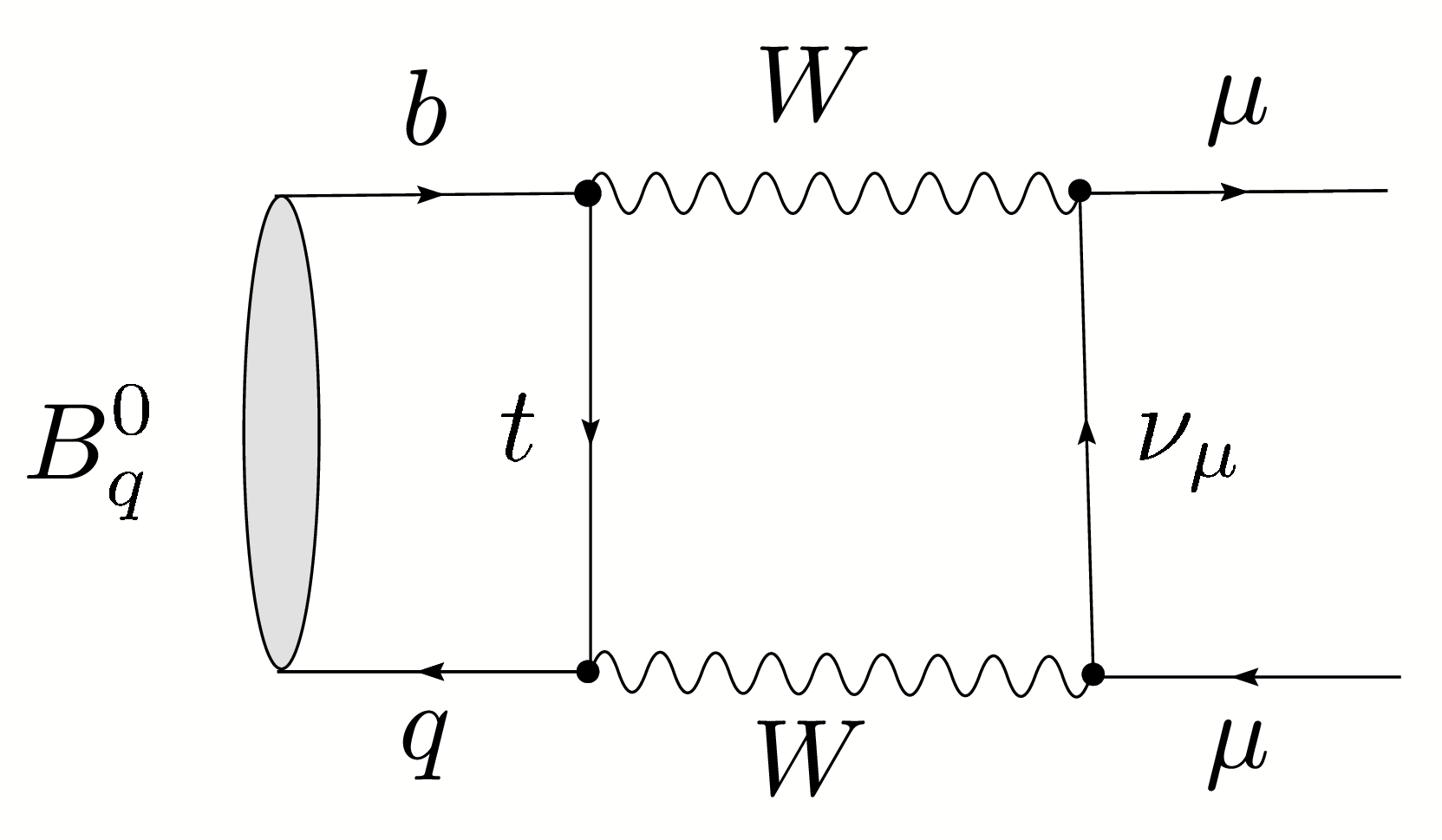}  
 }
\caption{Feynman diagrams contributing to 
$B^0_q\to \mu^+\mu^-$ ($q\in\{s,d\}$).}\label{fig:Bqmumu}
\end{figure}

\subsection{$B^0_s\to\mu^+\mu^-$ and $B^0_d\to\mu^+\mu^-$}\label{ssec:Bmumu}
Let us finally have a closer look at the rare decay $B^0_s\to\mu^+\mu^-$,
which we encountered already briefly in Subsection~\ref{ssec:rareKB}. 
As can be seen in Fig.~\ref{fig:Bqmumu}, this decay and its $B_d$-meson
counterpart $B^0_d\to\mu^+\mu^-$ originate from $Z^0$-penguin and
box diagrams in the SM. The corresponding low-energy effective Hamiltonian 
is given as follows \cite{B-LH98}:
\begin{equation}\label{Heff-Bmumu}
{\cal H}_{\rm eff}=-\frac{G_{\rm F}}{\sqrt{2}}\left[
\frac{\alpha}{2\pi\sin^2\Theta_{\rm W}}\right]
V_{tb}^\ast V_{tq} \eta_Y Y_0(x_t)(\bar b q)_{\rm V-A}(\bar\mu\mu)_{\rm V-A} 
\,+\, {\rm h.c.},
\end{equation}
where $\alpha$ denotes the QED coupling and $\Theta_{\rm W}$ is the
Weinberg angle. The short-distance physics is described by 
$Y(x_t)\equiv\eta_Y Y_0(x_t)$, where $\eta_Y=1.012$ is a perturbative 
QCD correction \cite{BB-Bmumu}--\cite{MiU}, and the Inami--Lim function
$Y_0(x_t)$ describes the top-quark mass dependence. We observe that
only the matrix element $\langle 0| (\bar b q)_{\rm V-A}|B^0_q\rangle$ 
is required. Since here the vector-current piece vanishes, as
the $B^0_q$ is a pseudoscalar meson, this matrix element is simply
given by the decay constant $f_{B_q}$, which is defined in analogy 
to (\ref{decay-const-def}). Consequently, we arrive at a very favourable 
situation with respect to the hadronic matrix elements. Since, moreover, 
NLO QCD corrections were calculated, and long-distance contributions are 
expected to play a negligible r\^ole \cite{BB-Bmumu}, the $B^0_q\to\mu^+\mu^-$ 
modes belong to the cleanest rare $B$ decays. The SM branching ratios
can then be written in the following compact form \cite{Brev01}:
\begin{eqnarray}
\lefteqn{\mbox{BR}( B_s \to \mu^+ \mu^-) = 4.1 \times 10^{-9}}\nonumber\\
&&\qquad\qquad\times \left[\frac{f_{B_s}}{0.24 \, \mbox{GeV}} \right]^2 \left[
\frac{|V_{ts}|}{0.040} \right]^2 \left[
\frac{\tau_{B_s}}{1.5 \, \mbox{ps}} \right] \left[ \frac{m_t}{167 
\, \mbox{GeV} } \right]^{3.12}\label{BR-Bsmumu}\\
\lefteqn{\mbox{BR}(B_d \to \mu^+ \mu^-) = 1.1 \times 10^{-10}}\nonumber\\
&&\qquad\qquad\times\left[ \frac{f_{B_d}}{0.20 \, \mbox{GeV}} \right]^2 \left[
\frac{|V_{td}|}{0.008} \right]^2
\left[ \frac{\tau_{B_d}}{1.5 \, \mbox{ps}} \right] \left[
\frac{m_t}{167 \, \mbox{GeV} } \right]^{3.12}.\label{BR-Bdmumu}
\end{eqnarray}
The most recent upper bounds (90\% C.L.) from CDF read as follows
\cite{CDF-Bmumu}:
\begin{equation}\label{Bmumu-exp-CDF}
\mbox{BR}(B_s\to\mu^+\mu^-)<1.5\times10^{-7}, \quad
\mbox{BR}(B_d\to\mu^+\mu^-)<3.9 \times10^{-8},
\end{equation}
while the D0 collaboration finds the following (95\% C.L.) upper limit \cite{D0-Bmumu}:
\begin{equation}\label{Bmumu-exp-D0}
\mbox{BR}(B_s\to\mu^+\mu^-)<3.7\times10^{-7}.
\end{equation}

Using again relation (\ref{Rt-simple-rel}), we find that the measurement of the 
ratio
\begin{equation}\label{RT1-rare}
\frac{\mbox{BR}(B_d\to\mu^+\mu^-)}{\mbox{BR}(B_s\to\mu^+\mu^-)}=
\left[\frac{\tau_{B_d}}{\tau_{B_s}}\right]
\left[\frac{M_{B_d}}{M_{B_s}}\right]
\left[\frac{f_{B_d}}{f_{B_s}}\right]^2
\left|\frac{V_{td}}{V_{ts}}\right|^2
\end{equation}
would allow an extraction of the UT side $R_t$. Since the short-distance
function $Y$ cancels, this determination does not only work in the SM,
but also in the NP scenarios with MFV \cite{buras-MFV}. This
strategy is complementary to that offered by (\ref{RT2-DM}), using
$\Delta M_d/\Delta M_s$. If we look at
(\ref{RT2-DM}) and  (\ref{RT1-rare}), we see that these expressions 
imply another relation \cite{Buras-rel}:
\begin{equation}\label{Bmumu-DM-rel}
\frac{\mbox{BR}(B_s\to\mu^+\mu^-)}{\mbox{BR}(B_d\to\mu^+\mu^-)}=
\left[\frac{\tau_{B_s}}{\tau_{B_d}}\right]
\left[\frac{\hat B_{B_d}}{\hat B_{B_s}}\right]
\left[\frac{\Delta M_s}{\Delta M_d}\right],
\end{equation}
which holds again in the context of MFV models, including the SM. 
Here the advantage is that the dependence on $(f_{B_d}/f_{B_s})^2$ cancels. 
Moreover, we may also use the (future) experimental data for 
$\Delta M_{(s)d}$ to reduce the hadronic uncertainties 
of the SM predictions of the $B_q\to\mu^+\mu^-$ branching ratios \cite{Buras-rel}:
\begin{eqnarray}
\mbox{BR}(B_s \to \mu^+ \mu^-) &=& (3.42 \pm 0.53)\times
\left[\frac{\Delta M_s}{18.0\, {\rm ps}^{-1}}\right]\times 10^{-9}\\
\mbox{BR}(B_d\to \mu^+ \mu^-) &=& (1.00 \pm 0.14)\times 10^{-10}.
\end{eqnarray}

The current experimental upper bounds in (\ref{Bmumu-exp-CDF}) 
and  (\ref{Bmumu-exp-D0}) are still about two
orders of magnitude away from these numbers. 
Consequently, should the $B_q \to \mu^+ \mu^-$ decays 
be governed by their SM contributions, we could only 
hope to observe them at the LHC \cite{LHC-Book}.
On the other hand, since the $B_q \to \mu^+ \mu^-$ transitions originate from 
FCNC processes, they are sensitive probes of NP. In particular, 
the branching ratios may be dramatically enhanced in specific NP (SUSY) 
scenarios, as was recently reviewed in Ref.~\cite{buras-NP}. Should this 
actually be the case, these decays may already be seen at run II of the 
Tevatron, and the $e^+e^-$ $B$ factories could observe $B_d\to \mu^+ \mu^-$. 
Let us finally emphasize that the experimental bounds on 
$B_s\to\mu^+\mu^-$ can also be converted into bounds on NP parameters
in specific scenarios. In the context of the constrained minimal 
supersymmetric extension 
of the SM (CMSSM) with universal scalar masses, such constraints were
recently critically  discussed by the authors of Ref.~\cite{EOS}.

\section{Conclusions and Outlook}\label{sec:concl}
CP violation is now well established in the $B$-meson system, thereby 
complementing the neutral $K$-meson system, where this phenomenon
was discovered more than 40 years ago. The data of the $e^+e^-$ $B$ factories 
have provided valuable insights into the physics of strong and weak interactions. Concerning the former aspect, which is sometimes only considered as a
by-product, the data give us important evidence for large non-factorizable 
effects in non-leptonic $B$-decays, so that the challenge for a reliable theoretical 
description within dynamical QCD approaches remains, despite interesting 
recent progress. As far as the latter aspect is concerned, the description of
CP violation through the KM mechanism has successfully passed its first
experimental tests, in particular through the comparison between the 
measurement of $\sin 2\beta$ with the help of $B^0_d\to J/\psi K_{\rm S}$ and 
the CKM fits. However, the most recent average for $(\sin2\beta)_{\psi K_{\rm S}}$ 
is now somewhat on the lower side, and there are a couple of puzzles in the
$B$-factory data. It will be very interesting to monitor these effects, which
could be first hints for physics beyond the SM, as the data improve. Moreover,
it is crucial to refine the corresponding theoretical analyses further, to have a 
critical look at the underlying working assumptions and to check them through
independent tests, and to explore correlations with other flavour probes. 

Despite this impressive progress, there are still regions of the 
$B$-physics landscape left that are essentially unexplored. 
For instance, $b\to d$ penguin processes are now entering the 
stage, since lower bounds for the corresponding branching ratios 
that can be derived in the SM turn out to be very close to
the corresponding experimental upper limits. Indeed, we have now
evidence for the $B_d\to K^0\bar K^0$ and $B^\pm\to K^\pm K$ channels,
and the first signals for the radiative $B\to\rho\gamma$ transitions
were recently reported, representing one of the hot topics of this summer. 
These modes have now to be explored in much more detail, and several other
decays are waiting to be observed. 

Moreover, also the $B_s$-meson system, which cannot be studied
with the BaBar and Belle experiments, is still essentially unexplored.
The accurate measurement of the mass difference $\Delta M_s$ is a key element
for the testing of the quark-flavour sector of the SM, and the width difference
$\Delta \Gamma_s$ may be sizeable, thereby offering studies with
``untagged" $B_s$ decay rates. Moreover, the $B_s$-meson
system provides sensitive probes to search for CP-violating NP contributions
to $B^0_s$--$\bar B^0_s$ mixing, allows several determinations the angle 
$\gamma$ of the UT in an essentially unambiguous way, and offers
further tests of the SM through strongly suppressed rare decays. After
new results from run II of the Tevatron, the promising physics potential of
the $B_s$-meson system can be fully exploited at the LHC, in particular by 
the LHCb experiment. 

These studies can nicely be complemented through the kaon system, which 
governed the stage of CP violation for more than 35 years. The future lies now 
on rare decays, in particular on the $K^+\to\pi^+\nu\bar\nu$ and
$K_{\rm L}\to\pi^0\nu\bar\nu$ modes; there is a new proposal to measure 
the former channel at the CERN SPS, and efforts to explore the latter 
at KEK/J-PARC in Japan. Furthermore, flavour physics offers several other
exciting topics. Important examples are top-quark physics, the $D$-meson 
system, the anomalous magnetic moment of the muon, electric dipole moments
and the flavour violation in the charged lepton and neutrino sectors. 

The established neutrino oscillations as well as the evidence for dark matter and the baryon asymmetry of the Universe tell us that the SM is incomplete, and specific 
extensions contain usually also new sources of flavour and CP violation, which 
may manifest themselves at the flavour factories. Fortunately, the LHC is expected 
to go into operation in the autumn of 2007. This new accelerator will provide 
insights  into electroweak symmetry breaking and, hopefully, also give us direct 
evidence for physics beyond the SM through the production and subsequent 
decays of NP particles in the ATLAS and CMS detectors. It is obvious that there 
should be a very fruitful interplay between these ``direct" studies of NP, and the 
``indirect" information provided by flavour physics \cite{workshop}. I have no 
doubt that an exciting future is ahead of us!

\end{document}